\documentclass[]{JFM-FLM_Au}

\usepackage{epstopdf}

\lefttitle{A. V. S. Nath, P. Patra and A. Roy}
\righttitle{Journal of Fluid Mechanics}

\title{Shear-induced self-diffusivity in dilute suspensions with repulsive interactions}

\author{Anu V. S. Nath\aff{1}\footnote{Present address: Laboratoire de Physique, ENS de Lyon and CNRS, Lyon, F-69007 France}, \, Pijush Patra\aff{1}\footnote{Present address: Nordita, KTH Royal Institute of Technology and Stockholm University, Stockholm 10691, Sweden} \, \and Anubhab Roy\aff{1}}
\affiliation{\aff{1}Department of Applied Mechanics and Biomedical Engineering, Indian Institute of Technology Madras,
Chennai 600036, India}
\corresau{Anubhab Roy, \email{anubhab@iitm.ac.in}}

\begin{document}
\maketitle

\begin{abstract}
In a dilute non-Brownian suspension undergoing simple shear, pairwise hydrodynamic interactions are fore–aft symmetric at zero Reynolds number and produce no net cross-streamline displacement. A weak central repulsive force between particles breaks this symmetry, deflecting trajectories and generating irreversible transverse displacements that cumulatively yield a shear-induced self-diffusivity. We derive, via matched asymptotic expansions in the limit of weak repulsion, closed-form scaling laws for the gradient and vorticity components of this diffusivity. The gradient component exhibits a logarithmic enhancement relative to the vorticity component, a structural anisotropy that persists for all monotonically decaying repulsive potentials. The specific interaction enters only through integral functionals of the force profile weighted by hydrodynamic mobility functions, establishing that the scaling is universal across physically distinct mechanisms, such as electrical double-layer repulsion, steric interactions, or any other short-range central force. We validate the asymptotic predictions against full numerical trajectory integration for the representative case of electrostatic repulsion, modelled using the Gouy–Chapman description of the electrical double layer, and find excellent agreement in the expected regime.
\end{abstract}

\begin{keywords}
Authors should not enter keywords on the manuscript, as these must be chosen by the author during the online submission process and will then be added during the typesetting process (see \href{https://www.cambridge.org/core/journals/journal-of-fluid-mechanics/information/list-of-keywords}{Keyword PDF} for the full list).  Other classifications will be added at the same time.
\end{keywords}


\section{Introduction}\label{Introduction}

Shear-induced self-diffusion in non-Brownian suspensions arises from the irreversible lateral displacements that particles experience during hydrodynamic encounters in a sheared flow \citep{davis1996hydrodynamic}. Unlike Brownian diffusion, which originates in thermal fluctuations, this mechanism is purely mechanical and persists even for arbitrarily large particles, provided the suspension is sufficiently concentrated for multi-particle interactions to occur. Understanding its origins and scaling is central to predicting cross-streamline transport in suspension flows, from viscous resuspension \citep{leighton1986viscous} to particle migration in microfluidic devices \citep{lopez2008enhancement}.

In this work, we address the dilute limit, where only pairwise interactions contribute. We derive, for the first time, closed-form asymptotic scaling laws for the shear-induced self-diffusivity arising from a general central repulsive potential between smooth spheres. The analysis reveals that the scaling is universal, identical in structure to that obtained previously for surface roughness \citep{da1996shear} and particle inertia \citep{subramanian2006trajectory}, with the specific interaction entering only through integral functionals of the force profile. This universality, which we establish formally here, unifies several earlier results within a single asymptotic framework.

Shear-induced dispersion is distinct from Brownian diffusion, as it arises from hydrodynamic interactions among particles rather than thermal fluctuations. In a sheared suspension, many-body hydrodynamic interactions lead to irreversible particle displacements that can be modeled as a random walk and characterized by a shear-induced self-diffusivity, $D$. Dimensional analysis suggests that $D$ is proportional to $\dot{\gamma}a^2$, where $\dot{\gamma}$ is the shearing rate and $a$ is the particle radius, with the proportionality constant being a function of the particle volume fraction $\Phi_v$. The pioneering work of \citet{eckstein1977self} marked the first experimental measurements of shear-induced self-diffusivity. They tracked the motion of a radioactively labeled particle in a suspension of otherwise identical particles in a Couette device and observed that $D/(\dot{\gamma}a^2)$ increases approximately linearly from $0$ to $0.02$ as $\Phi_v$ increases from $0$ to $0.2$. Subsequently, \citet{leighton1987measurement} improved this experimental technique and measured the anisotropic components of the self-diffusion tensor, particularly in the velocity-gradient and vorticity directions, in more concentrated suspensions. In a related study, \citet{leighton1987shear} analyzed particle migration in viscometric flows and proposed a diffusion-flux model linking particle transport to gradients in particle concentration and shear rate. 
 Simulation studies investigating shear-induced diffusion in monodispersed suspensions of smooth, non-Brownian spheres have predominantly used Stokesian dynamics and related hydrodynamic simulation techniques to analyze many-body interactions in the Stokes flow regime \citep{bossis1987self,marchioro2001shear,sierou2004shear}. Collectively, these investigations have established several significant findings: a scaling relationship where diffusivity $D$ is proportional to $\dot{\gamma} a^2$; a pronounced dependence of diffusivity on volume fraction; the emergence of anisotropic diffusion in concentrated systems; and the crucial role of near-contact hydrodynamics and microstructural characteristics in determining transport phenomena. For example, it has been verified that, with purely hydrodynamic interactions, symmetry considerations suppress the leading-order contribution from binary-interactions, resulting in an enhanced scaling $D \sim \dot{\gamma}\, a^2\, \Phi_v^2$, due to many-body interactions \citep[see][]{wang1996transverse,drazer2002deterministic}.

Recent computational and experimental studies have demonstrated the significant role of solid-solid contact forces, alongside viscous hydrodynamic interactions, in influencing shear-induced diffusion in concentrated suspensions \citep{zhang2023effect,zhang2024frictional,zhang2026frictional}. Notably, the incorporation of solid-solid contact forces, even in the Stokesian limit (zero Reynolds number limit), successfully addresses critical discrepancies between theoretical insights and empirical observations, including the asymmetry of pair distribution functions \citep{blanc2011experimental} and the loss of reversibility under oscillatory shear \citep{metzger2013irreversibility,pham2015particle}. The findings of these studies provide a clearer understanding of how particle roughness, friction, confinement, and polydispersity collectively govern shear-induced diffusivity in non-Brownian suspensions at moderate to high volume fractions. 

In the current analysis, we focus on dilute suspensions of smooth, rigid spherical particles. In an unbounded sheared suspension, the interactions between two smooth spheres are reversible, provided that inertia (of both fluid and particle) and non-hydrodynamic forces remain negligible. Under these conditions, the particles retrace their original trajectories after each encounter, implying that binary interactions alone do not give rise to diffusive behavior. In contrast, interactions among three or more particles can disrupt this retracing process, preventing the particles from returning to their initial streamlines and, consequently, leading to diffusion. \citet{acrivos1992longitudinal} derived an analytical expression for the self-diffusivity in the flow direction of a simple shear by considering far-field interactions between two pairs of four spheres. \citet{wang1996transverse} calculated the shear-induced self-diffusivity in the lateral directions arising from hydrodynamic interactions among three particles.

Two-particle interactions can exhibit asymmetry in the presence of symmetry-breaking mechanisms, such as surface roughness on the interacting particles \citep[see][]{da1996shear,zarraga2002measurement}, finite particle inertia \citep[see][]{subramanian2006trajectory}, or wall effects in bounded shear flows \citep[see][]{rohatgi2012shear}. Using these mechanisms, earlier studies have estimated the shear-induced self-diffusivity of non-colloidal particles in dilute suspensions. It is also well established that repulsive forces between particles can lead to asymmetric interactions and, consequently, diffusive behavior. These repulsive forces are often electrostatic in nature; for instance, in colloidal suspensions, electrical double-layer repulsion can render particle interactions irreversible, thereby facilitating shear-induced diffusion. In this study, we examine a dilute suspension of rigid, smooth spherical particles subjected to a simple shear flow, with pairwise particle interactions mediated by continuum hydrodynamics and a repulsive central potential. We derive asymptotic expressions for transverse self-diffusivities in the limit where the repulsive forces are much weaker than the flow effects. We then validate our asymptotic results by choosing the repulsive potential to be the electric double-layer repulsion.

The rest of the paper is organized as follows. We describe the problem statement in \S \ref{Problem_formulation}. In \S \ref{inplane-numerics}, we provide a comprehensive derivation of asymptotic expressions for transverse self-diffusivities in the limit of a weak repulsive force. We then validate our asymptotic results for the case of an electric double-layer potential in \S \ref{electrical-potential-results}. Finally, we conclude in \S \ref{conclusion}.

\begin{figure}
\centering
\includegraphics[width=0.5\textwidth]{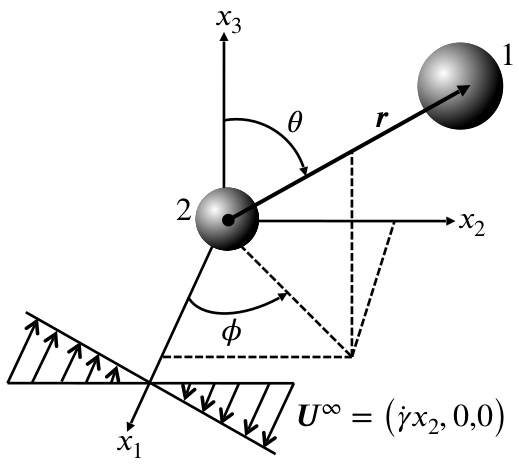}
\caption{The schematic representation of binary interactions. `1' indicates the satellite sphere of radius $a_1$ and `2' indicates the reference  sphere of radius $a_2$.}
\label{Schematic}
\end{figure}
\section{Problem Formulation}\label{Problem_formulation}

We consider a dilute suspension of rigid spherical particles subjected to a simple shear flow, $\textbf{\textit{U}}^\infty = (\dot{\gamma}x_2,0,0)$, with the coordinate axes $x_1$, $x_2$, and $x_3$, correspond to the flow, velocity-gradient, and (negative-)vorticity directions, respectively (see figure \ref{Schematic}). For our analysis, we assume that both the particles and the surrounding fluid have negligible inertia due to the small particle sizes. Additionally, the particles are sufficiently large that we can ignore Brownian diffusion. Given that the Reynolds number based on the particle length scale, defined as $Re_p = \rho_f \dot{\gamma} a^2/\mu_f$ (where $\rho_f$ and $\mu_f$ are the density and dynamic viscosity of the suspending fluid), is sufficiently small, we can accurately describe the motion of the surrounding fluid phase using the Stokes equations for creeping flow. Furthermore, we assume that the particles are neutrally buoyant and that we neglect surface roughness, so that all non-hydrodynamic interactions enter solely through a central repulsive potential.

In a dilute suspension, multi-particle interactions are rare, and thus it is sufficient to consider only pairwise interactions. Consequently, we analyze the relative motion of two particles by tracking the trajectory of a satellite particle (sphere `1' with radius $a_1$) with respect to a reference  particle (sphere `2' with radius $a_2$), as shown in figure \ref{Schematic}. We define the characteristic length, velocity, and time scales as $a^*=(a_1+a_2)/2$, $\dot{\gamma} a^*$, and $\dot{\gamma}^{-1}$, respectively. Two key quantities that describe the relative geometry of the two-sphere system are the size ratio $\beta = a_2/a_1$ and the dimensionless (non-dimensionalised by $a^*$) centre-to-centre distance $r$. This dimensionless radial distance spans the interval from $2$, referred to as the collision sphere, to $\infty$, where one particle exerts no influence on the other. We denote the nondimensional spatial coordinates with an overbar, such that $\overline{x}_1 = x_1/a^*$, $\overline{x}_2 = x_2/a^*$, and $\overline{x}_3 = x_3/a^*$. Although our primary focus will be on monodisperse suspensions, where $a_2 = a_1 = a$ (i.e., $\beta = 1$), the formulation we outline herein remains valid for arbitrary size ratios.

With these assumptions and characteristic scales, the nondimensional relative trajectory equations of a pair of spheres in simple shear flow, interacting through continuum hydrodynamics and a repulsive central potential, are given by \citep{batchelor1972hydrodynamic,wang1994collision}:
\begin{subequations}
\begin{align}
    \frac{dr}{dt} &=  (1-A)\, r \sin^2\theta\, \sin\phi\, \cos\phi + G \, \varepsilon\, F(r)~, \label{r_equation}\\
\frac{d\theta}{dt} &=  (1-B) \sin\theta\, \cos\theta\, \sin\phi\, \cos\phi~, \label{theta_equation}\\
\frac{d\phi}{dt} &=  - \Big[\sin^2\phi + \frac{1}{2}\,B\, \left(\cos^2\phi - \sin^2\phi \right) \Big]~.\label{phi_equation}
\end{align}
\label{gov-eqns}
\end{subequations}
Here $(r,\theta,\phi)$ are the standard spherical coordinates, as shown in figure \ref{Schematic}, and $t$ is the dimensionless time. The non-dimensional repulsion force $F(r)$ acts along the line connecting the centers of the two particles, allowing us to express this force as $F(r)=-dV(r)/dr$, where $V(r)$ is the dimensionless repulsive central potential. The non-dimensional parameter $\varepsilon$ in equation \eqref{r_equation} measures the relative strength of the repulsive force to the background flow. In \S \ref{electrical-potential-results}, we will show the expression for $\varepsilon$ pertinent to the electrical double-layer potential. The mobility functions $A$, $B$, and $G$ describe the hydrodynamic interactions. Specifically, $A$ and $B$ are the radial and tangential mobilities for hydrodynamic interactions between two spherical particles in a linear flow, while $G$ is the radial mobility due to a central potential. These mobilities depend on the size ratio $\beta$ and the dimensionless centre-to-centre distance $r$. The methods for calculating these mobilities, as well as their asymptotic expressions for small and large separation distances, are given in \citet{batchelor1972hydrodynamic,batchelor1976brownian,kim1985resistance,jeffrey1992calculation,wang1994collision}.

The shear-induced self-diffusivity of a tagged particle is obtained by averaging the squared transverse displacements over all upstream pairwise configurations, weighted by the rate at which encounters occur. For a dilute suspension with volume fraction $\Phi_v$, the dimensionless self-diffusivity is \citep{acrivos1992longitudinal,wang1996transverse,da1996shear}
\begin{eqnarray}
            \hat{D}_i = \frac{3}{8\, \pi}\, \int_{-\infty}^{\infty} \int_{-\infty}^{\infty} d\overline{x}_3^{-\infty}\, d\overline{x}_2^{-\infty}\, \lvert \overline{x}_2^{-\infty} \rvert\, \left( \upDelta \overline{x}_i\right)^2~,
        \label{Di-defenition}
\end{eqnarray}
where, $\hat{D}_i$ ($i=2, 3$) denotes the dimensionless self-diffusivity, defined as $\hat{D}_i = D_i/(\dot{\gamma}\, a^{*2}\, \Phi_v)$, with $D_i$ being the corresponding dimensional diffusivity. In the current configuration, $\hat{D}_2$ and $\hat{D}_3$ correspond to the dimensionless diffusivities in the velocity-gradient and vorticity directions, respectively. The quantities $\overline{x}_2^{-\infty}$ and $\overline{x}_3^{-\infty}$ denote the dimensionless coordinates of the pairwise relative particle trajectories far upstream of an encounter, along the velocity-gradient and vorticity directions. The term $\upDelta \overline{x}_i =\overline{x}_i^{+\infty}-\overline{x}_i^{-\infty}$ represents the net displacement in the $i$-th direction, where the superscripts \(+\infty\) and \(-\infty\) signify far downstream and far upstream positions, respectively. In the next section, we will derive the scaling of these diffusivities for weak repulsive interactions using an asymptotic analysis.

\section{Analytical evaluation of the diffusivity}
\label{inplane-numerics}

To evaluate self-diffusivity, we need to comprehensively understand the topology of the relative trajectories and determine how $\upDelta \overline{x}_i$ depends on $\overline{x}_2^{-\infty}$, $\overline{x}_3^{-\infty}$, and $\varepsilon$. This requires solving the governing equations \eqref{gov-eqns}. In this section, we analyze the problem in the weak-interaction limit, $\varepsilon \ll 1$, which allows us to use asymptotic methods. Even in this limit, the relative trajectory topology can be quite complex in three dimensions. Therefore, we first focus on the in-plane case, corresponding to trajectories confined to the shearing plane ($\overline{x}_1$--$\overline{x}_2$). We then use the insights gained from this reduced problem to extend the formulation to fully three-dimensional trajectories later in this section.



\subsection{In-plane analytical solutions for $\varepsilon \ll 1$} \label{inplane-analytical_small_NR}

In the shearing plane, $\theta = \pi/2$ and therefore equation~\eqref{theta_equation} becomes $d\theta/dt = 0$, implying that trajectories remain confined to this plane. Consequently, the governing system reduces to the two remaining evolution equations, \eqref{r_equation} and \eqref{phi_equation}. Since these equations are autonomous, we can combine them to eliminate time, yielding the following differential equation:
\begin{eqnarray}
\frac{d\phi}{dr} = -\frac{\Big[\sin^2\phi + \frac{1}{2} B \left(\cos^2\phi - \sin^2\phi \right) \Big]}{(1-A)\,r\,\sin\phi\,\cos\phi + \varepsilon\, G \, F}~. 
\label{In-plane_trajectory_equation}
\end{eqnarray}



\begin{figure}
    \centering
\includegraphics[width=1\linewidth]{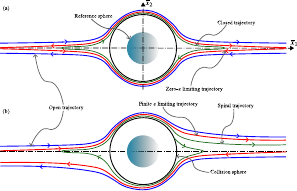}
    \caption{Schematic of in-plane trajectories for (a) $\varepsilon = 0$ and (b) $\varepsilon > 0$ in the presence of a repulsive interaction. Blue curves denote open trajectories, green curves denote closed trajectories in (a) and spiral trajectories in (b), and the red curve represents the limiting trajectory (separatrix) that separates the two types. The sphere at the center represents the reference sphere, and the black circle represents the collision sphere.}
    \label{fig:inplane-schematic3}
\end{figure}
Before solving the equations, it is helpful first to develop a qualitative understanding of the in-plane relative trajectories. Consider the fore–aft symmetric trajectories described by \citet{batchelor1972hydrodynamic} for the case $\varepsilon = 0$. In the absence of any repulsive interaction, the upstream part of the trajectory (as the particles approach the collision sphere) and the downstream part (as they move away from it) are mirror images of each other.
These trajectories can be classified into two types: open and closed (blue and green lines in figure \ref{fig:inplane-schematic3}(a). In the open case, the spheres approach one another, pass by, and then separate to infinity. In the closed case, the spheres relatively follow bounded closed curves and continue to loop around each other indefinitely. The two types are separated by limiting trajectories, known as separatrices, one in the upper half-plane and one in the lower half-plane. These separatrices have zero $\overline{x}_2$ coordinate in the far upstream and downstream. In addition, the trajectories in the upper and lower half-planes are antisymmetric, reflecting the inherent anti-symmetry of the background shear flow.

For any nonzero value of $\varepsilon$, the repulsive interaction breaks the fore--aft symmetry of the trajectories (see figure \ref{fig:inplane-schematic3}(b)). Far upstream and far downstream, the trajectories still approach the streamlines of simple shear flow, since both hydrodynamic and non-hydrodynamic interactions become negligible at large separations. However, due to the repulsive interaction between particles when they are close, the downstream trajectory may lie on a different streamline than the upstream one. The open trajectories can remain open, but they should now show a shift in the gradient direction due to the repulsion between spheres. In particular, the downstream position should be farther from the collision sphere than the upstream position; that is, in the upper (lower) half-plane, the far-downstream gradient
coordinate $\overline{x}_2^{+\infty}$ of the open trajectories becomes larger (smaller) than its far-upstream counterpart $\overline{x}_2^{-\infty}$. The closed trajectories that exist when $\varepsilon = 0$ can no longer remain closed in the presence of repulsion. Instead, they should open up into spiral trajectories. Even a weak repulsive interaction is expected to destabilize these trajectories, forming spirals that start near the sphere and extend to infinity downstream. As a result, no truly closed trajectories are expected when repulsion is present.
These spiral trajectories cannot have corresponding upstream counterparts and therefore do not contribute to the self-diffusion of a tagged sphere. For this reason, we focus only on the open trajectories, which determine the self-diffusion.
The separatrices that divide the open and spiral trajectories may still exist, but they lose their fore--aft symmetry and shift away from the sphere in the downstream direction. Far-upstream, they still have a zero $\overline{x}_2$ coordinate, but far-downstream, this coordinate takes a finite positive (negative) value in the upper (lower) half-plane. Finally, the repulsive interaction does not affect the anti-symmetry between the upper and lower half-planes. This can be seen by replacing $\phi \to \phi + \pi$, which leaves the governing equation \eqref{In-plane_trajectory_equation} unchanged. Therefore, without loss of generality, we will be dealing with trajectories in the upper-half plane in the following derivations. These expectations will be confirmed later using computed trajectories for the specific case of a repulsive electrical double-layer potential (see \S~\ref{electrical-potential-results}).

For $\varepsilon \ll 1$, we perform a regular perturbation expansion of the variable $\phi$ in powers of $\varepsilon$ as $\phi = \phi_0 + \varepsilon\,  \phi_1 + \mathcal{O}(\varepsilon^2) $. Substituting this expansion into equation~(\ref{In-plane_trajectory_equation}), we obtain the governing equation accurate up to $\mathcal{O}(\varepsilon)$ as 
\begin{subequations}
    \begin{align}
    \mathcal{O}(1) :  \frac{d\phi_0}{dr} &= -\frac{\Big[\sin^2\phi_0 + \frac{1}{2} B \left(\cos^2\phi_0 - \sin^2\phi_0 \right) \Big]}{(1-A)\,r\,\sin\phi_0\,\cos\phi_0}~, \label{O(1)_equation_for_phi}\\
\mathcal{O}(\varepsilon) :  \frac{d\phi_1}{dr} &+ \Big\{\frac{2(1-B)}{r(1-A)}+\frac{(1-\cot^2\phi_0)\left(\sin^2\phi_0 + \frac{1}{2} B \left(\cos^2\phi_0 - \sin^2\phi_0 \right)\right)}{r(1-A)\cos^2\phi_0}\Big\}\phi_1 \nonumber \\ &= \frac{\left(\sin^2\phi_0 + \frac{1}{2} B \left(\cos^2\phi_0 - \sin^2\phi_0 \right)\right)\, G\, F}{r^2\,(1-A)^2\,\sin^2\phi_0\,\cos^2\phi_0}~. \label{O(NR)_equation_for_phi}
\end{align}
\end{subequations}
Because of the anti-symmetry between the upper and lower halves of the $\overline{x}_1$--$\overline{x}_2$ plane, it is sufficient to restrict our analysis to the trajectories in the upper half-plane. To solve the governing equations, we employ a perturbative approach: $\phi_0(r)$ represents the fore-aft symmetric trajectory corresponding to $\varepsilon = 0$, while $\phi_1(r)$ denotes the first-order correction arising from a small but finite $\varepsilon$. 
\begin{figure}
    \centering
\includegraphics[width=1\linewidth]{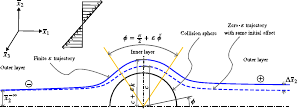}
    \caption{Schematic of in-plane open trajectories in the upper half-plane for $\varepsilon = 0$ (blue, dashed) and $\varepsilon > 0$ (blue, solid), both originating with the same $\mathcal{O}(1)$ upstream offset $\overline{x}_2^{-\infty}$. In the far-downstream, the $\varepsilon>0$ trajectory is displaced relative to the zero-$\varepsilon$ trajectory by an amount $\upDelta \overline{x}_2$.}
    \label{fig:inplane-schematic1}
\end{figure}
The situation can be visualized as in the schematic in figure~\ref{fig:inplane-schematic1}, where two trajectories are compared: one for $\varepsilon = 0$ (dashed) and another for $0<\varepsilon \ll 1$ (solid). Without loss of generality, both are assumed to originate from the same far-upstream lateral position. The boundary condition is thus specified as $r\sin\phi \rightarrow \overline{x}_2^{-\infty}$ when $\overline{x}_1^{-\infty} \to -\infty$, ensuring a common upstream offset for the zero and finite $\varepsilon$ trajectories. At each order, the boundary condition becomes: as $\phi_0 \to \pi$ and $r \to \infty$
\begin{subequations}
\begin{align}
\mathcal{O}(1) &:  r\, \sin\phi_0 \to \overline{x}_2^{-\infty}~, \label{O(1)_boundary_condition_for_phi} \\
\mathcal{O}(\varepsilon) &:  r\, \phi_1^- \to 0~.\label{O(NR)_boundary_condition_for_phi}
\end{align}
\end{subequations}
We denote the branch of $\phi$ within the interval $(\pi/2,\pi)$ by the superscript `-', and the corresponding branch within $(0,\pi/2)$ by the superscript `+'. As discussed earlier, the asymmetry of the finite--$\varepsilon$ open trajectories in the shearing plane can be quantified through their net lateral displacement in the velocity-gradient direction. 
The lateral displacement in the gradient direction, denoted by $(\upDelta \overline{x}_2)_\textrm{ip}$, is thus expressed as
\begin{eqnarray}
(\upDelta \overline{x}_2)_\textrm{ip} &=&  r\sin\phi \big|_{\overline{x}_1=-\infty}^{\overline{x}_1=+\infty} \nonumber \\ &=&  r\sin (\phi_0 + \varepsilon \, \phi_1)\big|_{\overline{x}_1=-\infty}^{\overline{x}_1=+\infty} \nonumber \\ &=&  \varepsilon\, \left\{\left(r\phi_1\cos\phi_0\right)_{\phi_0 \rightarrow 0, r \rightarrow \infty} - \left(r\phi_1\cos\phi_0\right)_{\phi_0 \rightarrow \pi, r \rightarrow \infty}\right\} \nonumber \\
&=&  \varepsilon \lim_{r \rightarrow \infty} r\phi_1^+. \label{In-plane_displacement}
\end{eqnarray}
Note that we have used the fact that $\sin \phi_0$ takes the same value in both the positive and negative branches because of the fore–aft symmetry of the trajectories when $\varepsilon = 0$. In addition, $r\phi_1^{-} = 0$, since the zero--$\varepsilon$ trajectory with the same upstream gradient offset is taken as the reference (base) trajectory for calculating the gradient displacement.
As discussed earlier, if the downstream branch of the trajectory shows a net shift in the positive gradient direction relative to the upstream branch, then the gradient displacement $(\upDelta \overline{x}_2)_\textrm{ip}$ should be positive for finite $(\upDelta \overline{x}_2)_\textrm{ip}$ in the upper half-plane. Here, the subscript ``$\textrm{ip}$'' represents `in-plane' quantities.


The regular perturbation solution should provide an adequate approximation in most of the domain. However, from equation~\eqref{In-plane_trajectory_equation}, it is evident that when $\phi \to \pi/2$, the first term in the denominator vanishes, rendering the equation singular in the $\varepsilon \ll 1$ regime. Similarly, equation~\eqref{O(1)_equation_for_phi} becomes singular when $\phi_0 \to \pi/2$, i.e., in the vicinity of the collision sphere. This singularity does not arise as $\phi_0 \to 0$ or $\phi_0 \to \pi$, because in those cases the trajectory is infinitely far from sphere ($r \to \infty$), and the products $r\,\sin \phi_0 \to \overline{x}_2^{-\infty}$ and $r\,\sin \phi_0 \to \overline{x}_2^{+\infty}$ remain finite in general. In contrast, for $\phi_0 \to \pi/2$, $r$ in the trajectory remains of order one, and thus $r\,\sin \phi_0 \to 0$, leading to a true singularity. This indicates the presence of a boundary layer, or a region near the collision sphere, where the regular perturbation approach fails and a singular perturbation analysis is required. Consequently, we divide the domain into an `outer layer'--far from the collision sphere, where the regular perturbation remains valid, and an `inner layer'--close to the collision sphere, requiring a singular perturbation treatment. 
A subsequent matching procedure can then be used to determine the unknown parameters that connect the solutions in these two regions.

\subsubsection{Outer layer solution}
Following the approach of \citet{batchelor1972hydrodynamic}, the solution to the $\mathcal{O}(1)$ governing equation~\eqref{O(1)_equation_for_phi} along with the corresponding boundary condition~\eqref{O(1)_boundary_condition_for_phi}, can be expressed as
\begin{eqnarray}
r^2\, \sin^2\phi_0 = e^{Q(r)}\, \left[\left(\overline{x}_2^{-\infty}\right)^2 + \int_r^{\infty}e^{-Q(r')}\, \frac{B'\, r'}{\left(1-A'\right)}dr'\right], \label{O(1)_solution_for_phi}
\end{eqnarray}
where, the prime ($'$) indicates that the function is evaluated at the dummy integration variable $r’$. We also define
$Q(r) := \int_r^{\infty}q(r')\,dr'$, where $q(r) := 2\,(A-B)/(r\,(1-A))$. This solution is valid both upstream and downstream, provided that $\phi_0$ is appropriately chosen to lie within its relevant range when inverting. The $\mathcal{O}(\varepsilon)$ governing equation~\eqref{O(NR)_equation_for_phi} can be rearranged as
\begin{eqnarray}
    \frac{d}{dr}\left(r\,\phi_1 \right) + \left[\frac{\sin^2 \phi_0-\frac{1}{2}\, B}{r\,(1-A)\, \sin^2 \phi_0\, \cos^2 \phi_0 }-\frac{1}{r}\right]\, r\, \phi_1  = \frac{\left[(1-B)\,\sin^2\phi_0 + \frac{1}{2}\, B\right]\, G\, F}{r\,(1-A)^2\, \sin^2\phi_0\, \cos^2\phi_0}~,
    \label{O(NR)_equation_for_phi_ver2}
\end{eqnarray}
which is a first order ordinary differential equation for $r\,\phi_1$, can be easily solved using the $\mathcal{O}(\varepsilon)$ boundary condition in~\eqref{O(NR)_boundary_condition_for_phi} to obtain the upstream solution as
\begin{eqnarray}
r\, \phi_1^- = -\int_r^{\infty}\exp\Bigg[-\int_{r'}^r\left\{\frac{\sin^2\phi_0''-\frac{1}{2}\, B''}{r''\, \left(1-A''\right)\,\sin^2\phi_0''\,\cos^2\phi_0''}-\frac{1}{r''}\right\}dr''\Bigg] \nonumber \\
\times \frac{ \left(\left(1-B'\right)\, \sin^2\phi_0' + \frac{1}{2}\, B'\right)\, G'\,F'}{r'\, (1-A')^2\,\sin^2\phi_0'\, \cos^2\phi_0'}dr',
\label{upstream_solution_for_phi_1_ver1}
\end{eqnarray}
where, the double-prime ($''$) also indicates that the function is evaluated at the dummy integration variable $r''$. In equation~\eqref{upstream_solution_for_phi_1_ver1}, the integrand term outside the exponential, which involves $G'\, F'$, contains $\cos^2 \phi_0$ in the denominator, becoming singular as $\phi_0 \to \pi/2$. From the solution in equation~\eqref{O(1)_solution_for_phi}, it can be inferred that $\cos \phi_0 \sim (r-c)^{1/2}$, where $c$ represents the distance of closest approach for the zero--$\varepsilon$ trajectory, i.e., the value of $r$ at $\phi_0 = \pi/2$. An implicit expression for $c$ can also be obtained from equation~\eqref{O(1)_solution_for_phi} as
\begin{eqnarray}
c^2 = e^{Q(c)}\,\left[\left(\overline{x}_2^{-\infty}\right)^2 + \int_c^{\infty}e^{-Q'}\,\frac{B'\, r'}{\left(1-A'\right)}dr'\right]~,
\label{expression_for_c}
\end{eqnarray}
where, we denote $Q(r')$ simply as $Q'$. It is important to note that $c$ depends on the upstream lateral offset, $\overline{x}_2^{-\infty}$. The singularity arising from the scaling $\cos^2 \phi_0 \sim (r-c)$ causes the outer integral in equation~\eqref{upstream_solution_for_phi_1_ver1} to diverge logarithmically. However, a similar (but negative) logarithmic singularity also appears in the integral inside the exponential, which likewise contains $\cos^2 \phi_0$ in the denominator. This observation allows the integral to be simplified, ultimately yielding
\begin{eqnarray}
r\,\phi_1^- = -\frac{e^{Q(r)}}{r\,\sin\phi_0\, \cos\phi_0}\int_r^{\infty}e^{-Q'}\, \frac{ \left(\left(1-B'\right)\sin^2\phi_0' + \frac{1}{2}\, B'\right)\, G'\, F'}{(1-A')^2\, \sin\phi_0'\, \cos\phi_0'}dr'~, \label{upstream_solution_for_phi_1_ver2}
\end{eqnarray}
where, $\phi_0 = \phi_0(r)$ and $\phi_0' = \phi_0(r')$. In this form, the previously divergent singularity has been transferred to the prefactor, while the integral itself is now convergent. The $\cos \phi_0'$ term in the denominator of the integrand produces only a square-root singularity, which is integrable. Although the solution is expressed for the negative branch ($\phi^-$), it is structured to be computed using $\phi_0$ and $\phi_0'$ from the interval $(0, \pi/2)$, i.e., from the positive branch; the appropriate sign changes are already accounted for in \eqref{upstream_solution_for_phi_1_ver2}. For brevity, we define the following integral:
\begin{eqnarray}
K_I(r) := \int_r^{\infty}e^{-Q'}\, \frac{ \left(\left(1-B'\right)\sin^2\phi_0' + \frac{1}{2}\, B'\right)\, G'\, F'}{(1-A')^2\, \sin\phi_0'\, \cos\phi_0'}\,dr'~, \label{eqn-KIintegral}
\end{eqnarray}
with $\phi_0' \in (0,\pi/2)$, so that, the solution in equation \eqref{upstream_solution_for_phi_1_ver2} can be rewritten simply as
\begin{eqnarray}
    \phi_1^{-} = -\frac{2\, K_I(r)\, e^{Q(r)}}{r^2\, \sin (2\, \phi_0^+)}~.
    \label{eq-soln-phi1m}
\end{eqnarray}
Similarly, by solving equation~\eqref{O(NR)_equation_for_phi_ver2} over the downstream branch, we obtain an analogous expression for $\phi_1^{+}$; however, in this case, the integration constant is nonzero, leading to the following solution as
\begin{eqnarray}
    \phi_1^{+} = \frac{2\, e^{Q(r)}}{r^2\, \sin (2\, \phi_0^+)}\, \left[\overline{x}_2^{-\infty}\, I_{\phi_1}^{+}-K_I(r)\right]~,
    \label{eq-soln-phi1p}
\end{eqnarray}
where the integration constant $I_{\phi_1}^{+}$ must be determined and is directly related to the gradient displacement of the trajectory. For $\varepsilon = 0$, this constant vanishes, i.e., $I_{\phi_1}^{+} = 0$. However, for non-zero $\varepsilon$, the downstream branch of the trajectory corresponds to a perturbed path that deviates from the downstream part of the zero--$\varepsilon$ trajectory. Consequently, a net lateral offset arises between the upstream and downstream positions in the gradient direction. Having determined the outer-layer solutions, we now turn to the inner-layer analysis to obtain the additional information required to evaluate the unknown integration constant. This analysis is presented in the following subsection.
\subsubsection{Inner layer solution}
\label{Inner-layer-solution}
Since the singular behavior in the inner layer occurs when $\phi \sim \pi/2$ and $r \sim c$, we introduce a suitable rescaling of variables to regularize the problem. Defining the inner-layer variables as $\phi^{I} = \pi/2 + \varepsilon\,\tilde{\phi}$ and $r = c + \varepsilon\,k + \varepsilon^2\,\tilde{r}$ transforms the governing equation~\eqref{In-plane_trajectory_equation} in terms of the rescaled variables as
\begin{eqnarray}
    \frac{d \tilde{\phi}}{d \tilde{r}} = \frac{1-\frac{1}{2}\, B_0}{c\, (1-A_0)\, \tilde{\phi}-G_0\, F_0}~,
    \label{inner-layer-gvernong-eqn}
\end{eqnarray}
Here, $A_0$, $B_0$, $G_0$, and $F_0$ denote the values of the functions $A(r)$, $B(r)$, $G(r)$, and $F(r)$, respectively, evaluated at $r = c$. For this inner-layer scaling to remain consistent, the parameter $k$ must be treated as a constant like $c$; in contrast, $\tilde{r}$ is a function of $\tilde{\phi}$. The term $k$ appears just as a first-order correction in the expansion of $r$ in the inner region. Although it does not directly enter into the inner-layer governing equation or its solution, it will play an important role in matching the inner and outer solutions. Specifically, the inclusion of the matching term $\varepsilon\, k$ ensures that the gradient offset between the upstream and downstream trajectories remains at least of order $\mathcal{O}(\varepsilon)$. This can also be interpreted geometrically: the zero--$\varepsilon$ upstream reference trajectory (shown as the blue curve in figure~\ref{fig:inplane-schematic1}) approaches the collision sphere to a minimum separation distance $c$. In contrast, the actual trajectory for a finite $\varepsilon$ (shown as the red curve in figure~\ref{fig:inplane-schematic1}) attains a slightly larger closest approach distance of $c + \varepsilon\,k$, where $k>0$, as illustrated in the figure. The value of $k$ must therefore be determined to quantify the resulting gradient offset in the trajectory. Solving equation~\eqref{inner-layer-gvernong-eqn} gives,
\begin{eqnarray}
    \tilde{\phi}^{\mp} = \frac{G_0\, F_0 \pm \sqrt{G_0^2\, F_0^2 + c\, (1-A_0)\, (2-B_0)\, (\tilde{r}-I_i   )}}{c\, (1-A_0)}~,
    \label{eqn-inner-layer-soln}
\end{eqnarray}
where the correspondence between the branch of the solution and the sign of the square-root term is determined by the interval to which $\tilde{\phi}$ belongs. Specifically, $\tilde{\phi} > 0$ implies $\phi > \pi/2$, corresponding to the $-$ve branch, whereas $\tilde{\phi} < 0$ corresponds to the $+$ve branch. The term $I_i$ denotes another integration constant, which, as we will see later, does not play a role in the matching. With both the inner and outer layer solutions now determined, we proceed in the following subsection to match them, thereby obtaining the values of the unknown parameters.
\subsubsection{Intermediate matching}
In this section, we employ an intermediate matching approach to connect the inner and outer layer solutions. We introduce an intermediate, dependent-variable in the general form $\phi = \pi/2 + \varepsilon^\alpha \, \hat{\phi}$, which, according to the governing equation~\eqref{In-plane_trajectory_equation}, requires the independent variable to scale as $r = c + \varepsilon^{2\alpha} \, \hat{r}$, with $0 < \alpha \leq 1$. The goal is to express both the outer and inner solutions up to $\mathcal{O}(\varepsilon)$ in terms of these intermediate variables and then match them to determine the unknown parameters. For the outer layer, we first rewrite the leading-order term $\phi_0$ using the intermediate variables as $\phi_0 = \pi/2 + \varepsilon^\alpha \, \hat{\phi}_0(\hat{r})$, and then expand the outer solution~\eqref{O(1)_solution_for_phi} for $\varepsilon  \to 0$ to obtain
\begin{eqnarray}
        \hat{\phi}_0^{\mp} = \pm \sqrt{\frac{\hat{r}\, (2-B_0)}{c\, (1-A_0)}}~, \,\,\, \textrm{and thus} \,\,\, \lim_{r \to c+\hat{r}\, \varepsilon^{2\alpha}} \phi_0^{\mp} = \frac{\pi}{2} \pm \varepsilon^{\alpha}\, \sqrt{\frac{\hat{r}\, (2-B_0)}{c\, (1-A_0)}}+h.o.t.~,
        \label{matching-variable-soln-zeroth-order-inplane}
\end{eqnarray}
where, $h.o.t.$ is a shorthand for `higher-order terms'. Similarly, by applying the same scaling to equations~\eqref{eq-soln-phi1m} and \eqref{eq-soln-phi1p}, we can express the first-order correction terms for both the negative and positive branches as
\begin{subequations}
    \begin{align}
   \lim_{r \to c+\hat{r}\, \varepsilon^{2\alpha}} \varepsilon\, \phi_1^{-} &= -\frac{\varepsilon^{1-\alpha}\, e^{Q(c)}}{c\, \sqrt{c\, \hat{r}}}\, \sqrt{\frac{1-A_0}{2-B_0}}\, K_I(c)+\frac{\varepsilon\, G_0\, F_0}{\textcolor{white}{2}\, c\, (1-A_0)}+o(\varepsilon)~, \label{matching-variable-soln-first-order-inplane-negative}\\
   \lim_{r \to c+\hat{r}\, \varepsilon^{2\alpha}} \varepsilon\, \phi_1^{+} &= \frac{\varepsilon^{1-\alpha}\, e^{Q(c)}}{c\, \sqrt{c\, \hat{r}}}\, \sqrt{\frac{1-A_0}{2-B_0}}\, \left[ I_{\phi_1}^{+}\, \overline{x}_2^{-\infty}-K_I(c)\right]+\frac{\varepsilon\, G_0\, F_0}{\textcolor{white}{2}\, c\, (1-A_0)}+o(\varepsilon)~. \label{matching-variable-soln-first-order-inplane-positive}
\end{align}
\label{matching-variable-soln-first-order-inplane}
\end{subequations}
Thus, by combining the leading-order solution~\eqref{matching-variable-soln-zeroth-order-inplane} with the first-order correction~\eqref{matching-variable-soln-first-order-inplane}, the outer layer solution for each branch, expressed in terms of the intermediate matching variable, can be written as $ \lim_{r \to c+\hat{r}\, \varepsilon^{2\alpha}} \phi^{\mp}_\textrm{outer} = \lim_{r \to c+\hat{r}\, \varepsilon^{2\alpha}} \left( \phi_0^{\mp}+\varepsilon\, \phi_1^{\mp}\right)$. 

In the inner layer, we equate the two expressions for $r$, namely $r = c+\varepsilon\, k + \varepsilon^2 \, \tilde{r}$ and $r = c+\hat{r}\, \varepsilon^{2\alpha}$, so that to get the inner layer independent variable in terms of the matching variable as $\tilde{r} = \hat{r}\, \varepsilon^{2\alpha-2}-k\, \varepsilon^{-1}$. Substituting this relation into the inner layer solution for $\phi$ from equation~\eqref{eqn-inner-layer-soln}, multiplying with $\varepsilon$ and taking the limit $\varepsilon \to 0$, we obtain
\begin{eqnarray}
    \lim_{r \to c+\hat{r}\, \varepsilon^{2\alpha}}\varepsilon\, \tilde{\phi}^{\mp} = \lim_{\varepsilon \to 0}\frac{\varepsilon\, G_0\, F_0}{c\, (1-A_0)}\, \left\{1\pm\sqrt{1+\frac{c\, (1-A_0)\, (2-B_0)\, (\hat{r}\, \varepsilon^{2\alpha-2}-k\, \varepsilon^{-1}-I_i)}{G_0^2\, F_0^2}} \right\}~.
    \label{eqn-inner-layer-matching-ver1}
\end{eqnarray}
Note that in the limit $\varepsilon \to 0$, only when $\alpha < 1/2$ does the term $\varepsilon^{2\alpha-2} \gg \varepsilon^{-1}$, and the expansion of equation~\eqref{eqn-inner-layer-matching-ver1} remains compatible with the outer layer expansion. Therefore, for $\alpha<1/2$, the inner layer solution $\phi^{\mp}_\textrm{inner} = \pi/2+\varepsilon\, \tilde{\phi}^{\mp}$ expressed in terms of the intermediate matching variables becomes
\begin{eqnarray}
    \lim_{r \to c+\hat{r}\, \varepsilon^{2\alpha}}\phi^{\mp}_\textrm{inner} = \frac{\pi}{2} \pm \varepsilon^\alpha\, \sqrt{\frac{\hat{r}\, (2-B_0)}{c\, (1-A_0)}} \mp \varepsilon^{1-\alpha}\,\frac{k}{2}\, \sqrt{\frac{2-B_0}{\hat{r}\, c\, (1-A_0)}}+\frac{\varepsilon\, G_0\, F_0}{c\, (1-A_0)}+h.o.t~.
    \label{inner-soln-matching-expansion}
\end{eqnarray}
Note that, as mentioned earlier, this expansion is independent of the order-one integration constant $I_i$. By comparing (or matching) the inner layer solution for the upstream -ve branch from equation~\eqref{inner-soln-matching-expansion} with the corresponding outer layer expansion given in equation~\eqref{matching-variable-soln-first-order-inplane-negative}, we obtain the constant $k$ as 
\begin{eqnarray}
    k = k(c) = \frac{2\, (1-A_0)}{c\, (2-B_0)}\, K_I(c)\, e^{Q(c)}~.
    \label{expression-for-k}
\end{eqnarray}
The above expression indicates that $k$ is a function of $c$, which in turn depends on $\overline{x}_2^{-\infty}$ through equation \eqref{expression_for_c}. In other words, although $k$ is constant for a given trajectory, its value differs from one trajectory to another; the same holds for $K_I(c)$ and $Q(c)$ as well. Similarly, by matching the inner and outer layer solutions for the downstream +ve branch, using equations~\eqref{inner-soln-matching-expansion} and~\eqref{matching-variable-soln-first-order-inplane-positive}, we obtain the integration constant $I_{\phi_1}^{+} = 2\, K_I(c)/\overline{x}_2^{-\infty}$.
Now, using the definition in equation~\eqref{In-plane_displacement}, along with the solution given in \eqref{eq-soln-phi1p}, the asymptotic form of in-plane gradient displacement can be obtained as
\begin{eqnarray}
    (\upDelta \overline{x}_2)_\textrm{ip} = \varepsilon \lim_{r \rightarrow \infty} r\phi_1^+ = \varepsilon\, I_{\phi_1}^{+} = \frac{2\, \varepsilon \, K_I(c; \overline{x}_2^{-\infty})}{\overline{x}_2^{-\infty}}~,
    \label{inplane-gradient-displacement-ver1}
\end{eqnarray}
The above formula is accurate up to errors of $\mathcal{O}(\varepsilon^2)$. One can show that, as expected, this predicted in-plane gradient displacement yields a positive (negative) value in the upper (lower) half-plane, when one accounts for the fact that $\varepsilon >0$ and \eqref{eqn-KIintegral} has an integrand that evaluates to positive in the desired domain. A direct numerical integration confirms this later for the specific case of double-layer repulsive potential in \S\S~\ref{inplane-numeric-eval}.

Note that this expression for gradient displacement is suited only for relatively large values of gradient offset $\overline{x}_2^{-\infty}$. To show this explicitly, let us assume $\overline{x}_2^{-\infty}$ scales as $\varepsilon^{\zeta}$ for some positive number $\zeta$; then from expression \eqref{inplane-gradient-displacement-ver1} one finds $(\upDelta \overline{x}_2)_\textrm{ip} \sim \varepsilon^{1-\zeta}$, as $K_I$ is independent of $\varepsilon$. In particular, when $\zeta=1/2$ both $\overline{x}_2^{-\infty}$ and $(\upDelta \overline{x}_2)_\textrm{ip}$ scale as $\mathcal{O}(\varepsilon^{1/2})$. Using \eqref{O(1)_solution_for_phi} and \eqref{eq-soln-phi1p} for $r \gg 1$ gives $\phi_0^+ \sim \overline{x}_2^{-\infty}/r$ and $\varepsilon\, \phi_1^+ \sim \varepsilon\, I_{\phi_1}^+/r$, so that in this regime $\phi_0 \sim \varepsilon\, \phi_1$. This violates the assumed asymptotic ordering, and hence \eqref{inplane-gradient-displacement-ver1} is not valid for $\overline{x}_2^{-\infty}$ values $\mathcal{O}(\varepsilon^{1/2})$ or smaller. In particular, for the limiting trajectory where $\overline{x}_2^{-\infty} \to 0$, expression \eqref{inplane-gradient-displacement-ver1} predicts that $(\upDelta \overline{x}_2)_\textrm{ip}$ becomes unbounded, which is neither physical nor expected. A different asymptotic approach is therefore needed, and this case is treated in the next subsection.
\subsubsection{Gradient displacement of trajectories with $\mathcal{O}(\varepsilon^{1/2})$ or smaller upstream gradient offsets}
\begin{figure}
    \centering
\includegraphics[width=1\linewidth]{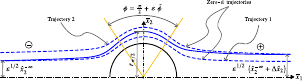}
   \caption{Schematic of an in-plane trajectory for finite $\varepsilon$ ($0 < \varepsilon \ll 1$), with $\mathcal{O}(\varepsilon^{1/2})$ initial offset, shown as a solid curve. The dashed trajectories 1 and 2 denote the corresponding zero-$\varepsilon$ reference trajectories that match the far-upstream and far-downstream offsets of the solid trajectory, and serve as reference paths in the perturbation analysis. For clarity, displacements are shown exaggerated.}
    \label{fig:inplane-schematic2}
\end{figure}

Consider the schematic in figure~\ref{fig:inplane-schematic2}. The $-$ve branch of the in-plane finite--$\varepsilon$ trajectory (shown as solid curve) is treated as a perturbation of the corresponding zero--$\varepsilon$ trajectory (trajectory 1) that has the same initial offset of $\hat{x}_2^{-\infty}\, \varepsilon^{1/2}$, where $\hat{x}_2 = \mathcal{O}(1)$. As noted from the scalings discussed at the end of the previous subsection, since the in-plane gradient offset is also expected to be of $\mathcal{O}(\varepsilon^{1/2})$, the $+$ve branch can similarly be viewed as a perturbation of another zero--$\varepsilon$ trajectory (trajectory 2) with a slightly larger initial offset, $\left(\hat{x}_2^{-\infty} +\upDelta \hat{x}_2\right)\, \varepsilon^{1/2}$, where we insist $\upDelta \hat{x}_2 > 0$ in the upper half-plane. From equation~\eqref{expression_for_c}, for $\varepsilon \ll 1$ we have
\begin{subequations}
    \begin{align}
        c &= e^{Q(c)/2}\,\left[\left(\hat{x}_2^{-\infty}\right)^2\, \varepsilon + J(c)\right]^{1/2} \nonumber \\
        &= J(c)^{1/2}\, e^{Q(c)/2}\, \left\{1+\frac{\varepsilon\, \left(\hat{x}_2^{-\infty}\right)^2}{2\, J(c)} \right\}+\mathcal{O}(\varepsilon^2)~,\\
        \tilde{c} &= e^{Q(\tilde{c})/2}\,\left[\left(\hat{x}_2^{-\infty}+\upDelta \hat{x}_2\right)^2\, \varepsilon + J(\tilde{c})\right]^{1/2} \nonumber \\
        &= J(\tilde{c})^{1/2}\, e^{Q(\tilde{c})/2}\, \left\{1+\frac{\varepsilon\, \left(\hat{x}_2^{-\infty} +\upDelta \hat{x}_2\right)^2}{2\, J(\tilde{c})} \right\}+\mathcal{O}(\varepsilon^2)~,
    \end{align}
    \label{expressions-for-c-andc-prime}
\end{subequations}
where $c$ and $\tilde{c}$ denote the radial coordinates at $\phi = \pi/2$ corresponding to trajectories 1 and 2, respectively. For convenience, we have defined the integral $J(r):=\int_r^{\infty}B'\, r'\, e^{-Q'}/(1-A')\,dr'$. If we denote $d$ as the radial coordinate at $\phi = \pi/2$ of the in-plane zero--$\varepsilon$ limiting trajectory, then, using equation~\eqref{expression_for_c}, we have $d = J(d)^{1/2}\, e^{Q(d)/2}$ as this trajectory has both $\hat{x}_2^{-\infty} = \upDelta \hat{x}_2 = 0$. It follows that, in leading order, the quantities $c$ and $\tilde{c}$ can be written in terms of $d$ as $c = d+\varepsilon\, p$ and $\tilde{c} = d+\varepsilon\, \tilde{p}$. Substituting these ansätze into equations~\eqref{expressions-for-c-andc-prime} and performing an asymptotic expansion, we can solve for the correction terms $p$ and $\tilde{p}$ as
\begin{subequations}
    \begin{align}
        p &= \frac{\left(\hat{x}_2^{-\infty}\right)^2\, (1-A_*)}{ (2-B_*)\, d}\, e^{Q_*}~,\\
         \tilde{p} &= \frac{\left(\hat{x}_2^{-\infty}+\upDelta \hat{x}_2\right)^2\, (1-A_*)}{ (2-B_*)\, d}\, e^{Q_*}~,
    \end{align}
    \label{expression-for-p-and-p'-ver1}
\end{subequations}
where the subscript `*' indicates that the corresponding functions are evaluated at the radial position $r = d$. To obtain the asymptotic form of the trajectory for $\varepsilon \ll 1$, the upstream and downstream parts of the finite--$\varepsilon$ trajectory (red) need to be matched appropriately at the inner layer, specifically when $\phi \to \pi/2$. On the upstream side, the radial position of the red trajectory at $\phi = \pi/2$ is related to that of the corresponding zero--$\varepsilon$ trajectory (blue trajectory 1) by an $\mathcal{O}(\varepsilon)$ correction as $r_{\pi/2}^{-} = c + \varepsilon\, k(c)$ (see the discussion in \S~\ref{Inner-layer-solution}). Similarly, on the downstream side, the corresponding relation with the second zero--$\varepsilon$ trajectory (blue trajectory 2) is $r_{\pi/2}^{+} = \tilde{c} - \varepsilon\, k(\tilde{c})$, where the negative sign appears because $k(\tilde{c})$ is a positive quantity, reflecting that the downstream trajectory lies slightly closer to the collision sphere compared to its zero--$\varepsilon$ counterpart. Imposing the matching condition $r_{\pi/2}^{-} = r_{\pi/2}^{+}$ in the limit $\varepsilon \to 0$ then gives
\begin{equation}
    \tilde{c} = c+ \varepsilon\, \left(k(c)+k(\tilde{c}) \right) = c+2\, \varepsilon\, k(d)+\mathcal{O}(\varepsilon^2)~,
    \label{rel-btw-cprime-and-c-and-d}
\end{equation}
where the expression for $k$ evaluated at $c$ is given in \eqref{expression-for-k}. The corresponding expressions for $k$ evaluated at $\tilde{c}$ or $d$ have the same functional form and can be obtained directly by replacing the evaluation point with $\tilde{c}$ or $d$, respectively. Using these relations in \eqref{rel-btw-cprime-and-c-and-d} together with the earlier ansätze gives the connection between $p$ and $\tilde{p}$ as $\tilde{p} = p + 2\, k(d)$. Substituting this into the expressions in \eqref{expression-for-p-and-p'-ver1} leads to a quadratic equation for $\upDelta\hat{x}_2$ as $\left(\upDelta\hat{x}_2\right)^2+2\, \hat{x}_2^{-\infty}\, \left(\upDelta\hat{x}_2\right)-4\, K_I(d) = 0$. Solving for $\upDelta\hat{x}_2$ and multiplying by $\varepsilon^{1/2}$ gives the in-plane gradient displacement for trajectories whose upstream gradient offset is $\mathcal{O}(\varepsilon^{1/2})$ or smaller as
\begin{eqnarray}
    \left(\upDelta \overline{x}_2\right)_\textrm{ip} &=&\upDelta\hat{x}_2\, \varepsilon^{1/2} = \varepsilon^{1/2}\, \left(- \hat{x}_2^{-\infty} + \sqrt{ \left(\hat{x}_2^{-\infty}\right)^2+4\, K_I(d)}\right) \nonumber \\
    &=& -\overline{x}_2^{-\infty} + \sqrt{ \left(\overline{x}_2^{-\infty}\right)^2+4\,\varepsilon\,  K_I(d)}~.
    \label{inplane-gradient-displacement-ver2}
\end{eqnarray}
The appropriate solution branch is selected such that $\left(\upDelta \overline{x}_2\right)_\textrm{ip}$ remains positive for positive $\overline{x}_2^{-\infty}$ (i.e., in the upper half-plane) and should also reduce to the expression in \eqref{inplane-gradient-displacement-ver1} in the limit of large $\overline{x}_2^{-\infty}$. Hence, equation~\eqref{inplane-gradient-displacement-ver2} represents a more general form of the in-plane gradient displacement. The displacement of the limiting trajectory can be obtained as a special case of this result. Setting $\overline{x}_2^{-\infty} \to 0$ in \eqref{inplane-gradient-displacement-ver2} gives $\left(\upDelta \overline{x}_2\right)_\textrm{ip}^\textrm{sep} =\overline{x}_2^{+\infty}= 2\, \varepsilon^{1/2}\, \sqrt{K_I(d)}$, which shows that the displacement is finite and nonzero, and scales as $\mathcal{O}(\varepsilon^{1/2})$. Note that this scaling is indicated in the schematic in figure \ref{fig:inplane-schematic3}.
\subsubsection{Evaluation of the in-plane self-diffusivity}
\label{repuls-in-plane-diff-scaling}
For in-plane trajectories, the particle undergoes a lateral displacement only in the gradient direction, with no displacement in the vorticity direction. Therefore, the in-plane contribution to the self-diffusivity in the gradient direction can be obtained by evaluating only the inner integral in equation \eqref{Di-defenition} for $i=2$, which gives
\begin{eqnarray}
    \hat{D}_2^\textrm{ip} = \frac{3}{8\, \pi} \times 2 \times \int_{0}^{\infty} d\overline{x}_2^{-\infty}\, \lvert \overline{x}_2^{-\infty} \rvert\, \left( \upDelta \overline{x}_2\right)_\textrm{ip}^2~,
    \label{D2-inplane-integral}
\end{eqnarray}
where, the integral is restricted to the upper half-plane by exploiting the anti-symmetry of the system in the $\overline{x}_1$--$\overline{x}_2$ plane. The factor of two that would arise from including the lower half-plane is accounted for in capturing the full contribution to the in-plane diffusivity. 

In the upper half-plane, the integration over $\overline{x}_2^{-\infty}$ extends from $0$ to $\infty$, since all upstream trajectories in this region contribute to the diffusivity. In the presence of a repulsive interaction, every upstream trajectory remains open downstream, undergoes a net lateral shift away from the collision sphere, and therefore contributes to self-diffusion. This situation is different for attractive interactions. In that case, some upstream trajectories may spiral inward toward the collision sphere and do not re-emerge downstream. As a result, the limiting trajectory for an attractive potential will have a non-zero value of $\overline{x}_2^{-\infty}$. Qualitatively, the trajectories in the attractive case can be viewed as a reversal of those in the repulsive case, with the upstream and downstream portions effectively exchanged, consistent with changing the sign of $F$. Only those upstream trajectories that remain open and lie outside the limiting trajectory should be included in the integration. Therefore, for attractive potentials, the lower limit of integration in \eqref{D2-inplane-integral} must be replaced by the non-zero value of $\overline{x}_2^{-\infty}$ associated with the separatrix. 
A similar behavior can be observed for inertial spheres studied by \citet{subramanian2006trajectory}. For the repulsive interaction considered here, however, the separatrix corresponds to $\overline{x}_2^{-\infty} = 0$, and therefore the full integration over the interval $[0, \infty)$ is required.

To obtain an analytical asymptotic expression for in-plane diffusivity in the limit of small $\varepsilon$, the integral in \eqref{D2-inplane-integral} may be evaluated by splitting the integration domain into two regions, each treated with appropriate asymptotic expressions for displacements that are valid there. Specifically, we write
\begin{eqnarray}
    \int_{0}^{\infty} d\overline{x}_2^{-\infty}\, \lvert \overline{x}_2^{-\infty} \rvert\, \left( \upDelta \overline{x}_2\right)_\textrm{ip}^2 = \int_{0}^{b\, \varepsilon^{1/2-h}} d\overline{x}_2^{-\infty}\,  \overline{x}_2^{-\infty} \, \left( \upDelta \overline{x}_2\right)_\textrm{ip}^2+\int_{b\, \varepsilon^{1/2-h}}^{\infty} d\overline{x}_2^{-\infty}\,  \overline{x}_2^{-\infty} \, \left( \upDelta \overline{x}_2\right)_\textrm{ip}^2~,
    \label{D2-inplane-integral-expanded}
\end{eqnarray}
The cutoff $b\, \varepsilon^{1/2-h}$ is introduced so that \eqref{inplane-gradient-displacement-ver1} is valid when $\overline{x}_2^{-\infty} \gg b\, \varepsilon^{1/2-h}$, while \eqref{inplane-gradient-displacement-ver2} describes the behavior when $\overline{x}_2^{-\infty} \ll b\, \varepsilon^{1/2-h}$. Here, $0<h<1/2$, and $b$ is an arbitrary quantity of order $\mathcal{O}(1)$. As one would expect, neither $h$ nor $b$ affects the final scaling, as shown below. When $\varepsilon \ll 1$, the conditions ($b\, \varepsilon^{-h} \gg 1$ and $b\, \varepsilon^{1/2-h} \ll 1$ ensure that the integral can be divided into two separate regions where the corresponding asymptotic expressions are valid. For the first integral on the right-hand side of \eqref{D2-inplane-integral-expanded}, we rescale the integration variable with $\varepsilon^{1/2}$ by writing $\overline{x}_2^{-\infty} = \varepsilon^{1/2}\, \hat{x}_2^{-\infty}$, which allows us to use \eqref{inplane-gradient-displacement-ver2} as
\begin{eqnarray}
    \mathscr{I}_1 &=& \int_{0}^{b\, \varepsilon^{1/2-h}} d\overline{x}_2^{-\infty}\,  \overline{x}_2^{-\infty} \, \left( \upDelta \overline{x}_2\right)_\textrm{ip}^2 = \varepsilon\, \int_{0}^{b\, \varepsilon^{-h}} d\hat{x}_2^{-\infty}\,  \hat{x}_2^{-\infty} \, \left( \upDelta \overline{x}_2\right)_\textrm{ip}^2 \nonumber\\
    &=& \varepsilon^2\, \int_{0}^{b\, \varepsilon^{-h}} d\hat{x}_2^{-\infty}\,  \hat{x}_2^{-\infty} \, \left( - \hat{x}_2^{-\infty} + \sqrt{ \left(\hat{x}_2^{-\infty}\right)^2+4\, K_I(d)}\right)^2 \nonumber \\
    &=& \frac{\varepsilon^2}{2}\, \Bigg[4\, K_I(d)\,\left( \hat{x}_2^{-\infty}\right)^2+\left( \hat{x}_2^{-\infty}\right)^4-\left( \hat{x}_2^{-\infty}\right)\, \left(2\, K_I(d)+\left( \hat{x}_2^{-\infty}\right)^2\right)\, \sqrt{4\, K_I(d)+\left( \hat{x}_2^{-\infty}\right)^2} \nonumber \\
    &&+8\, K_I(d)^2\, \tanh^{-1}\left(\frac{ \hat{x}_2^{-\infty}}{\sqrt{4\, K_I(d)+\left( \hat{x}_2^{-\infty}\right)^2}}\right)  \Bigg]_{\hat{x}_2^{-\infty} = b\, \varepsilon^{-h}} \nonumber \\
    &=& \varepsilon^2\, K_I(d)^2\, \left[4\, (\log b -h\, \log \varepsilon)-(1+2\, \log K_I(d)) \right]~,
    \label{inplane-diff-integral-1-evaluation}
\end{eqnarray}
where in the last step, we take the limit of $\hat{x}_2^{-\infty} = b\, \varepsilon^{-h} \gg 1$ to get the asymptotic form of the integral. The second integral on the right-hand side of \eqref{D2-inplane-integral-expanded} can be evaluated directly using the asymptotic form of $\left(\upDelta \overline{x}_2\right)_\textrm{ip}$ from \eqref{inplane-gradient-displacement-ver1}. However, care is needed near the lower limit, where a logarithmic divergence appears as $b\, \varepsilon^{1/2-h} \to 0$. To regularize this singularity, we add and subtract an auxiliary integral that carries the same divergent part, and then group the terms so that each resulting integral is individually convergent, as follows:
\begin{eqnarray}
    \mathscr{I}_2 &=& \int_{b\, \varepsilon^{1/2-h}}^{\infty} d\overline{x}_2^{-\infty}\,  \overline{x}_2^{-\infty} \, \left( \upDelta \overline{x}_2\right)_\textrm{ip}^2 = 4\, \varepsilon^2\,\int_{b\, \varepsilon^{1/2-h}}^{+\infty}K_I(c;\overline{x}_2^{-\infty})^2\,   \frac{d\overline{x}_2^{-\infty}}{\overline{x}_2^{-\infty}} \nonumber\\
    &=& 4\, \varepsilon^2\,\left[ \int_{b\, \varepsilon^{1/2-h}}^{\infty}K_I(c;\overline{x}_2^{-\infty})^2\,   \frac{d\overline{x}_2^{-\infty}}{\overline{x}_2^{-\infty}} -\int_{b\, \varepsilon^{1/2-h}}^{1}K_I(d)^2\,   \frac{d\overline{x}_2^{-\infty}}{\overline{x}_2^{-\infty}}+\int_{b\, \varepsilon^{1/2-h}}^{1}K_I(d)^2\,   \frac{d\overline{x}_2^{-\infty}}{\overline{x}_2^{-\infty}}\right]  \nonumber \\
    &=& 4\, \varepsilon^2\,\left[ \int_{b\, \varepsilon^{1/2-h}}^{\infty}\left\{K_I(c;\overline{x}_2^{-\infty})^2-K_I(d)^2\, \mathcal{H}(1-\overline{x}_2^{-\infty})\right\}\,   \frac{d\overline{x}_2^{-\infty}}{\overline{x}_2^{-\infty}}-K_I(d)^2\, \log\left(b\, \varepsilon^{1/2-h} \right) \right]  \nonumber \\
    &=&4\, \varepsilon^2\, \left[\int_{0}^{\infty}\left\{K_I(c;\overline{x}_2^{-\infty})^2-K_I(d)^2\, \mathcal{H}(1-\overline{x}_2^{-\infty})\right\}\,   \frac{d\overline{x}_2^{-\infty}}{\overline{x}_2^{-\infty}}\right] \nonumber\\
    && - \varepsilon^2\, K_I(d)^2\, \left[ 4\, (\log b-h\, \log \varepsilon)+2\, \log \varepsilon\right] + o(\varepsilon)~,
    \label{inplane-diff-integral-2-evaluation}
\end{eqnarray}
where $\mathcal{H}(\cdot)$ denotes the Heaviside step function, and the lower limit of the first integral has been extended to zero, introducing only an $o(\varepsilon)$ error in the final step. Substituting $\mathscr{I}_1$ and $\mathscr{I}_2$ back into \eqref{D2-inplane-integral-expanded} shows that, as anticipated, all terms involving the arbitrary parameters $h$ and $b$ cancel exactly. Using \eqref{D2-inplane-integral}, we then obtain the asymptotic form of the in-plane self-diffusivity for $\varepsilon \ll 1$ as
\begin{eqnarray}
    \hat{D}_2^\textrm{ip} = \frac{3}{2\, \pi}\, \varepsilon^2\, \left[\mathcal{A}_0-K_I(d)^2\, \log \varepsilon \right]~,
    \label{in-plane-diff-asymptotic}
\end{eqnarray}
with the constant $\mathcal{A}_0$ is given by
\begin{eqnarray}
    \mathcal{A}_0 = 2\, \left[\int_{0}^{\infty}\left\{K_I(c;\overline{x}_2^{-\infty})^2-K_I(d)^2\, \mathcal{H}(1-\overline{x}_2^{-\infty})\right\}\,   \frac{d\overline{x}_2^{-\infty}}{\overline{x}_2^{-\infty}} \right]-K_I(d)^2\, \left\{\frac{1}{2}+\log K_I(d) \right\}~.
    \label{integral-alpha}
\end{eqnarray}
Although we have derived an asymptotic expression, it still involves an improper integral that must be evaluated to obtain the exact value of the in-plane diffusivity. However, the dependence on $\varepsilon$ is now clear in the limit $\varepsilon \ll 1 $. This result will be validated later in \S\S~\ref{inplane-numeric-eval} by comparing it with the numerically calculated in-plane diffusivity obtained from \eqref{D2-inplane-integral} for the case of a repulsive electrical double-layer interaction.
The convergence of the improper integral in \eqref{integral-alpha} requires that $K_I(c;\overline{x}_2^{-\infty})$ approach $K_I(d)$ sufficiently rapidly as $\overline{x}_2^{-\infty}\rightarrow 0$. For any repulsive potential with $F(r)$ decaying faster than algebraically — including the exponentially decaying EDL force — this condition is satisfied. For potentials with slow algebraic decay, convergence must be verified on a case-by-case basis.

\subsection{Diffusivity accounting for the trajectories outside the shearing plane for $\varepsilon \ll 1$} 
\label{out-of-plane-section}
In the previous section, we focused only on in-plane trajectories and therefore obtained only the in-plane self-diffusivity. In this section, we consider the general set of trajectories, in particular, the off-plane trajectories, which are not confined to the shearing plane. These trajectories can have displacements in both the gradient and vorticity directions, and therefore contribute to both $\hat{D}_2$ and $\hat{D}_3$.

\begin{figure}
    \centering
\includegraphics[width=1\linewidth]{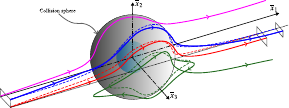}
    \caption{Schematic of representative off-plane relative trajectories in the first quadrant ($\overline{x}_2>0$ and $\overline{x}_3>0$). Dashed curves correspond to $\varepsilon = 0$ and solid curves to $\varepsilon > 0$. Blue curves denote open trajectories, green curves denote closed (for $\varepsilon = 0$) or spiral (for $\varepsilon > 0$) trajectories, and the red curve represents the limiting trajectory (separatrix) for the off-plane case. The magenta curve is shown for reference and represents the in-plane limiting trajectory for $\varepsilon > 0$. The parallelograms at the left and right ends indicate the upstream and downstream offsets of the trajectories, shown for visual clarity.}
    \label{fig:off-schematic1}
\end{figure}

We use the in-plane case as a guide for the off-plane analysis. As before, it is useful to first develop a qualitative understanding of the trajectory topology. When $\varepsilon = 0$, trajectories in the neighborhood of the shearing plane exhibit behavior analogous to the in-plane case in terms of their overall classification and structure. However, unlike the strictly planar dynamics, these trajectories are not confined to a single plane and instead evolve in three dimensions. In particular, a trajectory that originates in a given plane (parallel to the shearing plane) undergoes a finite out-of-plane excursion in the vicinity of the collision sphere before returning to the same plane. Far downstream, it recovers the same gradient and vorticity coordinates that it had upstream. Examples are shown by the dashed trajectories in figure \ref{fig:off-schematic1}.
As in the in-plane case, the trajectory topology consists of open, closed, and limiting trajectories. The essential difference is that these paths are no longer planar; they deviate out of their initial plane during their evolution, reaching a maximum out-of-plane displacement near the collision sphere before returning.

For $\varepsilon > 0$, the repulsive interaction induces a displacement of the trajectories away from the collision sphere. In contrast to the in-plane case, this displacement now has components in both the gradient and vorticity directions. The open trajectories that exist for $\varepsilon = 0$ remain open, but they lose their fore–aft symmetry as they evolve downstream. For example, in the first quadrant ($\overline{x}_2>0$ and $\overline{x}_3>0$), a trajectory acquires positive shifts $\upDelta \overline{x}_2$ and $\upDelta \overline{x}_3$ due to the repulsive interaction. Likewise, trajectories that are closed when $\varepsilon = 0$ open up and develop a spiraling motion with an out-of-plane component. Representative behavior is illustrated as solid curves in figure \ref{fig:off-schematic1}. The limiting trajectory in the off-plane case, as well, already exhibits three-dimensional excursions when $\varepsilon = 0$. For $\varepsilon > 0$, it is further deformed by the repulsive interaction, acquiring net shifts in both the gradient and vorticity directions. It continues to act as the boundary separating distinct classes of motion, although its far-upstream and far-downstream locations will no longer match.

For trajectories outside the shearing plane, all three equations in \eqref{gov-eqns} must be considered, as none of them can be neglected. However, similar to the in-plane case, the explicit time dependence in the governing equations can be eliminated, yielding two coupled equations for the trajectory evolution as
\begin{subequations}
\begin{align}
    \frac{d\phi}{dr} &=  -\frac{\sin^2\phi + \frac{1}{2} B \left(\cos^2\phi - \sin^2\phi \right)}{r\, (1-A)\, \sin^2\theta\, \sin\phi\, \cos\phi + \varepsilon \,G \, F}, \label{dphidr_equation}\\
\frac{d\theta}{dr} &=  \frac{(1-B)\, \sin\theta\, \cos\theta\, \sin\phi\, \cos\phi}{r\, (1-A)\, \sin^2\theta\, \sin\phi\, \cos\phi + \varepsilon \,G \, F}.
\label{dthetadr_equation}
\end{align}
\label{dphi_and_dtheta_by_dr_equations}
\end{subequations}
The fore--aft symmetry of the system is evident when $\varepsilon = 0$, since replacing $\phi$ with $\pi -\phi$ leaves the governing equations unchanged. Moreover, for any value of $\varepsilon$ , rotating $\phi$ by $ \pi$ also leaves the equations invariant, reflecting the antisymmetry from the background shear flow. By exploiting these symmetries, it is sufficient to restrict attention to trajectories in the range $\theta \in (0, \pi/2) $ and $ \phi \in (0, \pi) $.
In this regime, for finite $\varepsilon$, the trajectories behave similarly to the in-plane case in the gradient direction, with $\upDelta \overline{x}_2 = \overline{x}_2^{+\infty}-\overline{x}_2^{-\infty} > 0 $. However, unlike the in-plane case, there is also a nonzero displacement in the vorticity direction, $\upDelta \overline{x}_3 = \overline{x}_3^{+\infty}-\overline{x}_3^{-\infty} > 0 $. Both $\upDelta \overline{x}_2$ and $\upDelta \overline{x}_3$ will be positive in the regime considered due to the repulsive interaction.

We now proceed with the perturbation analysis of the governing equations \eqref{dphi_and_dtheta_by_dr_equations}. Before doing so, it is important to note that, from the right-hand side (RHS) of the equations, the $\mathcal{O}(1)$ terms in the denominator of \eqref{dphidr_equation}, as well as both the numerator and denominator in \eqref{dthetadr_equation}, vanish at $\phi = \pi/2$. However, the product $G F$, evaluated at any $r$ with $\theta = \theta_t$ and $\phi = \pi/2$, is in general nonzero, where $\theta_t$ denotes the value of $\theta$ when $\phi = \pi/2$.
For this reason, the perturbation becomes singular near $\phi = \pi/2$, similar to the in-plane trajectory case. A detailed analysis of the trajectories and the corresponding mathematical steps is provided in Appendix~\ref{full-analytical_small_NR}; only the essential elements are presented in this section, along with the results.
Let us assume that the trajectory for small $\varepsilon$ can be treated as a perturbation of a zero-$\varepsilon$ trajectory, with both starting from the same initial location far upstream. As in the in-plane case, the trajectory can be divided into three regions:
\begin{subequations}
\begin{align}
    \textrm{Outer layer O}_1:& \quad \phi \in \left(\frac{\pi}{2}+\mathcal{O}(1),\pi \right), \, \, \theta \in \left(\theta_t+\mathcal{O}(1),\frac{\pi}{2} \right), \, \, r > \frac{c}{\sin \theta_t}+\mathcal{O}(1)~, \\
\textrm{Inner layer I:}& \quad \phi = \frac{\pi}{2}+\varepsilon\, \tilde{\phi}, \, \, \theta = \theta_t+\varepsilon\, \theta_f+\varepsilon^2\, \tilde{\theta} , \, \, r = \frac{c}{\sin \theta_t}+\varepsilon\, k+\varepsilon^2\, \tilde{r}~,\\
\textrm{Outer layer O}_2:& \quad \phi \in \left(0,\frac{\pi}{2}-\mathcal{O}(1) \right), \, \, \theta \in \left(\theta_t+\mathcal{O}(1),\frac{\pi}{2} \right), \, \, r > \frac{c}{\sin \theta_t}+\mathcal{O}(1)~,
\end{align}
\end{subequations}
where regular perturbation arguments remain valid in the two outer regions, $\textrm{O}_1$ and $\textrm{O}_2$, and a singular perturbation treatment becomes necessary in the inner region $\textrm{I}$, as the trajectory changes its character near $\phi = \pi/2$. In the inner-layer scalings, $\tilde{\phi}$, $\theta_f$, $\tilde{\theta}$, $k$ and $\tilde{r}$ are all $\mathcal{O}(1)$ quantities; their explicit forms will be determined through the inner-layer analysis. The parameter $c$ is the same as in the in-plane case, where it represents the minimum radial separation of the zero-$\varepsilon$ trajectory from the collision sphere in the shearing plane.
For off-plane zero-$\varepsilon$ trajectories, however, the distance of nearest approach is modified to occur at $r = c/\sin \theta_t$ when $\phi_0 = \pi/2$, $\theta_0 = \theta_t$. This value is therefore used as the reference point in the inner-layer expansion. For convenience, here onwards, we use a short hand notation as $c_t = c/\sin \theta_t$.

We first focus on the outer region, where the trajectory remains sufficiently far from the singular region and can therefore be described as a smooth perturbation of the zero-$\varepsilon$ solution. In this region, we adopt the same regular perturbation expansion used earlier, $\phi = \phi_0+\varepsilon\, \phi_1+\cdots$
and similarly for $\theta = \theta_0+\varepsilon\, \theta_1+\cdots$. Substituting these expansions into \eqref{dphi_and_dtheta_by_dr_equations} yields the governing equations at leading order and at $\mathcal{O}(\varepsilon)$ (see Appendix~\ref{full-analytical_small_NR}, equations~\eqref{phi-expanded-eqns} and \eqref{theta-expanded-eqns}).
Physically, we impose that both the finite-$\varepsilon$ and zero-$\varepsilon$ trajectories originate from the same far-upstream location. This ensures that any differences between them arise solely from the interaction effects during the encounter. Accordingly, as $\overline{x}_1 \to -\infty$, the boundary conditions are
$r\, \cos \theta \to \overline{x}_3^{-\infty}$ and $r\, \sin \theta\, \sin \phi \to \overline{x}_2^{-\infty}$. Expanded in $\varepsilon$ in each order, these conditions can be written as:
\begin{subequations}
\begin{align}
\mathcal{O}(1) : && r\, \cos \theta_0 \to \overline{x}_3^{-\infty}~, \, \,  \textrm{and} \, \,  r\, \sin \theta_0\, \sin\phi_0 \rightarrow \overline{x}_2^{-\infty} \hspace{3mm} \textrm{as} \hspace{3mm} r \rightarrow \infty \label{O(1)_boundary_condition_for_phi_and_theta} \\
\mathcal{O}(\varepsilon) : && r\, \theta_1^- \rightarrow 0~, \, \, \textrm{and} \, \, r\, \phi_1^- \rightarrow 0 \hspace{3mm} \textrm{as} \hspace{3mm} \phi_0 \rightarrow \pi \,  (r \rightarrow \infty)~. \label{O(NR)_boundary_condition_for_phi_and_theta}
\end{align}
\label{boundary-conditions-full}
\end{subequations}
The expressions for lateral displacements in gradient and vorticity directions are then:
\begin{subequations}
\begin{align}
\upDelta \overline{x}_2 =  r\, \sin \theta\, \sin\phi |_{\phi \to \pi}^{\phi \to 0} = \varepsilon\, \lim_{r \to \infty} r\, \phi_1^{+}~,\\
\upDelta \overline{x}_3 =  r\, \cos \theta|_{\phi \to \pi}^{\phi \to 0} = -\varepsilon\, \lim_{r \to \infty} r\, \theta_1^{+}~,
\end{align}
\label{off-plane_displacement}
\end{subequations}
where, we have used $\theta^{\pm} = \pi/2$, $\phi^{-} = \pi$, and $\phi^{+} = 0$ as $r \to \infty$. Applying the boundary conditions in \eqref{O(1)_boundary_condition_for_phi_and_theta}, the leading-order governing equations can be solved to obtain the solutions $\theta_0(r)$ and $\phi_0(r)$, which is already available from \citet{batchelor1972hydrodynamic} as:
\begin{subequations}
\begin{align}
r\, \cos \theta_0 &= \overline{x}_3^{-\infty}\, e^{Q(r)/2}~, \label{theta0-solution}\\
r^2\, \sin^2 \phi_0 &= \frac{e^{Q(r)}}{\sin^2 \theta_0}\, \left[\left(\overline{x}_2^{-\infty}\right)^2+\int_{r}^{\infty}\frac{B'\, r'}{(1-A')}\, e^{-Q(r')}\, dr' \right]~. \label{phi0-solution}
\end{align}
\label{batchelor-solution-off-plane}
\end{subequations}
Note that the solution for $\phi_0$ differs slightly from the corresponding in-plane result in \eqref{O(1)_solution_for_phi}, due to the appearance of the $\sin^2 \theta_0$ term in the denominator on the RHS. Next, the boundary conditions in \eqref{O(NR)_boundary_condition_for_phi_and_theta} are applied to the $\mathcal{O}(\varepsilon) $ governing equations to determine the correction terms $\theta_1(r)$ and $\phi_1(r)$ for both the upstream and downstream branches as:
\begin{subequations}
\begin{align}
    \theta_1^{-} =& -\frac{e^{Q(r)/2}}{r\, \sin \theta_0}\, N_I(r)~, 
    \label{theta1m-solution} \\
    \theta_1^{+} =& \frac{e^{Q(r)/2}}{r\, \sin \theta_0}\, \left[ I_{\theta_1}^++N_I(r)\right]~,
    \label{theta1p-solution}  \\
        \phi_1^{-} =& -\frac{2\, e^{Q(r)}}{r^2\, \sin^2 \theta_0\, \sin 2\phi_0}\, M_I(r)~.
    \label{phi1m-solution} \\
       \phi_1^+ =& \frac{2\, e^{Q(r)}}{r^2\, \sin^2\theta_0\, \sin 2\phi_0}\left[\overline{x}_2^{-\infty}\,I_{\phi_1}^+-\overline{x}_3^{-\infty}\,I_{\theta_1}^+\, \sin^2\phi_0-M_I(r)\right]~.
    \label{phi1p-solution-outofplane}
\end{align}
\label{outer-layer-O(eps)-solns}
\end{subequations}
where, for compactness of notation, we have introduced a set of integral functionals, analogous to $K_I(r)$, obtained from $\mathcal{O}(1)$ solutions, defined as:
\begin{subequations}
\begin{align}
    N_I(r) :=& \int_{r}^{\infty}\frac{(1-B')\, \cos \theta_0'\, G'\, F'\, e^{-Q(r')/2}}{r'\, (1-A')^2\, \sin^2 \theta_0'\, \sin \phi_0'\, \cos \phi_0'}\, dr'~,\label{NI-definition} \\
    M_I(r) :=& \int_{r}^{\infty} \left[\frac{ \left(B'/2+(1-B')\, \sin^2\phi_0'\right)\, G'\, F'}{(1-A')^2\, \sin^2 \theta_0'\, \sin \phi_0'\, \cos \phi_0'}-\frac{ 2\, r'\, \left(B'/2+(1-B')\, \sin^2\phi_0'\right)\, \theta_1'^{-}}{(1-A')\, \tan \theta_0'}\right]\,e^{-Q(r')}\, dr'~, \label{MI-definition}
\end{align}
\end{subequations} 
with the integration is carried out over $\phi_0' \in (0,\pi/2)$, i.e., along the positive branch. We have also introduced the constants $I_{\theta_1}^+$ and $I_{\phi_1}^+$, which arise as integration constants and are to be determined through matching with the inner solution. We note that $M_I$, unlike $K_I$ and $N_I$, depends on the first-order correction $\theta_1'$ through the second term in its integrand. This means that evaluating $M_I$ requires first solving the $\mathcal{O}(\varepsilon)$ equation for $\theta_1'$ along the upstream branch — a preliminary numerical integration over the zero-$\varepsilon$ trajectory, before the quadrature in \eqref{MI-definition} can be performed. The off-plane gradient diffusivity thus involves a nested computation that is absent in the in-plane case.

Having obtained the outer-layer solutions, we now turn to the inner layer. As in the in-plane case, the inner region corresponds to the neighborhood of $\phi = \pi/2$, where the regular expansion breaks down and a different scaling is required. Consequently, we introduce the expansions $\phi = \pi/2+\varepsilon\, \tilde{\phi}(\tilde{r})$, $r = c_t+\varepsilon\, k+\varepsilon^2\, \tilde{r}$, and in addition $\theta = \theta_t+\varepsilon\, \theta_f+\varepsilon^2\, \tilde{\theta}(\tilde{r})$. Using \eqref{theta0-solution} and \eqref{phi0-solution}, we obtain relations that connect $c_t$, $\theta_t$ to the upstream trajectory initializations $\overline{x}_2^{-\infty}$ and $\overline{x}_3^{-\infty}$ as
\begin{subequations}
\begin{align}
c_t\, \cos \theta_t &= \overline{x}_3^{-\infty}\, e^{Q(c_t)/2}~, \label{c_t-theta_t-1}\\
        c^2 &= e^{Q(c_t)}\, \left[\left(\overline{x}_2^{-\infty}\right)^2+\int_{c_t}^{\infty}\frac{B'\, r'}{(1-A')}\, e^{-Q(r')}\, dr' \right]~.
        \label{c_t-theta_t-2}
\end{align}
\label{c_t-theta_t-12}
\end{subequations}
Inner-layer scalings introduced above are determined from the governing equations \eqref{dphi_and_dtheta_by_dr_equations} by requiring that its LHS and RHS retain terms to balance at the leading order when $\epsilon \to 0$. Specifically, we obtain the dependence on $\tilde{r}$ for $r$ and $\theta$ appear only at $\mathcal{O}(\varepsilon^2)$, while the variation in $\phi$ is from $\mathcal{O}(\varepsilon)$ itself. However, as in the in-plane case for $r$, here for $r$ and $\theta$, the variation needs to be about base values that already account for $\mathcal{O}(\varepsilon)$ constants, denoted $\varepsilon\, k$ and $\varepsilon\, \theta_f$ respectively. These $\mathcal{O}(\varepsilon)$ constants do not influence the leading-order inner-layer equations but will play a role later when matching the inner and outer solutions.
Substituting these inner-layer scalings into the governing equations \eqref{dphi_and_dtheta_by_dr_equations} then leads to differential equations for $\tilde{\phi}$ and $\tilde{\theta}$, which can be solved to obtain
\begin{subequations}
\begin{align}
    \tilde{\phi}^{\pm} =& \frac{G_0\, F_0}{c\, (1-A_0)\, \sin \theta_t}\, \left\{ 1\mp \sqrt{1+\frac{c\, (1-A_0)\, (2-B_0)\, (\tilde{r}-I_i^\textrm{off})\, \sin \theta_t}{G_0^2\, F_0^2}}\right\}~,
    \label{phi_tilde_pm_solution} \\
 \tilde{\theta} =& I_c+\frac{ \tilde{\phi}^2\,(1-B_0)}{2\, (2-B_0)}\, \sin 2\theta_t~,
    \label{theta_tilde_solution}
\end{align}
\end{subequations} 
Here, $I_i^\textrm{off}$ and $I_c$ are integration constants whose exact values do not affect the matching procedure, as in the in-plane case. The solution for $\tilde{\phi}$ links it directly to the corresponding $\tilde{\phi}$ in the appropriate regime, either upstream or downstream.

With both the outer and inner layer solutions now obtained, we proceed to match them following the same approach used for the in-plane trajectories. Assuming a scaling for the intermediate dependent variable $\phi$ as $\phi = \pi/2+\varepsilon^{\alpha}\, \hat{\phi}$, the governing equation \eqref{dphidr_equation} implies that the independent variable should scale as $r = c_t + \hat{r}\, \varepsilon^{2\alpha}$. In terms of these matching variables, equation \eqref{dphidr_equation} thus reduces to
\begin{equation}
    \frac{d \hat{\phi}}{d \hat{r}} = \frac{1-B_0/2}{c\, (1-A_0)\, (\sin \theta_t)\, \hat{\phi}-\varepsilon^{1-\alpha}\, G_0\, F_0}~.
\end{equation}
Here, for the general off-plane trajectories, $A_0, B_0, G_0$ and $F_0$ represent the values of $A(r), B(r), G(r)$ and $F(r)$ evaluated at $r = c_t$. To find the corresponding form of the dependent variable $\theta$, we start from \eqref{theta0-solution} and substitute the expansion of $r$ in terms of the matching variable $\hat{r}$. Using \eqref{c_t-theta_t-1} in this expression, we obtain $\cos \theta_0 = \cos \theta_t -(\cos \theta_t)\, (\hat{r}\, \varepsilon^{2\alpha})\, (1-B_0)/(c_t\, (1-A_0))+\mathcal{O}(\varepsilon^{4\alpha})$. This shows that in the outer layer, the matching variable expansion appears as $\varepsilon^{2\alpha}$ correction in $\theta$ from the leading term, as $\theta_0 = \theta_t+\hat{\theta}\, \varepsilon^{2\alpha}$, with $\hat{\theta} = (\cos \theta_t)\, \hat{r}\, (1-B_0)/(c\, (1-A_0)) $. Similarly, to obtain $\hat{\phi}$ for the outer layer matching variable expansion, we substitute the above scalings in \eqref{phi0-solution}, expand for $\varepsilon \ll 1$, and use \eqref{c_t-theta_t-2}. Comparing the left- and right-hand sides then yields $\hat{\phi} = ((2-B_0)\, \hat{r}/(c\, (1-A_0)\, \sin \theta_t))^{1/2}$. This shows that if $\phi$ varies at order $\varepsilon^\alpha$, then $\theta$ varies at order $\varepsilon^{2\alpha}$ in the matching region, consistent with the inner layer structure of the perturbation.

We can now express the $\mathcal{O}(\varepsilon)$ outer-layer solutions in \eqref{outer-layer-O(eps)-solns} using the matching variables $r = c_t + \hat{r}\, \varepsilon^{2\alpha}$, $\phi_0 = \pi/2 + \varepsilon^{\alpha} \hat{\phi}$, and $\theta_0 = \theta_t + \hat{\theta}\, \varepsilon^{2\alpha}$ in the limit $\varepsilon \to 0$. Combining these, the outer-layer solutions can be written as
\begin{equation}
\theta^{\textrm{O}_1} = \theta_0 +\varepsilon\, \theta_1^-, \, \, \theta^{\textrm{O}_2} = \theta_0 +\varepsilon\, \theta_1^+, \, \,  \phi^{\textrm{O}_1} = \phi_0 +\varepsilon\, \phi_1^-, \quad \textrm{and} \quad \phi^{\textrm{O}_2} = \phi_0 +\varepsilon\, \phi_1^+
\end{equation}
giving all outer-layer solutions in terms of the matching variables. For $\theta$, the expansion need to be retained up to terms of order $\varepsilon$, $\varepsilon^{1+\alpha}$, and $\varepsilon^{2\alpha}$, while for $\phi$, terms up to $\varepsilon$, $\varepsilon^{1-\alpha}$, and $\varepsilon^{\alpha}$ are kept, as these are needed for the matching.

For the inner layer, similar to the in-plane case, we start from $r = c_t + \varepsilon, k + \varepsilon^2 \tilde{r}$, while in terms of the matching variable, $r = c_t + \hat{r}\, \varepsilon^{2\alpha}$. Combining these two gives $\tilde{r} = \hat{r}\, \varepsilon^{2\alpha-2}-k\, \varepsilon^{-1}$. Substituting this into \eqref{phi_tilde_pm_solution} and expanding for $\varepsilon \ll 1$ with $\alpha > 1/2$ (so that $\varepsilon^{2\alpha-2} \gg \varepsilon^{-1} \gg I_i^\textrm{off}$) yields the inner-layer expansion for $\hat{\phi}$. This gives the inner-layer solution for $\phi$ in terms of the matching variable as $\phi^{I \mp} = \pi/2+\varepsilon\, \tilde{\phi}^{\mp}$. Similarly, using \eqref{theta_tilde_solution}, the inner-layer solution for $\theta$ can be written in terms of the matching variable by evaluating $ \theta^{I\mp} = \theta_t+\varepsilon\, \theta_f+\varepsilon^2\, \tilde{\theta}^{\mp}$. 

With both inner and outer-layer solutions now expressed in terms of the matching variables, we can directly perform the matching for $\phi$ in each branch. A direct comparison then leads to an equation connecting the unknown constants $I_{\phi_1}^+$ and $I_{\theta_1}^{+}$ as
\begin{equation}
    \overline{x}_2^{-\infty}\, I_{\phi_1}^+-\overline{x}_3^{-\infty}\, I_{\theta_1}^{+} = 2\, M_I(c_t)~.
\label{eqn-I-theta-I-phi-connection}
\end{equation}
In this process, we can also obtain the expression for the constant $k$ as a function of $c_t$. From \eqref{batchelor-solution-off-plane}, one can show that $\theta_0$, unlike $\phi_0$, stays real-valued even when $r$ falls below the zero-$\varepsilon$ minimum $c_t$. This allows us to determine the vorticity-direction displacement $\upDelta \overline{x}_3$ directly, without relying on the inner-layer solution for $\theta$, by simply matching the limiting expressions in the outer layers $\textrm{O}_1$ and $\textrm{O}_2$, as $I_{\theta_1}^+ = -2\, N_I(c_t)$; which can be then substituted in \eqref{eqn-I-theta-I-phi-connection} to solve for $I_{\phi_1}^+$ as $I_{\phi_1}^+ = 2\, \left( M_I(c_t)- \overline{x}_3^{-\infty}\, N_I(c_t)\right)/\overline{x}_2^{-\infty}$. Next, by matching the inner and outer-layer solutions for $\theta^-$, the unknown constant $\theta_f$ can also be determined. With all constants now fixed, we can finally evaluate both the vorticity- and gradient-direction displacements for a general off-plane trajectory in the small-$\varepsilon$ limit using \eqref{off-plane_displacement} as
\begin{subequations}
\begin{align}
    \upDelta\overline{x}_2 &= \varepsilon\, \lim_{r\to \infty}r\, \phi_1^+ = \varepsilon\, \lim_{r\to \infty, \theta_0 \to \pi/2, \phi_0 \to 0} \frac{2\, e^{Q(r)}}{r\, \sin^2 \theta_0\, \sin 2\phi_0}\, \left[\overline{x}_2^{-\infty}\, I_{\phi_1}^+-\overline{x}_3^{-\infty}\, I_{\theta_1}^{+}\, \sin^2\phi_0-M_I(r)\right] \nonumber \\
    &= \varepsilon\, \lim_{r\to \infty, \theta_0 \to \pi/2, \phi_0 \to 0}\left(\frac{\overline{x}_2^{-\infty}\, I_{\phi_1}^+}{\overline{x}_2^{-\infty}\, \sin \theta_0\, \cos \phi_0}\right) = \varepsilon\, I_{\phi_1}^+ = \frac{2\, \varepsilon}{\overline{x}_2^{-\infty}}\, M_I(c_t)-2\, \varepsilon\,\frac{\overline{x}_3^{-\infty}}{\overline{x}_2^{-\infty}} \, N_I(c_t) \nonumber \\
    &= \frac{2\, \varepsilon}{\overline{x}_2^{-\infty}}\, M_I(c_t)-\left(\upDelta\overline{x}_3\right)\,\frac{\overline{x}_3^{-\infty}}{\overline{x}_2^{-\infty}}~,\label{Deltax2-off-plane-1}\\
     \upDelta\overline{x}_3 &= -\varepsilon\, \lim_{r\to \infty}r\, \theta_1^+ = -\varepsilon\, \lim_{r\to \infty, \theta_0 \to \pi/2} \frac{e^{Q(r)/2}}{\sin \theta_0}\left[I_{\theta_1}^++N_I(r)\right] = -\varepsilon\, I_{\theta_1}^+ = 2\, \varepsilon\, N_I(c_t)~,\label{Deltax3-off-plane-1}
\end{align}
\label{Deltaxs-off-plane-1}
\end{subequations}
where we have used $Q(\infty) = N_I(\infty) = M_I(\infty) \to 0$. The expression for $\upDelta\overline{x}_3$ in \eqref{Deltax2-off-plane-1} depends explicitly on $\varepsilon$ and $c_t$, though it also indirectly depends on $\overline{x}_2^{-\infty}$ and $\overline{x}_3^{-\infty}$ through $c_t$ as per \eqref{c_t-theta_t-12}. This formula shows that $\upDelta \overline{x}_3$ remains of order $\mathcal{O}(\varepsilon)$ for all finite-$\varepsilon$ open trajectories and vanishes as $\theta_t \to \pi/2$ (i.e., $\overline{x}_3^{-\infty} \to 0$), independently of the upstream gradient offset $\overline{x}_2^{-\infty}$, since $N_I$ goes to zero in this limit. In this same limit, \eqref{Deltax2-off-plane-1} naturally reduces to the in-plane result \eqref{inplane-gradient-displacement-ver1} because $M_I$ simplifies to $K_I$.

By the same reasoning as in the in-plane case, when $\overline{x}_2^{-\infty} \sim \varepsilon^{1/2}$ or smaller, the expressions in \eqref{Deltaxs-off-plane-1} no longer remain consistent with the original assumption of regular perturbation ordering. In this regime, the correction terms become comparable to the leading-order contributions, and the expansions breaks down.
Therefore, a separate treatment is required for off-plane trajectories as well. The detailed derivation is given in Appendix~\ref{off-plane-closer-trajs}, and the main results are summarized below as:
\begin{subequations}
\begin{align}
    \upDelta \overline{x}_2 =& -\overline{x}_2^{-\infty}+\sqrt{\left(\overline{x}_2^{-\infty}\right)^2-2\, (\upDelta \overline{x}_3)\,\overline{x}_3^{-\infty}+4\, \varepsilon\,  M_I(d_t)}~,
    \label{Deltax2-off-plane-2}\\
    \upDelta \overline{x}_3 =& 2\, \varepsilon\,  N_I(d_t)~,
\label{Deltax3-off-plane-2}
    \end{align}
    \label{Deltaxs-off-plane-2}
\end{subequations}
where $d_t = d/\sin\theta_t$. Here, $d$, as defined earlier, denotes the minimum radial coordinate at $\phi = \pi/2$ for the in-plane zero-$\varepsilon$ limiting trajectory. In contrast, $d_t$ represents the corresponding minimum radial coordinate for the off-plane case, evaluated at $\phi = \pi/2$ and $\theta = \theta_t^{d}$.
Geometrically, the in-plane limiting trajectory lies entirely in the symmetry plane and attains its minimum separation $d$ at $\phi = \pi/2$. The off-plane limiting trajectory, however, is obtained by allowing a finite out-of-plane displacement while enforcing $\overline{x}_2^{-\infty} = 0$; its closest approach at $\phi = \pi/2$ therefore occurs at a shifted polar angle $\theta_t^{d}$, leading to the modified minimum distance $d_t$.
Unlike $c_t$, which is defined for a general trajectory and therefore depends on both $\overline{x}_2^{-\infty}$ and $\overline{x}_3^{-\infty}$, the quantity $d_t$ corresponds specifically to this limiting trajectory. Consequently, $d_t$ depends only on $\overline{x}_3^{-\infty}$. Note that when $\overline{x}_3^{-\infty} = 0$, the expression in \eqref{Deltax2-off-plane-2} reduces to the corresponding in-plane result in \eqref{inplane-gradient-displacement-ver2}. On the other hand, when $\overline{x}_2^{-\infty} \gg 1$, it reduces to \eqref{Deltax2-off-plane-1}.
Moreover, \eqref{Deltax3-off-plane-2} coincides with \eqref{Deltax3-off-plane-1} up to $\textit{o}(\varepsilon)$. i.e., unlike the gradient-direction displacement, the expression for the vorticity-direction displacement is uniformly valid everywhere. Any non-uniformity arises only in the gradient displacement $\upDelta \overline{x}_2$, similar to what occurs in the in-plane case.

\subsubsection{Self-diffusivity evaluation using off-plane trajectories}
The self-diffusivities in the gradient and vorticity directions can now be determined from the trajectory displacements using the definitions in \eqref{Di-defenition}, by substituting the expressions for $\upDelta \overline{x}_2$ and $\upDelta \overline{x}_3$ from \eqref{Deltaxs-off-plane-1} and \eqref{Deltaxs-off-plane-2}, as appropriate. First, let us evaluate $\hat{D}_2$, which follows the same general strategy as in the in-plane case, but with some additional features due to the off-plane geometry. As before, we split the integral for $\hat{D}_2$. Here, however, the diffusivity involves a double integral over $\overline{x}_2^{-\infty}$ and $\overline{x}_3^{-\infty}$. There is no need to split the integral with respect to $\overline{x}_3^{-\infty}$, since the expression for $\upDelta \overline{x}_2$ is uniformly valid for all $\overline{x}_3^{-\infty}$. In contrast, the integral over $\overline{x}_2^{-\infty}$ must be divided to account for the two forms \eqref{Deltax2-off-plane-1} and \eqref{Deltax2-off-plane-2}, which apply in different regimes of $\overline{x}_2^{-\infty}$. Accordingly, the diffusivity can be written as the sum of two contributions, $\hat{D}_2 = \mathscr{I}_3+\mathscr{I}_4$. Each is calculated as follows:
\begin{eqnarray}
    \mathscr{I}_3 &=& \frac{4 \times 3}{8\, \pi}\, \int_{0}^{b\, \varepsilon^{1/2-h}} d\overline{x}_2^{-\infty} \int_{0}^{\infty} d\overline{x}_3^{-\infty}\, \lvert \overline{x}_2^{-\infty} \rvert\, \left( \upDelta \overline{x}_2\right)^2 \nonumber \\
    &=& \frac{3}{2\, \pi}\, \int_{0}^{b\, \varepsilon^{1/2-h}}d\overline{x}_2^{-\infty}\int_{0}^{\infty} d\overline{x}_3^{-\infty}\, \overline{x}_2^{-\infty}\, \left(-\overline{x}_2^{-\infty}+\sqrt{\left(\overline{x}_2^{-\infty}\right)^2-2\, (\upDelta \overline{x}_3)\,\overline{x}_3^{-\infty}+4\, \varepsilon\,  M_I(d_t)}\right)^2 \nonumber \\
    &=& \frac{3\, \varepsilon^2}{2\, \pi}\,\int_{0}^{\infty} d\overline{x}_3^{-\infty} \int_{0}^{b\, \varepsilon^{-h}}d\hat{x}_2^{-\infty}\, \hat{x}_2^{-\infty}\, \bigg(-\hat{x}_2^{-\infty} \nonumber\\
    &&+\sqrt{\left(\hat{x}_2^{-\infty}\right)^2+4\,   \left[M_I(d_t;\overline{x}_3^{-\infty})-\overline{x}_3^{-\infty}\, N_I(d_t;\overline{x}_3^{-\infty})\right]}\bigg)^2 \nonumber \\
    &=& \frac{3\, \varepsilon^2}{2\, \pi}\, \int_{0}^{\infty} d\overline{x}_3^{-\infty}\left(M_I^*-\overline{x}_3^{-\infty}\, N_I^*\right)^2\, \left[4\, \left(\log b -h\, \log \varepsilon\right)-1-2\, \log \left(M_I^*-\overline{x}_3^{-\infty}\, N_I^*\right)\right] \nonumber \\
    &=& \frac{6\, \varepsilon^2}{\pi}\, \int_{0}^{\infty}d\overline{x}_3^{-\infty}\left(M_I^*-\overline{x}_3^{-\infty}\, N_I^*\right)^2\, \left[\log\left(b\, \varepsilon^{-h}\right)-\frac{1}{4}-\frac{1}{2}\, \log\left(M_I^*-\overline{x}_3^{-\infty}\, N_I^*\right)\right]~,\nonumber\\
\end{eqnarray}
and,
\begin{eqnarray}
    \mathscr{I}_4 &=&\frac{4 \times 3}{8\, \pi}\, \int_{b\, \varepsilon^{1/2-h}}^{\infty}d\overline{x}_2^{-\infty} \int_{0}^{\infty} d\overline{x}_3^{-\infty}\, \lvert \overline{x}_2^{-\infty} \rvert\, \left( \upDelta \overline{x}_2\right)^2 \nonumber \\
    &=& \frac{6\, \varepsilon^2}{\pi}\,\int_{0}^{\infty}d\overline{x}_3^{-\infty} \int_{b\, \varepsilon^{1/2-h}}^{\infty}d\overline{x}_2^{-\infty} \frac{\left(M_I(c_t;\overline{x}_2^{-\infty},\overline{x}_3^{-\infty})-\overline{x}_3^{-\infty}\, N_I(c_t;\overline{x}_2^{-\infty},\overline{x}_3^{-\infty})\right)^2}{\overline{x}_2^{-\infty}} \nonumber \\
    &=& \frac{6\, \varepsilon^2}{\pi}\, \int_{0}^{\infty} d\overline{x}_3^{-\infty} \bigg[ \int_{b\, \varepsilon^{1/2-h}}^{1}d\overline{x}_2^{-\infty}\, \frac{\left(M_I^*-\overline{x}_3^{-\infty}\, N_I^*\right)^2}{\overline{x}_2^{-\infty}} \nonumber \\
    &+& \int_{b\, \varepsilon^{1/2-h}}^{\infty}d\overline{x}_2^{-\infty} \frac{\left(M_I(c_t)-\overline{x}_3^{-\infty}\, N_I(c_t)\right)^2}{\overline{x}_2^{-\infty}}-\int_{b\, \varepsilon^{1/2-h}}^{1}d\overline{x}_2^{-\infty}\, \frac{\left(M_I^*-\overline{x}_3^{-\infty}\, N_I^*\right)^2}{\overline{x}_2^{-\infty}}\bigg] \nonumber \\
    &=& \frac{6\, \varepsilon^2}{\pi}\, \int_{0}^{\infty}d\overline{x}_3^{-\infty}\Bigg\{-\left(M_I^*-\overline{x}_3^{-\infty}\, N_I^* \right)^2\, \left[\frac{1}{2}\, \log \varepsilon +\log\left(b\, \varepsilon^{-h}\right)\right]\nonumber \\
    &+& \int_{0}^{\infty} d\overline{x}_2^{-\infty}\frac{\left(M_I(c_t)-\overline{x}_3^{-\infty}\, N_I(c_t)\right)^2-\left(M_I(d_t)-\overline{x}_3^{-\infty}\, N_I(d_t)\right)^2\, \mathcal{H}(1-\overline{x}_2^{-\infty})}{\overline{x}_2^{-\infty}} \Bigg\}~.\nonumber\\
\end{eqnarray}
Here, $M_I^*$ and $N_I^*$ are shorthand notations for the corresponding quantities evaluated at $d_t$. The factor of $4$ multiplying each integral in the first line arises from restricting the domain of integration of both $\overline{x}_2^{-\infty}$ and $\overline{x}_3^{-\infty}$ to $[0,\infty)$. Since the original integrals extend over the entire real line, symmetry of the integrand allows us to compute the contribution over the positive quadrant and multiply by four to recover the full-domain result. The subsequent evaluation follows the same procedure as in the in-plane diffusivity calculation (see \eqref{inplane-diff-integral-1-evaluation} and \eqref{inplane-diff-integral-2-evaluation}). The main difference in the present case is that the cutoff parameter $b$ may, in general, depend on $\overline{x}_3^{-\infty}$. Nevertheless, as in the in-plane analysis, both $b$ and $h$ are auxiliary parameters introduced to calculate the splitted integrals, and they must cancel in the final expression once we sum $\mathscr{I}_3$ and $\mathscr{I}_4$. Consequently, the resulting expression for the gradient-direction self-diffusivity is
\begin{equation}
       \hat{D}_2 = \frac{3}{\pi}\, \varepsilon^2\, \left[\mathcal{A}_1-\mathcal{A}_2\, \log \varepsilon \right]~,
    \label{full-diff2-asymptotic}
\end{equation}
where the constants $\mathcal{A}_1$ and $\mathcal{A}_2$ are,
\begin{subequations}
    \begin{align}
        \mathcal{A}_1 &= \int_{0}^{\infty}d\overline{x}_3^{-\infty} \bigg\{-\left(M_I^*-\overline{x}_3^{-\infty}\, N_I^* \right)^2\, \left[\frac{1}{2}+\log \left(M_I^*-\overline{x}_3^{-\infty}\, N_I^* \right)\right] \nonumber \\
        &\quad \quad \quad +2\, \int_{0}^{\infty} \frac{d\overline{x}_2^{-\infty}}{\overline{x}_2^{-\infty}}\left[ \left(M_I(c_t)-\overline{x}_3^{-\infty}\, N_I(c_t) \right)^2-\left(M_I^*-\overline{x}_3^{-\infty}\, N_I^* \right)^2\, \mathcal{H}(1-\overline{x}_2^{-\infty}) \right]\bigg\}~,\\
        \mathcal{A}_2 &= \int_{0}^{\infty}d\overline{x}_3^{-\infty}\,\left(M_I^*-\overline{x}_3^{-\infty}\, N_I^* \right)^2~.
    \end{align}
\end{subequations}
Note that $M_I(c_t)$ and $N_I(c_t)$ depend on both $\overline{x}_2^{-\infty}$ and $\overline{x}_3^{-\infty}$, since $c_t$ itself is a function of these two trajectory initializations. In contrast, $M_I^*$ and $N_I^*$, being evaluated at $d_t$, thus depend only on $ \overline{x}_3^{-\infty}$. We observe that the scaling of the gradient-direction diffusivity is identical to that of the in-plane case (see \eqref{in-plane-diff-asymptotic}), differing only in the associated coefficients. Indeed, the structure of the coefficients is formally the same. This becomes evident by comparing $\mathcal{A}_0$ in \eqref{integral-alpha} with $\mathcal{A}_1 $ and $\mathcal{A}_2$ obtained here. In particular, $\mathcal{A}_0$ has the same form as $ \mathcal{A}_1$, except that $K_I$ is replaced by ( $M_I - \overline{x}_3^{-\infty} N_I)$, and an additional outer integral over $\overline{x}_3^{-\infty}$ appears to account for cumulative off-plane effects. In this case, the expression for $\upDelta \overline{x}_3$ is uniformly valid throughout the entire domain, and therefore no splitting of the integral is required. Substituting \eqref{Deltax3-off-plane-1} into \eqref{Di-defenition} with $ i= 3$, we obtain:
\begin{eqnarray}
    \hat{D}_3 &=& \frac{3}{2\, \pi} \int_{0}^{\infty}d\overline{x}_3^{-\infty} \int_{0}^{\infty}d\overline{x}_2^{-\infty}\, \overline{x}_2^{-\infty}\, (\upDelta\overline{x}_3)^2 
    = \frac{3\, \varepsilon^2}{2\, \pi} \int_{0}^{\infty}d\overline{x}_3^{-\infty} \int_{0}^{\infty}d\overline{x}_2^{-\infty}\, \overline{x}_2^{-\infty}\, (\upDelta\hat{x}_3)^2 \nonumber \\
    &=& \frac{3\, \varepsilon^2}{2\, \pi} \int_{0}^{\infty}d\overline{x}_3^{-\infty} \int_{0}^{\infty}d\overline{x}_2^{-\infty}\, \overline{x}_2^{-\infty}\, (2\, N_I(c_t))^2\nonumber \\
    &=&\frac{6\, \varepsilon^2}{ \pi} \int_{0}^{\infty}d\overline{x}_3^{-\infty} \int_{0}^{\infty}d\overline{x}_2^{-\infty}\, \overline{x}_2^{-\infty}\,  N_I(c_t;\overline{x}_2^{-\infty},\overline{x}_3^{-\infty})^2~.
    \label{full-diff3-asymptotic}
\end{eqnarray}
Note that the diffusivity in the vorticity direction scales quadratically with $\varepsilon$, in contrast to the gradient direction, where an additional logarithmic dependence on $\varepsilon$ appears. The self-diffusivity is therefore anisotropic.

We have thus derived the asymptotic scaling of the self-diffusivities of tagged spheres undergoing pairwise interactions, mediated both hydrodynamically and through a non-hydrodynamic central repulsive potential, in a simple shear flow, in the limit of weak interactions.
In the following section, we evaluate the self-diffusivity from first principles and compare the analytical predictions with numerical results for a representative example involving electrical double-layer repulsion.


\section{Validation of asymptotic theory using electrical double-layer interactions}
\label{electrical-potential-results}
To validate the asymptotic theory presented in the preceding section, we consider a simplified model for the repulsive interactions arising from the electrical double layer surrounding colloidal particles. When a colloidal particle is suspended in an electrolyte solution with a sufficiently high concentration of free ions, its surface charge attracts counter-ions from the surrounding fluid, thereby facilitating the formation of an electrical double layer. The layer immediately adjacent to the particle surface, where ions may be strongly bound, is commonly referred to as the compact or Stern layer. Beyond this region lies the diffuse layer, where ions remain mobile and are distributed according to the combined influence of electrostatic forces and thermal motion. Together, the Stern and diffuse layers constitute the electrical double layer, which effectively screens the particle surface charge over a characteristic length scale known as the Debye length. When two particles approach one another at separations comparable to this screening length, the overlap of their diffuse layers results in an increase in electrostatic free energy and osmotic pressure within the intervening fluid. Consequently, this interaction generates a repulsive interaction between the particles.

To account for this interaction, we utilize the Gouy–Chapman model for the electrical double layer. A comprehensive discussion of the underlying assumptions and the limits of applicability of the various approximations used to derive this electrical double-layer potential can be found in \citet{russel1991colloidal}. In this study, we assume that the double layers are thin, characterized by the condition $\tilde{\kappa} = \kappa a^* > 2$, where $\tilde{\kappa}$ is the scaled inverse Debye length with $\kappa$ being the inverse Debye length. For particles with a constant surface potential $\psi$, the interaction potential under the linearised Derjaguin approximation is given by
\begin{equation}
\widehat{V}=\frac{a_1 \beta}{1+\beta} 4\, \pi\, \epsilon_D\, \epsilon_0\, \psi^2\,  \ln\left(1+e^{-\tilde{\kappa} (r-2)}\right)~,
\label{Electrical_double_layer_repulsion_dimensional}
\end{equation}
where $\widehat{V}$ is the dimensional interaction potential due to double layer repulsion, $\epsilon_D$ is the dielectric constant of the suspending medium, and $\epsilon_0$ is the permittivity of free space. Using this form of the double-layer potential, the dimensionless parameter $\varepsilon$, which captures the relative strength of the double-layer repulsion compared to the background flow effects, as well as the dimensionless repulsion force $F(r)$, can be expressed as follows:
\begin{eqnarray}
    \varepsilon = \dfrac{8\, \epsilon_0\, \epsilon_D\, \psi^2}{3\, \mu_f\, \dot{\gamma} a_1^2\, (1+\beta)^2}, \label{Expression_of_varepsilon} \\
    F(r) = \dfrac{\tilde{\kappa}}{1+e^{\tilde{\kappa}\,(r-2)}}. \label{expression_of_F}
\end{eqnarray}
Let us estimate the typical range of values of $\tilde{\kappa}$ and $\varepsilon$. For aqueous monovalent electrolytes at room temperature, typical colloidal ionic strengths of $1-100$ mM correspond to inverse Debye lengths $\kappa \sim 10^8-10^9$ m$^{-1}$, or equivalently Debye lengths $\sim 1-10$ nm. For particle pairs with average radii ($a^*$) in the range $10^{-1}-10$ \textmu m, this implies $\tilde{\kappa} \sim 10^1-10^4$, indicating that most colloidal systems operate well within the thin double-layer regime. Under such conditions, the parameter $\varepsilon$ may vary widely, typically ranging from $O(10^{-3})$ to $O(10^{2})$ or larger depending on factors such as particle sizes, surface potential, and the imposed shear rate. By employing this repulsion force model, we present in this section representative relative trajectories and the dependence of diffusivity on $\varepsilon$ for a monodisperse suspension. We note that the thin-double-layer assumption underlying \eqref{Electrical_double_layer_repulsion_dimensional} requires $\tilde{\kappa}\geq\mathcal{O}(1)$. In the validation that follows, we include the cases $\tilde{\kappa}= 0.1$ and $\tilde{\kappa}= 1$ to test the asymptotic predictions over a broader parameter range, even though these values fall outside the strict regime of validity of the Derjaguin approximation.
\color{black}

\subsection{In-plane trajectories and diffusivity} \label{inplane-numeric-eval}
\begin{figure}
\centering
\includegraphics[width=1.0\textwidth]{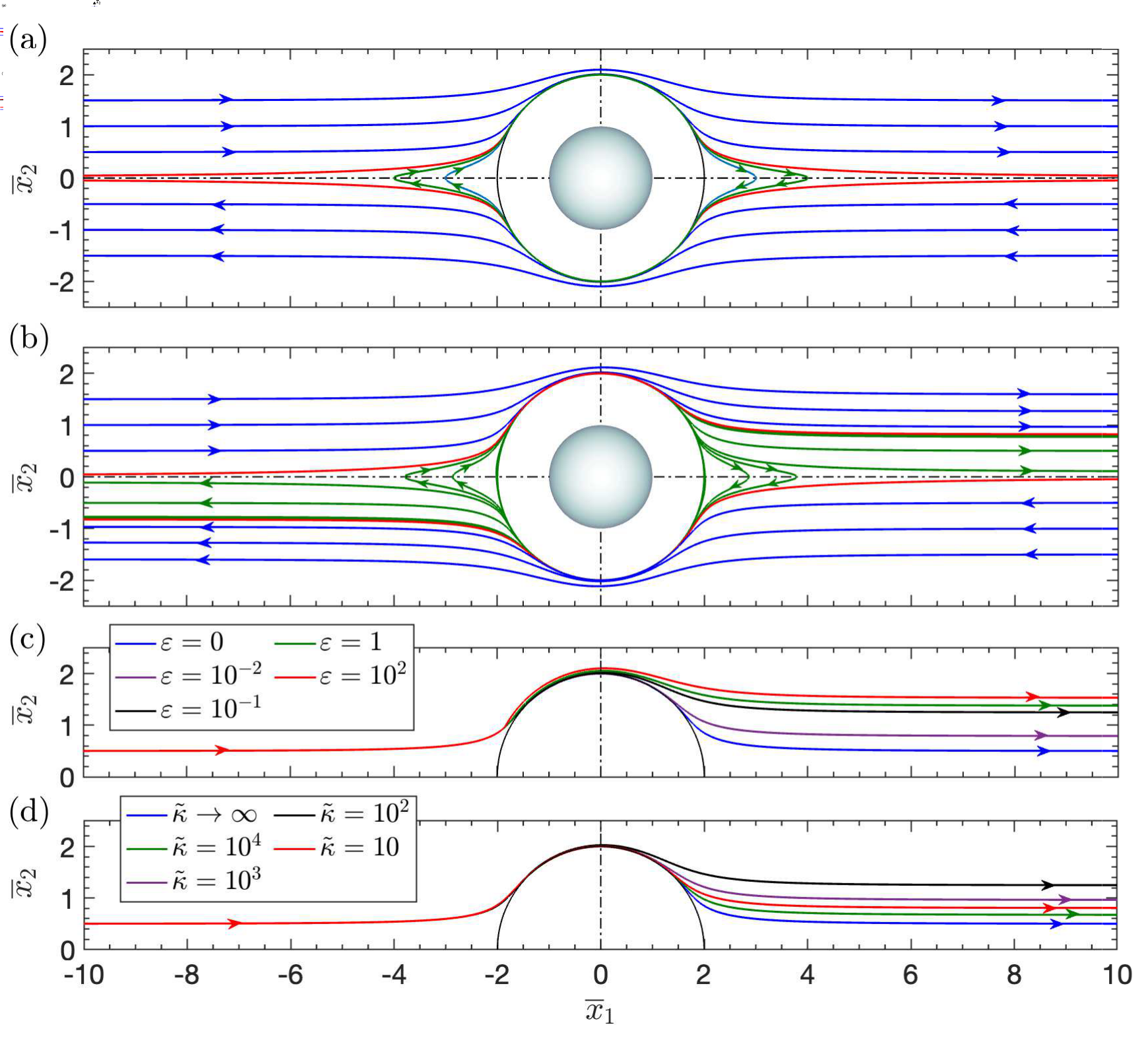}
\caption{Representative relative trajectories of two equal-sized spheres in the shearing plane ($\overline{x}_1$--$\overline{x}_2$ plane): (a) purely hydrodynamic interactions ($\varepsilon = 0$) and (b) in the presence of electrical double-layer repulsion with $\varepsilon = 10^{-1}$ and $\tilde{\kappa}=10$. The sphere at the centre represents the test sphere, and the black circle denotes the collision circle in the shearing plane. Blue curves correspond to open trajectories in both panels (a) and (b). Green curves represent closed trajectories in (a) and spiral trajectories in (b). Red curves represent the separatrices separating different classes of trajectories in both panels (a) and (b). Arrows along the curves denote their directions. Panels (c) and (d) show representative variations in the lateral displacements of in-plane trajectories in the upper half-plane: (c) for different values of $\varepsilon$ with $\tilde{\kappa}=10^{2}$, and (d) for different values of $\tilde{\kappa}$ with $\varepsilon=10^{-1}$. The initial conditions for the trajectories shown in (c) and (d) are $\overline{x}_1=-10$, $\overline{x}_2=0.5$, and $\overline{x}_3=0$.}
\label{sample_inplane_trajectories}
\end{figure}

We begin by analyzing the relative trajectories in the shearing plane (i.e., $\theta=\pi/2$) and then determine the in-plane self-diffusivity both numerically and using the asymptotic expression derived in the previous section. To compute the in-plane trajectories, we numerically integrate equation \eqref{In-plane_trajectory_equation} using a fourth-order Runge-Kutta method for various values of $\varepsilon$ and $\tilde{\kappa}$. Figure \ref{sample_inplane_trajectories}(a) shows in-plane relative trajectories in the absence of double-layer repulsion (i.e., when $\varepsilon=0$). In this scenario, continuum hydrodynamic interactions between the particle pairs result in two distinct categories of relative trajectories: open and closed \citep{batchelor1972hydrodynamic,patra2022collision}. Moreover, these trajectories exhibit fore-aft symmetry. The separatrices (in red color), which separate these closed and open trajectories, asymptotically approach the $\overline{x}_1$-axis as $r \rightarrow \infty$.


Incorporating repulsive interactions (i.e., $\varepsilon > 0$) breaks the fore-aft symmetry of relative trajectories, as illustrated in figure~\ref{sample_inplane_trajectories}(b). Under these conditions, open trajectories remain open but undergo a net lateral displacement downstream, such that $\overline{x}_2^{+\infty} > \overline{x}_2^{-\infty}$ in the upper half-plane and $\overline{x}_2^{+\infty} < \overline{x}_2^{-\infty}$ in the lower half-plane. Trajectories that were closed in the purely hydrodynamic case transform into spiral trajectories that extend downstream, while the separatrices persist but suffer net positive lateral displacements and thus widen the gap from the $\overline{x}_1$-axis at the downstream to accommodate the spiraling trajectories. Near $\phi \sim \pi/2$ in the vicinity of the collision sphere, the trajectories crowd together, reflecting the presence of the inner and outer regions discussed in \S~\ref{inplane-analytical_small_NR}. Figures~\ref{sample_inplane_trajectories}(c) and (d) show the variation in the downstream offsets of the in-plane trajectories in the upper half-plane for different values of $\varepsilon$ and $\tilde{\kappa}$, respectively, with the initial, upstream positions of these trajectories set at $\overline{x}_1=-10$ and $\overline{x}_2^{-\infty} = 0.5$. No lateral displacement occurs when $\epsilon=0$ or $\tilde{\kappa} \rightarrow \infty$. As expected, for a given $\tilde{\kappa}$, increasing $\varepsilon$ strengthens the repulsive interaction, leading to progressively larger lateral shifts of the trajectories downstream. In contrast, figure~\ref{sample_inplane_trajectories}(d) indicates that, for a given $\varepsilon$, the dependence of the lateral displacement on $\tilde{\kappa}$ is non-monotonic.

\begin{figure}
    \centering
\includegraphics[width=0.8\linewidth]{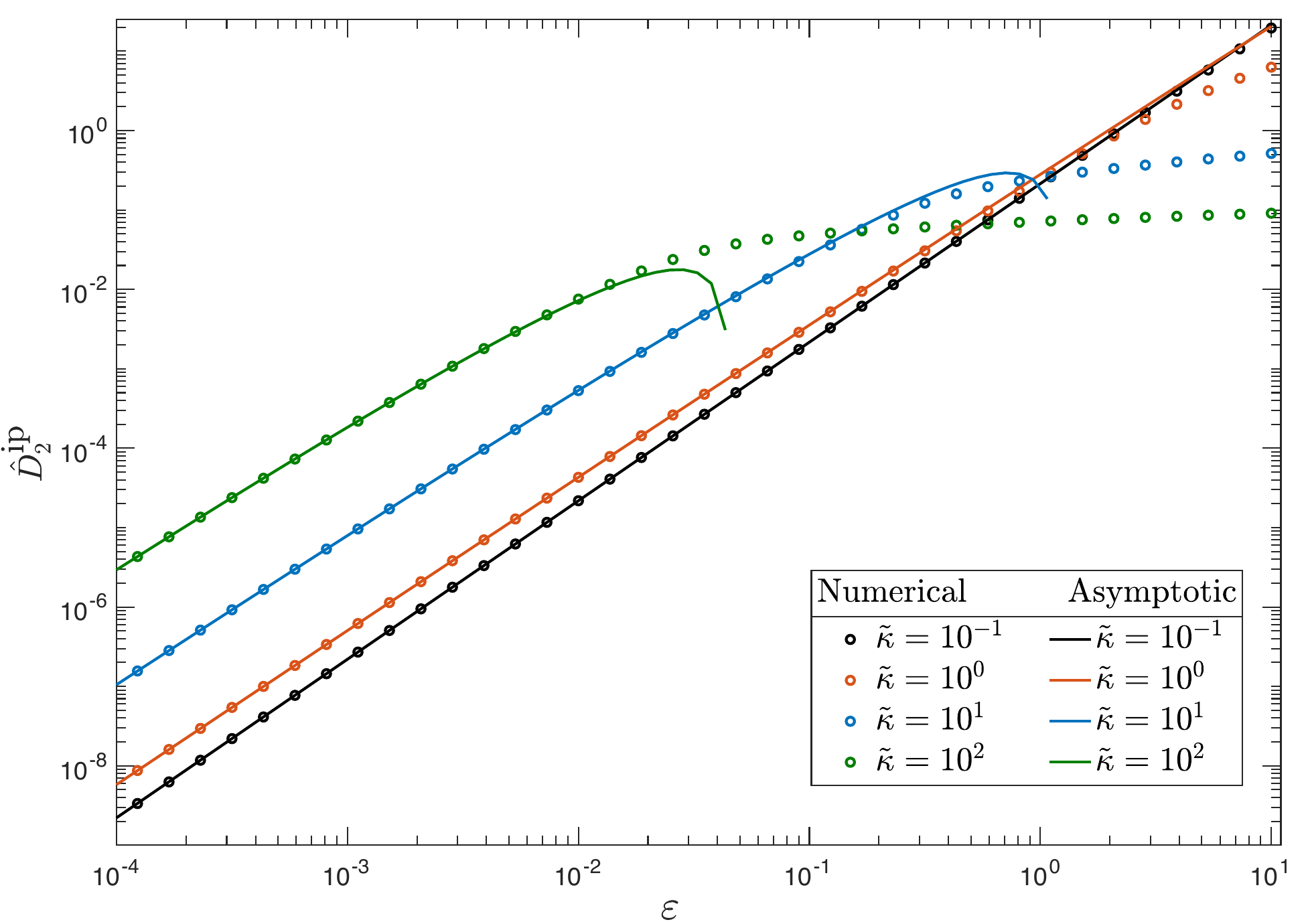}
    \caption{Comparison of numerical results (circle markers) and asymptotic predictions (solid lines) for the in-plane self-diffusivity of spheres with size ratio $\beta = 1$, shown for various values of $\tilde{\kappa}$ as a function of $\varepsilon$. Good agreement is observed in the regime $\varepsilon\, \tilde{\kappa} \ll 1$.}
    \label{fig:comparing-numeric-vs-asymptitic-inplane-diff}
\end{figure}

We numerically compute the in-plane self-diffusivity by evaluating the integral in \eqref{D2-inplane-integral} using simulated trajectory displacements. For a given set of parameters, $\varepsilon$ and $\tilde{\kappa}$, we first calculate $\left(\upDelta \overline{x}_2\right)_\textrm{ip}$ over a discrete set of initial upstream offsets $\overline{x}_2^{-\infty}$, with $\overline{x}_1^{-\infty}=-10^3$ for each initial offset. Since $\left(\upDelta \overline{x}_2\right)_\textrm{ip}$ decreases with increasing $\overline{x}_2^{-\infty}$ (as shown in figure~\ref{sample_inplane_trajectories} and in formula \eqref{inplane-gradient-displacement-ver1}), and given that the repulsive interaction decays exponentially for $\mathcal{O}(1)$ or larger values of $\tilde{\kappa}$, the integrand becomes negligible beyond a certain $\overline{x}_2^{-\infty}$. This allows us to truncate the integral at a finite upper limit while maintaining sufficient accuracy. To ensure an accurate representation of contributions from small values of $\overline{x}_2^{-\infty}$, we employ logarithmically spaced grids for sampling the upstream offsets, which extend from a small lower bound to a sufficiently large upper cutoff. Beyond this upper limit, the integrand is effectively negligible. For instance, when $\tilde{\kappa} = 10^{-1}$, we consider $\overline{x}_2^{-\infty}$ in the range $[10^{-5}, 10^3]$, discretizing it with $10^4$ logarithmically spaced points. Conversely, for $\tilde{\kappa} = 10^{2}$, we restrict the interval to $\overline{x}_2^{-\infty} \in [10^{-5}, 10^1]$ and discretize it using only $150$ points.

Figure \ref{fig:comparing-numeric-vs-asymptitic-inplane-diff} shows the numerically computed (circle markers) in-plane diffusivity as a function of $\varepsilon$ for different values of $\tilde{\kappa}$. In all cases, the diffusivity increases monotonically with $\varepsilon$, as stronger repulsive interactions produce larger lateral displacements (see figure \ref{sample_inplane_trajectories}(c)). However, this increase becomes slow at large $\varepsilon$, particularly when $\varepsilon\, \tilde{\kappa} > O(1)$, indicating a saturation regime in which further increases in interaction strength produce only marginal changes in the net trajectory displacements.

The dependence of diffusivity on $\tilde{\kappa}$ is more subtle and reflects a competition between the interaction range and the magnitude of the local repulsive force. From \eqref{expression_of_F}, the force scales as $F(r) \sim \tilde{\kappa}$ near contact $(r \to 2)$ and decays over a characteristic length scale $1/\tilde{\kappa}$. For small $\varepsilon$, the repulsive interaction weakly perturbs the hydrodynamic trajectories, and the dominant contribution to the diffusivity arises from near-contact interactions. In this regime, increasing $\tilde{\kappa}$ enhances the magnitude of the repulsive force in the near-contact region, thereby increasing the lateral displacement and diffusivity. In contrast, for larger $\varepsilon$, the repulsive interaction influences the trajectories over a finite separation distance determined by its range. Since the interaction range decreases as $1/\tilde{\kappa}$, smaller values of $\tilde{\kappa}$ allow particles to interact over a larger interparticle separation distance, resulting in greater cumulative displacements and higher diffusivities. This interplay between the near-contact force scaling and the interaction range leads to a non-monotonic dependence of the in-plane diffusivity on $\tilde{\kappa}$ for a given $\varepsilon$, consistent with the behavior observed in the trajectory displacements as well (see figure \ref{sample_inplane_trajectories}(d)).


\begin{figure}
\centering
\includegraphics[width=1.0\textwidth]{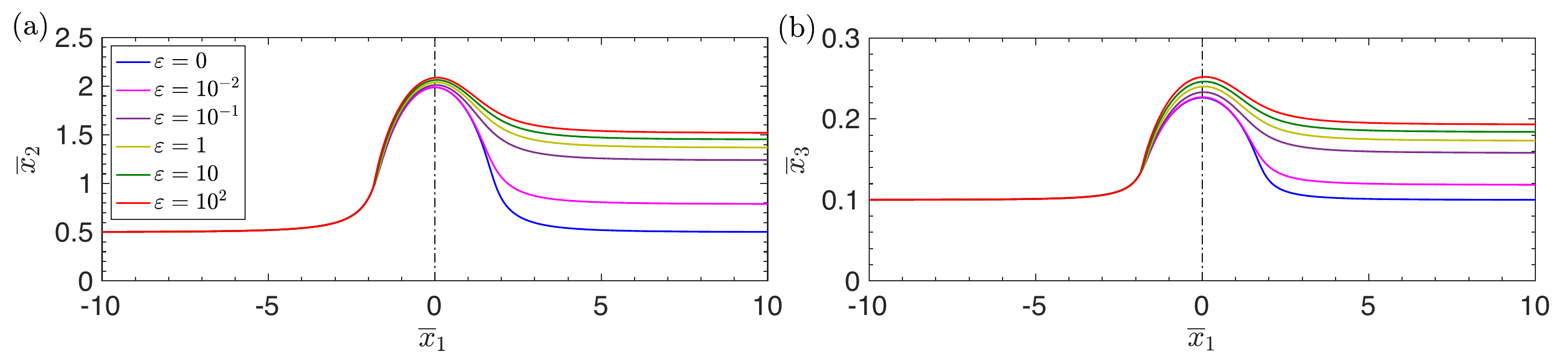}
\caption{Typical lateral displacements of trajectories in (a) $\overline{x}_1$--$\overline{x}_2$ and (b) $\overline{x}_1$--$\overline{x}_3$ plane for different values of the parameter $\varepsilon$ when $\tilde{\kappa} = 10^2$. The initial conditions used in all cases are $\overline{x}_1=-10$, $\overline{x}_2=0.5$ and $\overline{x}_3=0.1$.}
\label{Off-plane_trajectories_for_kappatilde_100_different_epsilon}
\end{figure}


\begin{figure}
\centering
\includegraphics[width=1.0\textwidth]{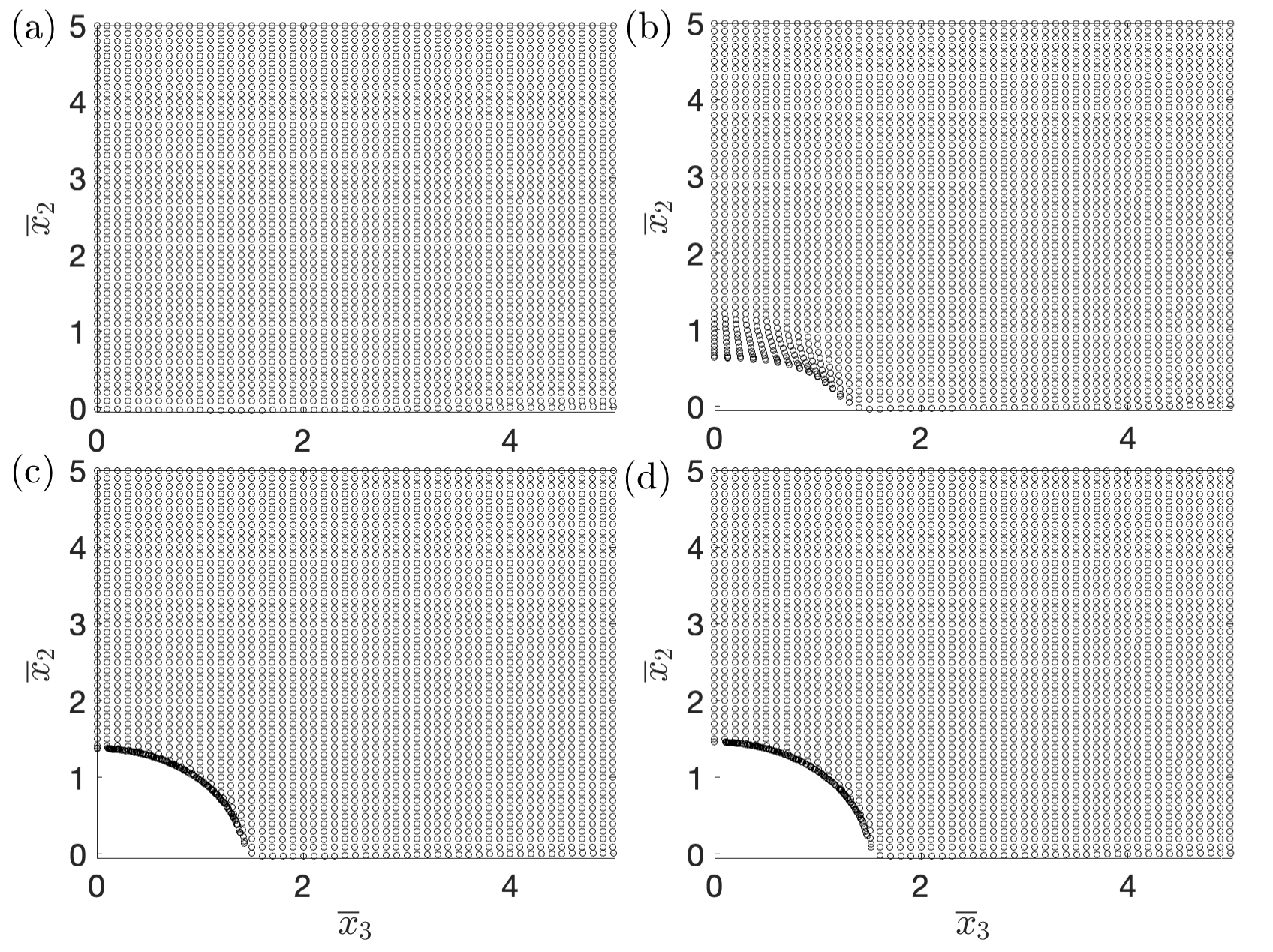}
\caption{Distribution of the far-downstream coordinates $(\overline{x}_3^{+\infty}, \overline{x}_2^{+\infty})$ of the relative trajectories of the satellite sphere, initialized far-upstream on a regular grid over $[0,5] \times [0,5]$. The reference sphere is initially located at the origin. The subfigures correspond to varying strength of the repulsion: (a) $\varepsilon = 10^{-5}$, (b) $\varepsilon = 10^{-2}$, (c) $\varepsilon = 1$, and (d) $\varepsilon = 10$, with $\tilde{\kappa} = 10^2$.}
\label{Cross_stream_displacement_x3_x2_kappatilde_100_NR_1e-2_1_10_100_dl_01}
\end{figure}

\begin{figure}
    \centering
\includegraphics[width=1\linewidth]{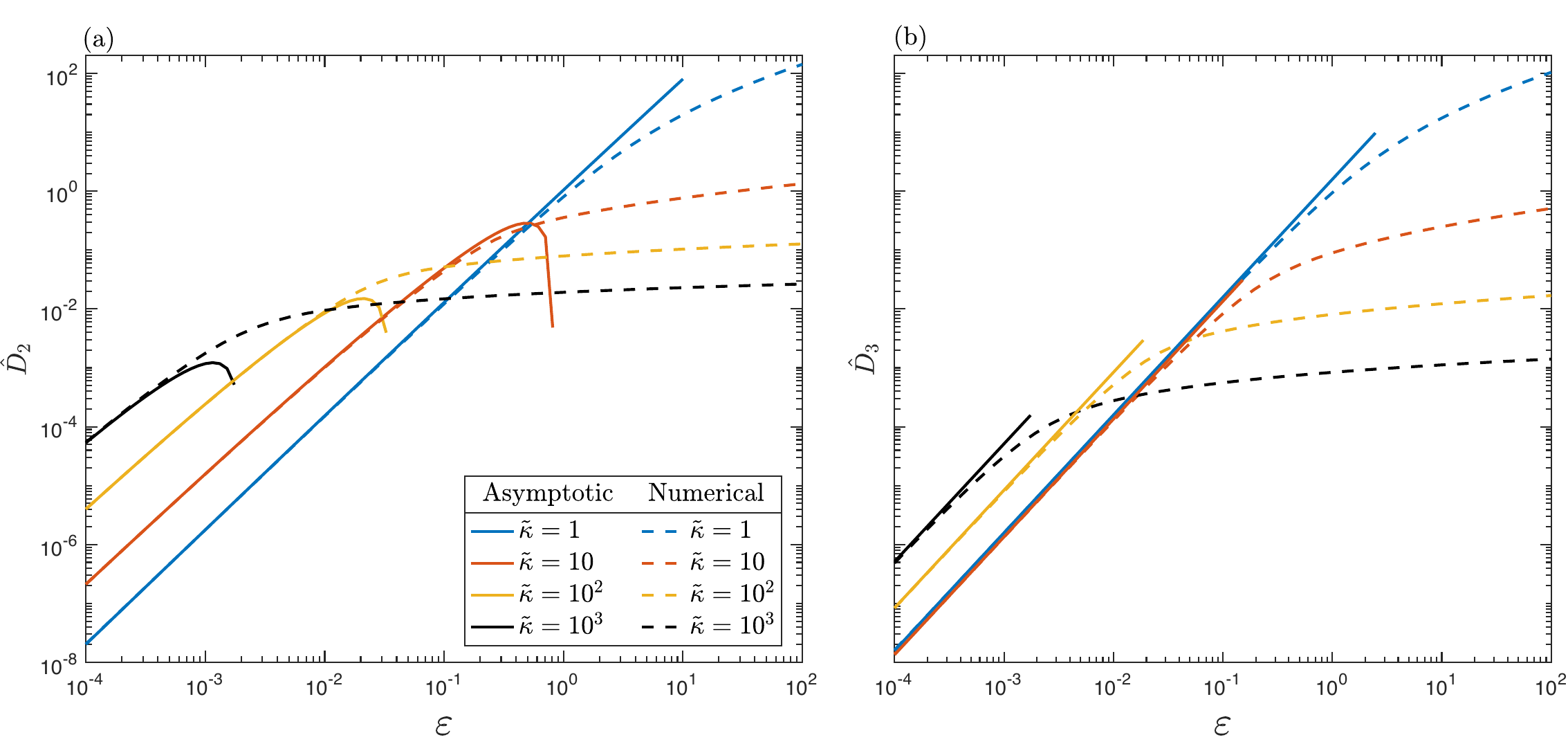}
    \caption{Comparison of numerical (dashed curves) and asymptotic (solid curves) predictions of (a) the gradient-direction self-diffusivity and (b) the vorticity-direction self-diffusivity for spheres with size ratio $\beta = 1$, shown for various values of $\tilde{\kappa}$. The asymptotic expressions agree well with the numerical results in the regime $\varepsilon\, \tilde{\kappa} \ll 1$.}
    \label{fig:comparing-numeric-vs-asymptitic-full-diff}
\end{figure}

We next evaluate the asymptotic expression \eqref{in-plane-diff-asymptotic} for the in-plane self-diffusivity. This requires the computation of $\mathcal{A}_0$ from the integral in \eqref{integral-alpha} over the zero-$\varepsilon$ trajectory. To accurately evaluate this integral, we first compute the full in-plane trajectory $\phi(r)$ numerically, since the available asymptotic representation of $\phi(r)$ is only valid in a layer-wise sense and is insufficient to resolve the near-contact region. We discretize $\phi(r)$ using uniformly spaced $4000$ points for all values of $\tilde{\kappa}$, while retaining the same discretization in $\overline{x}_2^{-\infty}$ as we used earlier in the numerical evaluation of diffusivity. We then evaluate the integral in \eqref{integral-alpha} using the trapezoidal rule and use the resulting value of $\mathcal{A}_0$ to compute the asymptotic in-plane diffusivity for the same set of $\tilde{\kappa}$ values considered previously. The asymptotic predictions, plotted as solid curves in figure~\ref{fig:comparing-numeric-vs-asymptitic-inplane-diff}, show excellent agreement with the fully numerical results in the limit $\varepsilon \ll 1$, thus validating the asymptotic result. However, the range of validity of asymptotic expression decreases as $\tilde{\kappa}$ increases. This trend reflects the structure of the governing equation \eqref{In-plane_trajectory_equation}, in which the perturbation term appears as $\varepsilon F(r)$, with $F(r) \sim \tilde{\kappa}$ in close separation distance. As a result, the effective small parameter that governs asymptotics is $\varepsilon \tilde{\kappa}$, rather than $\varepsilon$ alone. Consequently, larger values of $\tilde{\kappa}$ require proportionally smaller values of $\varepsilon$ for the asymptotic approximation to remain valid.

\subsection{Off-plane trajectories and transverse diffusivities}
In the off-plane case, particle trajectories undergo full three-dimensional excursions, resulting in net lateral displacements in both the velocity-gradient and vorticity directions. Figure \ref{Off-plane_trajectories_for_kappatilde_100_different_epsilon} illustrates the typical variation in the lateral displacements of the trajectories in the $(\overline{x}_1,\overline{x}_2)$- and $(\overline{x}_1,\overline{x}_3)$-planes for different values of $\varepsilon$ when $\tilde{\kappa}=10^2$. As expected, the lateral shifts of the particle trajectories, in both $\overline{x}_2$ and $\overline{x}_3$ directions, increase with the strengthening of repulsive interactions. 

In figure \ref{Cross_stream_displacement_x3_x2_kappatilde_100_NR_1e-2_1_10_100_dl_01}, we represent the lateral displacements by plotting the final coordinates $(\overline{x}_2^{+\infty},\overline{x}_3^{+\infty})$
of the trajectories that were initialized on a regular grid in $0 \leq \overline{x}_2^{-\infty} \leq 5$, $0 \leq \overline{x}_3^{-\infty} \leq 5$ far upstream, for various values of $\varepsilon$ with $\tilde{\kappa}=10^2$. For a small repulsive strength ($\varepsilon=10^{-5} $), almost all trajectories return to their original streamlines, indicating negligible net displacement (see figure \ref{Cross_stream_displacement_x3_x2_kappatilde_100_NR_1e-2_1_10_100_dl_01}(a)). However, as $\varepsilon$ increases, the trajectories exhibit progressively larger displacements, leading to a pronounced distortion of the initially uniform grid, particularly evident near the lower-left corner of the plots. This distortion clearly indicates displacements in the gradient and vorticity directions $(\upDelta \overline{x}_2$ and $\upDelta \overline{x}_3)$ of the trajectories. These distortions become more pronounced with higher values of $\varepsilon$, consistent with the notion that stronger repulsive interactions yield greater trajectory displacements. Additionally, Figure \ref{Cross_stream_displacement_x3_x2_kappatilde_100_NR_1e-2_1_10_100_dl_01} illustrates that the displacements decrease rapidly as initial upstream offsets $\overline{x}_2^{-\infty}$ and $\overline{x}_3^{-\infty}$ increase, for fixed $\tilde{\kappa}$, suggesting that significant contributions primarily arise from smaller values of $\overline{x}_2^{-\infty}$ and $\overline{x}_3^{-\infty}$ close to the collision sphere.

Following the same procedure outlined for calculating the in-plane diffusivity in \S~\ref{inplane-numeric-eval}, we numerically compute the diffusivities $\hat{D}_2$ and $\hat{D}_3$ using \eqref{Di-defenition}. In addition, we evaluate the integrals in equations \eqref{full-diff2-asymptotic} and \eqref{full-diff3-asymptotic} to obtain the corresponding asymptotic results. The comparison between the numerical and asymptotic results, shown in figure \ref{fig:comparing-numeric-vs-asymptitic-full-diff}, demonstrates good agreement in the limit $\varepsilon \ll 1$, thereby substantiating the validity of our asymptotic scalings. Consistent with the in-plane results, we observe that both diffusivity components exhibit qualitatively analogous dependencies on $\varepsilon$ and $\tilde{\kappa}$. This observation underscores the fundamental mechanisms that govern the dependence of trajectory displacements on both the interaction strength $\varepsilon$ and the interaction range $\tilde{\kappa}$, as discussed previously.
\color{black}
\section{Summary and Conclusion} \label{conclusion}
We investigated shear-induced self-diffusion in a dilute suspension of smooth, neutrally buoyant spherical particles interacting through hydrodynamics and a short-range repulsive central potential. In purely hydrodynamic interactions, pair trajectories in a simple shear flow exhibit fore-aft symmetry, resulting in no net displacement following each encounter. Consequently, binary hydrodynamic interactions alone cannot lead to diffusive behavior. However, even weak repulsive interactions between particles break this symmetry, leading to irreversible particle trajectories, with cumulative lateral displacements giving rise to shear-induced self-diffusion. 

Employing asymptotic analysis in the regime of weak repulsive interaction forces, we derived analytical expressions for the displacements of trajectories and the resulting self-diffusivities. The introduction of repulsive interactions fundamentally alters the topology of pair trajectories: open trajectories experience finite lateral shifts away from the collision sphere, whereas the stable closed trajectories typical of purely hydrodynamic interactions become unstable and transition into spiral trajectories. Notably, because spiral trajectories lack upstream counterparts, only open trajectories contribute to the self-diffusivity of a tagged sphere.

The perturbation analysis revealed two distinct spatial regions that govern the dynamics. In the outer region, where the particles are well separated, trajectory perturbations can be described using a regular perturbation expansion. Conversely, at close particle separation distances, the governing equations become singular, necessitating a boundary-layer analysis. By matching the inner and outer solutions, we derived expressions for the net lateral displacement. This displacement depends on the upstream offset of the interacting particles and exhibits two distinct asymptotic regimes: trajectories with large upstream offsets scale linearly with the interaction strength, while those with small offsets are primarily influenced by near-contact dynamics. These results suggest that experimental measurements of self-diffusivity could potentially be used to infer the nature or strength of particle interactions in suspensions.

Integrating the squared displacements across the ensemble of upstream configurations yields shear-induced self-diffusivity. A central conclusion of this work is that the $\varepsilon^2\,\lvert \log \varepsilon \rvert$ and $\varepsilon^2$ scalings for the gradient and vorticity diffusivities, respectively, are universal consequences of any weak, central perturbation to reversible Stokes-flow pair trajectories. The specific pair potential enters only through the integral functionals $K_I$, $M_I$, and $N_I$, which weight the perturbation force against the hydrodynamic mobility functions. The asymptotic structure itself is invariant.

Direct quantitative validation of the present results against existing data is not straightforward, since our analysis is based on binary interactions in the strict dilute limit $\Phi_v \ll 1$, a regime that is difficult to access experimentally while still obtaining statistically robust diffusivity measurements. Most simulations of shear-induced self-diffusion \citep{wang1996transverse,marchioro2001shear,sierou2004shear} and experiments \citep{eckstein1977self,leighton1987measurement,phan1999measurement,breedveld2001measuring} report data at $\Phi_v \gtrsim 0.1$, well outside the dilute regime. Nonetheless, several points of contact with the literature merit discussion. The predicted anisotropy $\hat{D}_2 > \hat{D}_3$ is qualitatively consistent with the anisotropy universally reported across these studies. Among the simulations, \citet{wang1996transverse} considered purely hydrodynamic interactions and reported $D_2 \sim 0.11\,\Phi_v^2$ and $D_3 \sim 0.005\,\Phi_v^2$; since fore-aft symmetry is preserved in the absence of non-hydrodynamic forces, the leading contribution arises from three-body interactions and scales as $\Phi_v^2$, a fundamentally different mechanism from the binary symmetry-breaking studied here. A more relevant comparison is offered by \citet{da1996shear}, who introduced surface roughness as a short-range non-hydrodynamic interaction and obtained $D_2 \approx 0.02\,\Phi_v\,\dot{\gamma}\,a^2$. The $\mathcal{O}(\Phi_v)$ scaling is consistent with our framework, since roughness, like repulsive forces, breaks fore-aft symmetry at the pairwise level. As a rough analogy, a roughness length $\delta/a \sim 0.1$ translates to an effective force range $1/\tilde{\kappa} \sim 0.1$, or $\tilde{\kappa} \sim 10$; for $\varepsilon \sim 1$, our numerical results in figure~\ref{fig:comparing-numeric-vs-asymptitic-full-diff}(a) give $\hat{D}_2 \approx 0.02$, broadly consistent with this estimate, though $\varepsilon \sim 1$ lies outside the asymptotic regime of the present theory. The experiments of \citet{zarraga2002measurement} are notable in this context: they measured shear-induced self-diffusivity at volume fractions as low as $\Phi_v = 0.05$ and found a linear dependence on $\Phi_v$, the signature of binary symmetry-breaking, with magnitudes at least an order of magnitude larger than predicted by existing theories. \citet{zarraga2002measurement} explored several possible explanations, including inertial lift and non-Newtonian effects of the suspending fluid, but none proved sufficient. The discrepancy was subsequently resolved by \citet{zurita2007swapping}, who showed that wall-induced swapping trajectories in the bounded Couette geometry quantitatively account for the measured diffusivities. This resolution underscores that in confined geometries, wall effects can dominate over bulk symmetry-breaking mechanisms, and highlights the importance of using sufficiently wide channels when attempting to isolate the unbounded binary-interaction contribution that the present theory describes. To our knowledge, no study has systematically measured shear-induced self-diffusivity in the truly dilute, unbounded limit with controlled, tuneable non-hydrodynamic repulsion, and such measurements, for instance using particles with adjustable surface charge in electrolytes of varying ionic strength, would provide a direct test of the $\varepsilon^2\,|\log\varepsilon|$ scaling derived here.

An important limitation of the present analysis is the neglect of Brownian motion. The shear-induced contribution to self-diffusivity scales as $D_\textrm{shear}\sim \Phi_v \varepsilon^2 \dot{\gamma}a^2$, while the Brownian diffusivity is $D_B = k_BT/(6\pi\mu_f a)$. The two become comparable when the Péclet number  $Pe= \dot{\gamma}a^2/D_B$ satisfies  $Pe \sim 1/(\Phi_v \,\varepsilon^2)$. For micrometre-scale colloids in water ($a \sim 1 \mu\textrm{m}, \, \mu_f \sim 10^{-3} \,\textrm{Pa.s}, T = 300 \,\textrm{K}$), $D_B \sim 2 \times 10^{-13} \textrm{m}^2/\textrm{s}$, and even at modest $\Phi_v \sim 0.01$ and $\varepsilon \sim 0.1$, the required $Pe$ is of order $10^4$, corresponding to $\dot{\gamma} \sim 10^{3} \, \textrm{s}^{-1}$. The present non-Brownian theory is therefore quantitatively applicable only at shear rates exceeding this threshold, or for larger particles where $D_B$ is smaller. Mapping the finite-$Pe$ crossover between Brownian and shear-dominated diffusion remains an important open problem.

Several directions for future work naturally follow from the current analysis. An important extension would be to incorporate Brownian motion, which would allow investigation of the interplay between thermal fluctuations and shear-driven hydrodynamic interactions in determining particle trajectories and diffusivity. The present theoretical framework could also be extended to time-dependent flows, such as oscillatory shear flows. In this context, periodic reversal of the flow may alter the symmetry properties of particle encounters, thereby affecting the resulting diffusive transport characteristics. Another promising direction is the extension to polydisperse suspensions. The present formulation already admits arbitrary size ratio $\beta=a_2/a_1$, so the pairwise diffusivity for any given $(\beta,\varepsilon,\kappa)$ can be computed without further theoretical development. The principal challenge lies in averaging over the size distribution to obtain the effective self-diffusivity of a tagged sphere in a polydisperse mixture, accounting for the $\beta$-dependence of both the encounter rate and the trajectory displacements. Lastly, a more realistic description of interparticle interactions could be achieved by incorporating the full Derjaguin-Landau-Verwey-Overbeek (DLVO) potential, which accounts for both the attractive van der Waals forces and the repulsive interactions arising from the electrical double layer. Together, these extensions would enhance the applicability of the present theoretical framework and provide deeper insights into the self-diffusivity in polydisperse suspensions.

\vspace{5mm}
\textbf{Acknowledgments.} A.R. acknowledges SERB project CRG/2023/008504 for funding. P.P. acknowledges the support from the Prime Minister’s Research Fellows (PMRF) scheme at the Ministry of Education, Government of India (project no. SB22230184AMPMRF008746). A.V.S.N. acknowledges the support from the Prime Minister’s Research Fellows (PMRF) scheme at the Ministry of Education, Government of India (project no. SB22230200AMPMRF008746).

\vspace{5mm}
\textbf{Declaration of interests.} The authors report no conflict of interest.

\appendix
\section{Asymptotic analysis of off-plane trajectories for $\varepsilon \ll 1$} \label{full-analytical_small_NR}
\subsection{Outer layer solution}
Substituting the regular perturbation expansions for $\theta$ and $\phi$ into \eqref{dphidr_equation}, we obtain the governing equations separated into their leading-order and first-order correction terms as 
\begin{subequations}
    \begin{align}
    \mathcal{O}(1) : && \frac{d\phi_0}{dr} &= -\frac{\Big[\sin^2\phi_0 + \frac{1}{2} B \left(\cos^2\phi_0 - \sin^2\phi_0 \right) \Big]}{(1-A)\,r\,\sin\phi_0\,\cos\phi_0\, \sin^2 \theta_0}~, \label{O(1)_equation_for_phi_outofplane}\\
\mathcal{O}(\varepsilon) : && \frac{d\phi_1}{dr} &=  \frac{\left(\frac{1}{2}\, B- \sin^2\phi_0\right)\, \phi_1}{r \, (1-A)\, \sin^2\phi_0\,\cos^2\phi_0\, \sin^2 \theta_0} + \frac{2\, \cos \theta_0\, \left(\frac{1}{2}\, B+(1-B)\, \sin^2\phi_0\right)\, \theta_1}{r\, (1-A)\, \sin^3 \theta_0\, \sin \phi_0\, \cos \phi_0}\nonumber \\ &&  &+ \frac{\left(\frac{1}{2}\, B+(1-B)\, \sin^2\phi_0\right)\, G\, F}{r^2\,(1-A)^2\,\sin^2\phi_0\,\cos^2\phi_0\, \sin^4 \theta_0}~.\label{O(NR)_equation_for_phi_outofplane}
\end{align}
\label{phi-expanded-eqns}
\end{subequations}
Similarly, from \eqref{dthetadr_equation} we get
\begin{subequations}
    \begin{align}
    \mathcal{O}(1) : && \frac{d\theta_0}{dr} &= \frac{(1-B)}{r\, (1-A)}\, \frac{\cos \theta_0}{\sin \theta_0}~, \label{O(1)_equation_for_theta_outofplane}\\
\mathcal{O}(\varepsilon) : && \frac{d\theta_1}{dr} &=  -\frac{(1-B)\, \theta_1}{r \, (1-A)\, \sin^2\theta_0} - \frac{(1-B)\, \cos \theta_0\, \, G\, F}{r^2\,(1-A)^2\,\sin \phi_0\,\cos \phi_0\, \sin^3 \theta_0}~.\label{O(NR)_equation_for_theta_outofplane}
\end{align}
\label{theta-expanded-eqns}
\end{subequations}
Note that the $\mathcal{O}(1)$ equation for $\theta_0$ depends on $\phi_0$, while the $\mathcal{O}(\varepsilon)$ equation for $\theta_1$ is independent of $\phi_1$. Consequently, in the outer layer, the evolution of $\phi$ at each order is effectively driven by the corresponding $\theta$ solution. The solutions to the leading-order equations \eqref{O(1)_equation_for_phi_outofplane} and \eqref{O(1)_equation_for_theta_outofplane}, subject to the boundary conditions \eqref{O(1)_boundary_condition_for_phi_and_theta}, are already known from \citet{batchelor1972hydrodynamic}, as expressed in \eqref{batchelor-solution-off-plane}. To address the $\mathcal{O}(\varepsilon)$ corrections, we first rewrite equation \eqref{O(NR)_equation_for_theta_outofplane} in the form:
\begin{equation}
    \frac{d}{d r}(r \, \theta_1)+r \, \theta_1\, \left[\frac{(1-B)}{r\, (1-A)\, \sin^2\theta_0}-\frac{1}{r}\right] = -\frac{(1-B)\, \cos \theta_0\, G\, F}{r\, (1-A)^2\, \sin \phi_0\, \cos \phi_0\, \sin^3 \theta_0}~.
    \label{O(NR)_equation_for_theta_outofplane-1}
\end{equation}
To solve this first-order ordinary differential equation, we first compute the integrating factor as
\begin{eqnarray}
    J_1(r) &=& \int^{r} \left[\frac{(1-B)}{r\, (1-A)\, \sin^2\theta_0}-\frac{1}{r}\right]\, dr 
    = \int^{r}\left[\frac{1}{\sin^2\theta_0}\, \frac{\sin \theta_0}{\cos \theta_0}\, \frac{d\theta_0}{dr}-\frac{1}{r}\right]\, dr \nonumber \\
    &=& \int^{r} \left[2\, \csc (2\, \theta_0)\, d\theta_0-\frac{dr}{r}\right] \quad \Leftarrow  \textrm{used \eqref{O(1)_equation_for_theta_outofplane}}  \nonumber\\
    &=& \log \left[\frac{\tan \theta_0}{r}\right]~.
\end{eqnarray}
Hence, the general solution of equation \eqref{O(NR)_equation_for_theta_outofplane-1} can be written as
\begin{eqnarray}
    r\, \theta_1 &=& e^{-J_1(r)}\, \left[\textrm{Const.}-\int^{r} \frac{(1-B')\, \cos \theta_0'\, G'\, F'\, e^{J_1(r')}}{r'\, (1-A')^2\, \sin \phi_0'\, \cos \phi_0'\, \sin^3 \theta_0'}\, dr'\right] \nonumber \\
    &=& \frac{r}{\tan \theta_0}\, \left[\textrm{Const.}+\int_{r}^{\infty} \frac{(1-B')\, \cos \theta_0'\, G'\, F'}{r'\, (1-A')^2\, \sin \phi_0'\, \cos \phi_0'\, \sin^3 \theta_0'}\, \frac{\tan \theta_0'}{r'}\, dr'\right] \nonumber\\
    &=& \frac{r\, \cos \theta_0}{\sin \theta_0}\, \left[\textrm{Const.}+ \int_{r}^{\infty} \frac{(1-B')\, \cos \theta_0'\, G'\, F'}{r'\, (1-A')^2\, \sin \phi_0'\, \cos \phi_0'\, \sin^2 \theta_0'}\, \frac{dr'}{r'\, \cos \theta_0'} \right] \nonumber \\
    &=& \frac{\overline{x}_3^{-\infty}\, e^{Q/2}}{\sin \theta_0}\, \left[ \textrm{Const.}+ \int_{r}^{\infty} \frac{(1-B')\, \cos \theta_0'\, G'\, F'}{r'\, (1-A')^2\, \sin \phi_0'\, \cos \phi_0'\, \sin^2 \theta_0'}\, \frac{e^{-Q'/2}\, dr'}{\overline{x}_3^{-\infty}}\right]~,
    \label{theta1-solution0}
\end{eqnarray}
where we have employed \eqref{theta0-solution} in the final step, allowing $\overline{x}_3^{-\infty}$ to be canceled inside and outside the integral by an appropriate choice of the integration constant. Applying the boundary condition from \eqref{O(NR)_boundary_condition_for_phi_and_theta} then sets this constant to zero, yielding $\theta_1$ in the outer layer $\textrm{O}_1$ as 
\begin{equation}
    r\, \theta_1^{-} = \frac{e^{Q/2}}{\sin \theta_0}\, \int_{r}^{\infty}\frac{(1-B')\, \cos \theta_0'\, G'\, F'\, e^{-Q'/2}}{r'\, (1-A')^2\, \sin^2 \theta_0'\, \sin \phi_0'\, \cos \phi_0'}\, dr'~,
    \label{theta1-solution-1}
\end{equation}
where, the integration is carried out over $\phi_0' \in (\pi/2, \pi)$, i.e., along the negative branch. By introducing the integral defined in \eqref{NI-definition}, the solution \eqref{theta1-solution-1} can be expressed more compactly as in \eqref{theta1m-solution}. For the solution $\theta_1^{+}$, as in the in-plane case, the integration constant does not vanish. Moreover, the integral in \eqref{theta1-solution0} is inherently taken over the positive branch, i.e., $\phi_0' \in (0, \pi/2)$. Consequently, the outer-layer solution in region $\textrm{O}_2$ can be written in general as \eqref{theta1p-solution} with the introduction of an integration constant $I_{\theta_1}^+$. To proceed with solving for $\phi_1$, we re-write \eqref{O(NR)_equation_for_phi_outofplane} as\begin{eqnarray}
        \frac{d}{d r}\left(r\, \phi_1\right)+\left[\frac{\sin^2 \phi_0-B/2}{r\, (1-A)\, \sin^2 \theta_0\, \sin^2 \phi_0\, \cos^2\phi_0}-\frac{1}{r}\right]\, r\, \phi_1 &=& \frac{2\, \cos \theta_0\, \left(B/2+(1-B)\, \sin^2\phi_0\right)}{(1-A)\, \sin^3 \theta_0\, \sin \phi_0\, \cos \phi_0}\, \theta_1 \nonumber \\
    &+& \frac{\left(B/2+(1-B)\, \sin^2\phi_0\right)\, G\, F}{r\, (1-A)^2\, \sin^4\theta_0\, \sin^2 \phi_0\, \cos^2\phi_0}~.
    \label{O(NR)_equation_for_phi_outofplane-1}
\end{eqnarray}
Let us now evaluate the integrating factor for this first order, ordinary differential equation as
\begin{eqnarray}
    J_2(r) &=& \int^{r}\left[\frac{\sin^2 \phi_0-B/2}{r\, (1-A)\, \sin^2 \theta_0\, \sin^2 \phi_0\, \cos^2\phi_0}-\frac{1}{r}\right]\, dr \nonumber \\
    &=& \log \left[r\, \sin \phi_0\, \cos 
    \phi_0\, \sin^2\theta_0\, e^{-Q(r)}\right]~,
\end{eqnarray}
which is correct up to an integration constant. Here we have used expressions in \eqref{O(1)_equation_for_phi_outofplane}, \eqref{O(1)_equation_for_theta_outofplane}, and \eqref{theta0-solution} for the evaluation of the integral. Using this integrating factor, the general solution to \eqref{O(NR)_equation_for_phi_outofplane-1} can then be written as
\begin{eqnarray}
    r\, \phi_1 &=& e^{-J_2(r)}\, \Bigg\{\textrm{Const.}+\int^r\Bigg[\frac{2\, \cos \theta_0'\, \left(B'/2+(1-B')\, \sin^2\phi_0'\right)}{(1-A')\, \sin^3 \theta_0'\, \sin \phi_0'\, \cos \phi_0'}\, \theta_1' \nonumber \\
    && \quad \quad \quad \quad \quad \quad \quad +\frac{\left(B'/2+(1-B')\, \sin^2\phi_0'\right)\, G'\, F'}{r'\, (1-A')^2\, \sin^4\theta_0'\, \sin^2 \phi_0'\, \cos^2\phi_0'}\Bigg]\, e^{J_2(r')}\, dr'\Bigg\} \nonumber \\
    &=& \frac{4\, e^{Q(r)}}{r\, \sin 2\phi_0\, \sin^2 \theta_0}\, \Bigg\{ \textrm{Const.} - \int_{r}^{\infty}\Bigg[\frac{ r'\, \left(B'/2+(1-B')\, \sin^2\phi_0'\right)\, \theta_1'}{(1-A')\, \tan \theta_0'} \nonumber \\
    && \quad \quad \quad \quad \quad \quad \quad \quad +\frac{ \left(B'/2+(1-B')\, \sin^2\phi_0'\right)\, G'\, F'}{(1-A')^2\, \sin^2 \theta_0'\, \sin 2\phi_0'}\Bigg]\, e^{-Q(r')\, dr'}\Bigg\}~.
\end{eqnarray}
Applying the corresponding boundary condition from \eqref{O(NR)_boundary_condition_for_phi_and_theta}, the integration constant for the negative branch is found to be zero. Consequently, the solution in the outer layer $\textrm{O}_1$ becomes
\begin{equation}
    r\, \phi_1^{-} = -\frac{4\, e^{Q(r)}}{r\, \sin 2\phi_0\, \sin^2 \theta_0}\,  \int_{r}^{\infty}\left[\frac{ r'\, \left(B'/2+(1-B')\, \sin^2\phi_0'\right)\, \theta_1'^{-}}{(1-A')\, \tan \theta_0'}+\frac{ \left(B'/2+(1-B')\, \sin^2\phi_0'\right)\, G'\, F'}{(1-A')^2\, \sin^2 \theta_0'\, \sin 2\phi_0'}\right]\, e^{-Q(r')}\, dr'~.
    \label{phi1_solution-1}
\end{equation}
Here, all terms are evaluated with $\phi_0 \in (\pi/2, \pi)$, i.e., along the negative branch. As before, we introduce an integral as in \eqref{MI-definition}, allowing the solution to be expressed compactly as in \eqref{phi1m-solution}. For the positive branch, the integration constant is generally nonzero; we take it as $\overline{x}_2^{-\infty}\times I_{\phi_1}^+$, leading to the solution as
\begin{eqnarray}
    r\, \phi_1^{+} = \frac{2\, e^{Q(r)}}{r\, \sin 2\phi_0\, \sin^2 \theta_0}\,\bigg\{\overline{x}_2^{-\infty}\,I_{\phi_1}^+  &-&2\,\int_{r}^{\infty}\bigg[\frac{ r'\, \left(B'/2+(1-B')\, \sin^2\phi_0'\right)\, \theta_1'^{+}}{(1-A')\, \tan \theta_0'} \nonumber \\
    &+&\frac{ \left(B'/2+(1-B')\, \sin^2\phi_0'\right)\, G'\, F'}{(1-A')^2\, \sin^2 \theta_0'\, \sin 2\phi_0'}\bigg]\, e^{-Q(r')}\, dr' \bigg\}~.
    \label{phi1p_solution-1}
\end{eqnarray}
From \eqref{theta1m-solution} and \eqref{theta1p-solution}, note that $\theta_1^+ = -\theta_1^-+I_{\theta_1}^+\, e^{Q/2}/(r\, \sin \theta_0)$, which can be substituted in \eqref{phi1p_solution-1} to modify it as
\begin{eqnarray}
        r\, \phi_1^{+} &=& \frac{2\, e^{Q(r)}}{r\, \sin 2\phi_0\, \sin^2 \theta_0}\,\Bigg\{
         \overline{x}_2^{-\infty}\,I_{\phi_1}^+ -2\, I_{\theta_1}^+\, \int_{r}^{\infty}\frac{\cos \theta_0'\, \left(B'/2+(1-B')\, \sin^2\phi_0'\right)}{(1-A')\, \sin^2\theta_0'}\, e^{-Q(r')/2}\, dr' \nonumber \\
         &-&2\,\int_{r}^{\infty}\bigg[-\frac{ r'\, \left(B'/2+(1-B')\, \sin^2\phi_0'\right)\, \theta_1'^{-}}{(1-A')\, \tan \theta_0'} 
    +\frac{ \left(B'/2+(1-B')\, \sin^2\phi_0'\right)\, G'\, F'}{(1-A')^2\, \sin^2 \theta_0'\, \sin 2\phi_0'}\bigg]\, e^{-Q(r')}\, dr' 
   \Bigg\}~.
    \label{phi1p_solution-2}
\end{eqnarray}
Here the integral in the first line can be evaluated exactly to $\frac{1}{2}\,\overline{x}_3^{-\infty}\, \sin^2\phi_0$ with the help of expressions \eqref{O(1)_equation_for_phi_outofplane} and \eqref{theta0-solution}, and the integral in second line is related to $M_I(r)$. Thus, the solution \eqref{phi1p_solution-2} can be simplified to the final form as in \eqref{phi1p-solution-outofplane}.
\subsection{Inner layer solution}
After expressing $r$, $\theta$, and $\phi$ in terms of the inner layer expansions, the governing equations \eqref{dphi_and_dtheta_by_dr_equations} can be rewritten in terms of the intermediate variables $\tilde{r}$, $\tilde{\theta}$, and $\tilde{\phi}$. From \eqref{dphidr_equation}, the leading-order equation for $\tilde{\phi}$ then takes the form:
\begin{equation}
    \frac{d \tilde{\phi}}{d \tilde{r}} = \frac{1-B_0/2}{c\, (1-A_0)\, (\sin \theta_t)\, \tilde{\phi}-G_0\, F_0}~.
    \label{layerI-phitilde-diff-eqn}
\end{equation}
The general solution of \eqref{layerI-phitilde-diff-eqn} for $\tilde{\phi}(r)$ can then be readily written as in \eqref{phi_tilde_pm_solution}. To obtain the inner-layer expression for $\tilde{\theta}$, it is convenient to first combine the governing equations \eqref{dphi_and_dtheta_by_dr_equations} in the following manner:
\begin{equation}
    \frac{d \phi}{d \theta} = -\frac{\left[\sin^2 \phi+B/2\, \left(\cos^2 \phi-\sin^2 \phi\right)\right]}{(1-B)\, \sin \theta\, \cos \theta\, \sin \phi\, \sin \phi}~,
\end{equation}
which is independent of $\varepsilon$. Substituting the inner-layer ansatz and taking the limit $\varepsilon \ll 1$, it reduces to the simplified form
\begin{equation}
\tilde{\phi}\, d\tilde{\phi} = \frac{(2-B_0)\, d\tilde{\theta}}{(1-B_0)\, \sin 2\theta_t}~.
\label{phidphi_eqn}
\end{equation}
Since $\tilde{\phi}(r)$ is already known, we can integrate equation \eqref{phidphi_eqn} to obtain $\tilde{\theta}(r)$ explicitly in terms of $\tilde{\phi}(r)$, yielding the solution shown in \eqref{theta_tilde_solution}.
\subsection{Asymptotic matching}
In the outer layer, the leading-order solutions from \eqref{batchelor-solution-off-plane} can be expressed in terms of the matching variables in the limit $\varepsilon \ll 1$ as
\begin{subequations}
    \begin{align}
            \lim_{r \to c_t+\hat{r}\, \varepsilon^{2\alpha}} \theta_0 = \theta_t+\varepsilon^{2\alpha}\, \frac{\hat{r}\, (1-B_0) }{c\, (1-A_0)}\, \cos \theta_t + h.o.t.,
    \label{theta0-matching-expression1}\\
        \lim_{r \to c_t+\hat{r}\, \varepsilon^{2\alpha}} \phi_0^{\mp} = \frac{\pi}{2} \pm \varepsilon^\alpha\, \sqrt{\frac{(2-B_0)\, \hat{r}}{c\, (1-A_0)\, \sin \theta_t}}+h.o.t.
    \label{phi0-matching-expression1}
    \end{align}
\end{subequations}
Here, the branches are chosen such that for the negative branch, $\phi_0 \in (\pi/2, \pi)$, and for the positive branch, $\phi_0 \in (0, \pi/2)$. Note that the leading-order solution $\theta_0$ is independent of the branch, as expected. We now proceed to evaluate the outer-layer $\mathcal{O}(\varepsilon)$ solutions in terms of the matching variables. Starting with \eqref{theta1m-solution}, we substitute $r$ and $\theta_0$ in terms of the matching-variable expressions, and using \eqref{theta0-matching-expression1} and \eqref{phi0-matching-expression1} gives
\begin{eqnarray}
    \lim_{r \to c_t+\hat{r}\, \varepsilon^{2\alpha}} \theta_1^- &=& \frac{e^{Q(c_t)/2}}{c}\, \int_{c_t}^{\infty} \frac{2\, (1-B')\, \cos \theta_0'\, G'\, F'\, e^{-Q'/2}}{r'\, (1-A')^2\, \sin^2\theta_0\, \sin 2\phi_0'}\,dr'-\frac{\varepsilon^{\alpha}\, (1-B_0)\, G_0\, F_0\, \cos \theta_t\, \sqrt{\hat{r}}}{c^{3/2}\, (1-A_0)^{3/2} \sqrt{(2-B_0)\, \sin \theta_t}} \nonumber \\
    &-& \frac{e^{Q(c_t)}}{c^2}\, \frac{\hat{r}\, (1-B_0)\, \varepsilon^{2\alpha}}{(1-A_0)\, \sin \theta_t}\, \int_{c_t}^{\infty} \frac{2\, (1-B')\, \cos \theta_0'\, G'\, F'\, e^{-Q'/2}}{r'\, (1-A')^2\, \sin^2\theta_0\, \sin 2\phi_0'}\, dr'+\mathcal{O}(\varepsilon^{3\alpha})~.
    \label{theta1m-matching-expansion}
\end{eqnarray}
Thus, by combining \eqref{theta0-matching-expression1} and \eqref{theta1m-matching-expansion}, we obtain the $\textrm{O}_1$ layer solution for $\theta$ expressed in terms of the matching variables as
\begin{eqnarray}
    \lim_{r \to c_t+\hat{r}\, \varepsilon^{2\alpha}} \theta^{\textrm{O}_1} &=& \lim_{r \to c_t+\hat{r}\, \varepsilon^{2\alpha}} \left( \theta_0+\varepsilon\, \theta_1^{-}\right) = \theta_t + \varepsilon^{2\alpha}\, \frac{\hat{r}\, (1-B_0)\, \cos\theta_t}{c\, (1-A_0)}-\varepsilon^{1+\alpha}\, \frac{(1-B_0)\, G_0\, F_0\, \cos \theta_t\, \sqrt{\hat{r}}}{c^{3/2}\, (1-A_0)^{3/2}\, \sqrt{(2-B_0)\, \sin \theta_t}} \nonumber \\
    &+& \varepsilon\, \frac{e^{Q(c_t)/2}}{c}\, \int_{c_t}^{\infty}\frac{2\, (1-B')\, \cos \theta_0'\, G'\, F'\, e^{-Q'/2}}{r'\, (1-A')^2\, \sin^2 \theta_0\, \sin 2\phi_0'}\, dr'+h.o.t.,
    \label{theta-solution-outer1-matching-variable-expansion}
\end{eqnarray}
with $\phi_0'\in (0,\pi/2)$ inside the integral. Similarly, by expanding \eqref{phi1m-solution} in terms of the matching variables and combining it with \eqref{phi0-matching-expression1}, we obtain
\begin{eqnarray}
    \lim_{r \to c_t+\hat{r}\, \varepsilon^{2\alpha}} \phi^{\textrm{O}_1} &=& \lim_{r \to c_t+\hat{r}\, \varepsilon^{2\alpha}} \left( \phi_0^- + \varepsilon\, \phi_1^{-}\right) = \frac{\pi}{2}+\varepsilon^{\alpha}\, \sqrt{\frac{(2-B_0)\, \hat{r}}{c\, (1-A_0)\, \sin \theta_t}}+\frac{\varepsilon\, G_0\, F_0}{\textcolor{white}{2}\, c\, (1-A_0)\, \sin \theta_t} \nonumber \\
    &-& \frac{2\, \varepsilon^{1-\alpha}\, e^{Q(c_t)}}{c\, \sqrt{c_t\, \hat{r}}}\, \sqrt{\frac{1-A_0}{1-B_0}}\, \int_{c_t}^{\infty}\bigg[-\frac{r'\,\left(B'/2+(1-B')\, \sin^2\phi_0'\right)\, \theta_1'^{-}}{(1-A')\, \tan \theta_0'} \nonumber \\
    && \quad \quad \quad \quad \quad \quad \quad \quad \quad \quad \quad \quad \quad +\frac{ \left(B'/2+(1-B')\, \sin^2\phi_0'\right)\, G'\, F'}{(1-A')^2\, \sin^2\theta_0'\, \sin 2\phi_0'}\bigg]\,dr' \nonumber \\
    &+&h.o.t. 
    \label{phi-solution-outer1-matching-variable-expansion}
\end{eqnarray}
Here, it should be noted that the last integral term on the RHS is related to $M_I(c_t)$. Using the same procedure with \eqref{phi1p-solution-outofplane} appropriately, we get
\begin{eqnarray}
       \lim_{r \to c_t+\hat{r}\, \varepsilon^{2\alpha}} \phi^{\textrm{O}_2} &=& \lim_{r \to c_t+\hat{r}\, \varepsilon^{2\alpha}} \left( \phi_0^+ + \varepsilon\, \phi_1^{+}\right) = \frac{\pi}{2}-\varepsilon^{\alpha}\, \sqrt{\frac{(2-B_0)\, \hat{r}}{c\, (1-A_0)\, \sin \theta_t}}+\frac{\varepsilon\, G_0\, F_0}{\textcolor{white}{2}\, c\, (1-A_0)\, \sin \theta_t} \nonumber \\
    &+& \frac{\varepsilon^{1-\alpha}\, e^{Q(c_t)}}{c\, \sqrt{c_t\, \hat{r}}}\, \sqrt{\frac{1-A_0}{1-B_0}}\, \left[\overline{x}_2^{-\infty}\, I_{\phi_1}^+-\overline{x}_3^{-\infty}\, I_{\theta_1}^{+}-M_I(c_t)\right] + h.o.t. 
    \label{phi-solution-outer2-matching-variable-expansion}
\end{eqnarray}
Now, let's write the solution in the inner layer in terms of matching variables. As in the case of in-plane trajectories, we have in the inner layer $r = c_t+\varepsilon\, k+\varepsilon^2\, \tilde{r}$, and in terms of the matching variable, $r = c_t + \hat{r}\, \varepsilon^{2\alpha}$. Thus combining both, we get $\tilde{r} = \hat{r}\, \varepsilon^{2\alpha-2}-k\, \varepsilon^{-1}$. Substituting this into \eqref{phi_tilde_pm_solution}, and and expanding for $\varepsilon \ll 1$ with $\alpha >1/2$ (so that $\varepsilon^{2\alpha-2} \gg \varepsilon^{-1} \gg I_i^\textrm{off})$, we get
\begin{eqnarray}
    \lim_{r \to c_t+\hat{r}\, \varepsilon^{2\alpha}} \phi^{I \mp} = \frac{\pi}{2}+\varepsilon\, \tilde{\phi}^{\mp} &=& \frac{\pi}{2}+\frac{\varepsilon\, G_0\, F_0}{c\, (1-A_0)\, \sin \theta_t} \pm \varepsilon^{\alpha}\, \sqrt{\frac{(2-B_0)\, \hat{r}}{c\, (1-A_0)\, \sin \theta_t}} \nonumber \\
    &\mp&\varepsilon^{1-\alpha}\,\frac{k}{2}\, \sqrt{\frac{2-B_0}{\hat{r}\, c\, (1-A_0)\, \sin \theta_t}}+h.o.t.
    \label{phi-solution-inner-matching-variable-expansion}
\end{eqnarray}
By comparing the inner-layer and outer-layer solutions for $\phi^{-}$, i.e., the inner layer -ve branch solution in \eqref{phi-solution-inner-matching-variable-expansion} with the outer layer solution in \eqref{phi-solution-outer1-matching-variable-expansion}, we see that the $\mathcal{O}(\varepsilon^{1-\alpha})$ terms must be matched. This gives
\begin{equation}
    k := k(c_t) =  \frac{2\, (1-A_0)}{c_t\, (2-B_0)}\, M_I(c_t)\, e^{Q(c_t)}~,
    \label{expression-for-k-1}
\end{equation}
where we note that, as mentioned earlier, $A_0$ and $B_0$ correspond to the values of $A(r)$ and $B(r)$ evaluated at $r = c_t$. A similar comparison for the +ve branch solutions of $\phi$, i.e., between $\phi^{I+}$ from \eqref{phi-solution-inner-matching-variable-expansion} and \eqref{phi-solution-outer2-matching-variable-expansion}, yields
\begin{equation}
    k = \frac{2\, e^{Q(c_t)}}{c_t}\, \left(\frac{1-A_0}{2-B_0}\right)\, \left[\overline{x}_2^{-\infty}\, I_{\phi_1}^+-\overline{x}_3^{-\infty}\, I_{\theta_1}^{+}-M_I(c_t)\right]~.
    \label{expression-for-k-2}
\end{equation}
By combining \eqref{expression-for-k-1} and \eqref{expression-for-k-2}, we arrive at \eqref{eqn-I-theta-I-phi-connection}. Since $\theta_0$, unlike $\phi_0$, remains real-valued even for $r$ smaller than $c_t$, the vorticity-direction displacement $\upDelta\overline{x}_3$ can be determined without resorting to the inner-layer solution for $\theta$; that is, it can be obtained directly by matching the limiting expressions in the outer layers $\textrm{O}_1$ and $\textrm{O}_2$, as
\begin{equation}
    \lim_{r \to c_t+\hat{r}\, \varepsilon^{2\alpha}} \theta^{\textrm{O}_1} = \lim_{r \to c_t+\hat{r}\, \varepsilon^{2\alpha}} \theta^{\textrm{O}_2} \Rightarrow \lim_{r \to c_t+\hat{r}\, \varepsilon^{2\alpha}} \theta_0+\varepsilon\, \theta_1^{-} = \lim_{r \to c_t+\hat{r}\, \varepsilon^{2\alpha}} \theta_0+\varepsilon\, \theta_1^{+}~.
\end{equation}
After directly substituting the expressions \eqref{theta1m-solution} and \eqref{theta1p-solution} into this matching equation, and performing the necessary simplifications, we obtain one of the integration constant as
\begin{equation}
    I_{\theta_1}^+ = \lim_{r \to c_t+\hat{r}\, \varepsilon^{2\alpha}} \left(-2\, N_I(r)\right) = -2\, N_I(c_t)~.
\end{equation}
Substituting this solution into \eqref{eqn-I-theta-I-phi-connection}, we can then determine the other integration constant as
\begin{equation}
    I_{\phi_1}^+ = \frac{2\, M_I(c_t)}{\overline{x}_2^{-\infty}}-\frac{2\, \overline{x}_3^{-\infty}}{\overline{x}_2^{-\infty}}\, N_I(c_t)~.
\end{equation}
We now proceed to evaluate the expansion of the polar angle $\theta$ in terms of the matching variable. For the inner layer,
\begin{eqnarray}
    \lim_{r \to c_t+\hat{r}\, \varepsilon^{2\alpha}} \theta^{I^-} &=& \theta_t+\varepsilon\, \theta_f+\varepsilon^2\, \tilde{\theta}^{-} \\
    &=& \theta_t+\varepsilon\, \theta_f+\varepsilon^2\, I_c+\frac{(1-B_0)}{2\, (2-B_0)}\, \sin 2\theta_t\,\bigg\{\frac{G_0^2\, F_0^2\, \varepsilon^2}{c^2\, (1-A_0)^2\, \sin^2 \theta_t}+\frac{(2-B_0)\, \hat{r}\, \varepsilon^{2\alpha}}{c\, (1-A_0)\, \sin \theta_t} \nonumber \\
    &-&\frac{k^2\, (2-B_0)\, \varepsilon^{2-2\alpha}}{4\, c\, \hat{r}\, (1-A_0)\, \sin \theta_t}+\frac{\varepsilon^{1+\alpha}\, 2\, G_0\, F_0\, \sqrt{(2-B_0)\, \hat{r}}}{\left(c\, (1-A_0)\, \sin \theta_t\right)^{3/2}}-\frac{\varepsilon\, k\, (2-B_0)}{c\, (1-A_0)\, \sin \theta_t} \\
    &-&\frac{\varepsilon^{2-\alpha}\, k\, G_0\, F_0\, \sqrt{2-B_0}}{\sqrt{\hat{r}}\, \left(c\, (1-A_0)\, \sin \theta_t\right)^{3/2}}\bigg\} + h.o.t.,
    \end{eqnarray}
where, we have employed the expansion of $\tilde{\phi}^{-}$, as used in \eqref{phi-solution-inner-matching-variable-expansion}. By matching this inner-layer expansion with the outer-layer expansion of $\theta^{-}$ from \eqref{theta-solution-outer1-matching-variable-expansion}, we can determine the unknown constant $\theta_f$:
\begin{eqnarray}
    \lim_{r \to c_t+\hat{r}\, \varepsilon^{2\alpha}} \theta^{\textrm{O}_1} &=& \lim_{r \to c_t+\hat{r}\, \varepsilon^{2\alpha}} \theta^{I^-} \nonumber \\
    \Rightarrow \varepsilon\, \left[\theta_f-\frac{(1-B_0)\, \sin 2\theta_t}{2\, (2-B_0)}\, \frac{k\, (2-B_0)}{c\, (1-A_0)\, \sin \theta_t}\right] &=& -\varepsilon\, \frac{e^{Q(c_t)/2}}{c}\, N_I(c_t) \nonumber \\
    \Rightarrow \theta_f := \theta_f(c,\theta_t^c) &=& \frac{k\, (1-B_0)}{c\, (1-A_0)}\, \cos \theta_t -\frac{e^{Q(c_t)/2}}{c}\, N_I(c_t)~.
\end{eqnarray}
Similarly, as a special case, we can also write
\begin{equation}
    \theta_f^* := \theta_f(d,\theta_t^d) = \frac{k^*\, (1-B^*)}{d\, (1-A^*)}\, \cos \theta_t^d -\frac{e^{Q^*/2}}{d}\, N_I(d_t)~.
    \label{theta_f_star}
\end{equation}
\section{Off-plane open trajectories with gradient displacement $\mathcal{O}(\varepsilon^{1/2})$ or smaller}
\label{off-plane-closer-trajs}
\begin{figure}
    \centering
\includegraphics[width=1\linewidth]{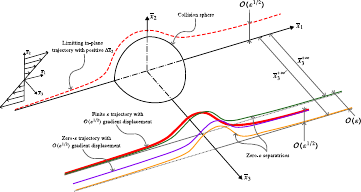}
\caption{Schematic of an off-plane trajectory with $\mathcal{O}(\varepsilon^{1/2})$ gradient direction displacement and $\mathcal{O}(\varepsilon)$ vorticity direction displacement (red, solid, thick line). The violet curve shows the corresponding zero-$\varepsilon$ trajectory with the same far-downstream offset. The green curve denotes the zero-$\varepsilon$ limiting trajectory sharing the same far-upstream vorticity location as the red trajectory, while the orange curve represents the zero-$\varepsilon$ limiting trajectory with the same downstream vorticity location. The red dashed curve is shown for reference and corresponds to the in-plane limiting trajectory for $\varepsilon > 0$.}
    \label{fig:offplane-schematic2}
\end{figure}

Analogous to the in-plane trajectories, for off-plane trajectories with gradient displacements of order $\mathcal{O}(\varepsilon^{1/2})$ or smaller, we match the $-\text{ve}$ branch ($\phi \in (\pi/2, \pi)$) and $+\text{ve}$ branch ($\phi \in (0, \pi/2)$) of the $\varepsilon>0$ trajectory at $\phi = \pi/2$. In figure \ref{fig:offplane-schematic2}, the solid, thick red curve represents a $\varepsilon>0$ trajectory that originates from a zero-$\varepsilon$ trajectory with zero initial gradient offset (not shown) and terminates on another zero-$\varepsilon$ trajectory (violet), which has a gradient direction displacement of order $\mathcal{O}(\varepsilon^{1/2})$. This violet zero-$\varepsilon$ trajectory is characterized by parameters $(\tilde{c}, \theta_t^{\tilde{c}})$ and lies close to a zero-$\varepsilon$ separatrix with parameters $(d', \theta_t^{d'})$. The corresponding zero-$\varepsilon$ trajectory from which the $\varepsilon>0$ trajectory originates has parameters $(c, \theta_t^c)$, with its nearest zero-$\varepsilon$ separatrix having parameters $(d, \theta_t^d)$. The far-upstream off-plane coordinates of these separatrices (orange and green in the figure) are denoted by $\overline{x}_3^{-\infty'}$ and $\overline{x}_3^{-\infty}$, which coincide with the far-downstream coordinates for the zero-$\varepsilon$ case. It is important to note that the parameters defining the zero-$\varepsilon$ separatrices are not independent, as we will show later in equations \eqref{d-prime-square} and \eqref{d-square}. For finite-$\varepsilon$ trajectories, $\overline{x}_3^{-\infty'}$ is larger than $\overline{x}_3^{-\infty}$ by $\mathcal{O}(\varepsilon)$, since $\overline{x}_3^{+\infty'} = \overline{x}_3^{+\infty} + \varepsilon \, (\upDelta \hat{x}_3)$, which will be verified later. Applying equation \eqref{batchelor-solution-off-plane} appropriately for each trajectory scenario described above, we have:  
\begin{subequations}
\begin{align} 
d' &= \overline{x}_3^{-\infty'}\, e^{Q(d_t')/2}\, \tan \theta_t^{d'}, \quad d'^2 = e^{Q(d_t')}\, \int_{d_t'}^{\infty} \frac{B'\, r'\, e^{-Q'}}{1-A'}\, dr'~, \label{d-prime-square}\\
d &= \overline{x}_3^{-\infty}\, e^{Q(d_t)/2}\, \tan \theta_t^{d}, \quad d^2 = e^{Q(d_t)}\, \int_{d_t}^{\infty} \frac{B'\, r'\, e^{-Q'}}{1-A'}\, dr'~, \label{d-square}\\
c &= \overline{x}_3^{-\infty}\, e^{Q(c_t)/2}\, \tan \theta_t^{c}, \quad c^2 = e^{Q(c_t)}\, \left[m_1^2\, \varepsilon+\int_{c_t}^{\infty} \frac{B'\, r'\, e^{-Q'}}{1-A'}\, dr'\right]~,\label{c-square}\\
\tilde{c} &= \overline{x}_3^{-\infty'}\, e^{Q(c_t')/2}\, \tan \theta_t^{\tilde{c}}, \quad \tilde{c}^2 = e^{Q(c_t')}\, \left[(m_1+m_1')^2\, \varepsilon+\int_{c_t'}^{\infty} \frac{B'\, r'\, e^{-Q'}}{1-A'}\, dr'\right]~.
\label{c-prime-square}
\end{align}
\label{d1dcctilde}
\end{subequations}
For instance, considering the trajectory with $r = d_t'$, we have $\phi_0 = \pi/2$ and $\overline{x}_2^{-\infty} = 0$. Here, we adopt the shorthand notations: $d_t' = d'/\sin \theta_t^{d'}$, $d_t = d/\sin \theta_t^d$, $c_t = c/\sin \theta_t^c$, and $c_t' = \tilde{c}/\sin \theta_t^{\tilde{c}}$. In equation \eqref{c-square}, the zero-$\varepsilon$ trajectory characterized by $(c, \theta_t^c)$ has a far-upstream (and thus far-downstream) gradient offset $\overline{x}_2^{-\infty} = m_1\, \varepsilon^{1/2}$. Anticipating the same $\mathcal{O}(\varepsilon^{1/2})$ scaling with $\varepsilon$ for the $(\tilde{c}, \theta_t^{\tilde{c}})$ trajectory, we take $\overline{x}_2^{-\infty'} = (m_1 + m_1')\, \varepsilon^{1/2}$. The goal is to determine $m_1'$ in terms of $m_1$, $d$, and $\theta_t^d$, so that the corresponding gradient displacement can be expressed as $\upDelta \overline{x}_2 = m_1'\, \varepsilon^{1/2}$. Because of the $\mathcal{O}(\varepsilon)$ perturbation, the preceding relations in \eqref{d1dcctilde} then are expected to yield:
\begin{eqnarray}
    d' &=& d + \varepsilon\, a_1~, \quad \theta_t^{d'} = \theta_t^d+\varepsilon\, a_2~, \nonumber \\
    c &=& d+\varepsilon\, b_1~, \quad \theta_t^c = \theta_t^d+\varepsilon\, b_2~, \nonumber \\
    \tilde{c} &=& d+\varepsilon\, b_1'~, \quad \theta_t^{\tilde{c}} = \theta_t^d+\varepsilon\, b_2'~,
    \label{expansion-ansatz}
\end{eqnarray}
for some $\mathcal{O}(1)$ constants $a_1$, $b_1$, $a_2$, $b_2$, $b_1'$, and $b_2'$. From these relations, we can also deduce that $\tilde{c} = d' + \varepsilon \, (b_1' - a_1)$ and $\theta_t^{\tilde{c}} = \theta_t^{d'} + \varepsilon \, (b_2' - a_2)$. Accordingly, the short-hand parameters can be represented as
\begin{eqnarray}
    d_t' &=& d_t+\varepsilon\, d_t\, \left(\frac{a_1}{d}-\frac{a_2}{\tan \theta_t^d}\right)~,\\
    c_t &=& d_t+\varepsilon\, d_t\, \left(\frac{b_1}{d}-\frac{b_2}{\tan \theta_t^d}\right)~,\\
    c_t' &=& d_t+\varepsilon\, d_t\, \left(\frac{b_1'}{d}-\frac{b_2'}{\tan \theta_t^d}\right)= d_t' +\varepsilon\, d_t'\, \left(\frac{b_1'-a_1}{d}-\frac{b_2'-a_2}{\tan \theta_t^d}\right)~.
\end{eqnarray}
By substituting these ansatz into the right-set of equations in \eqref{d-prime-square}, \eqref{c-square}, and \eqref{c-prime-square}, together with \eqref{d-square}, we obtain
\begin{subequations}
    \begin{align}
        a_1 &= -\frac{d}{2\, (1-A^*)}\, \left(\frac{a_1}{d}-\frac{a_2}{\tan \theta_t^d}\right)\, \left[2\, A^*+\frac{B^*\, \cos 2\theta_t^d}{\sin^2 \theta_t^d}\right]~,\label{a1-eqn-1}\\
        b_1 &= \frac{m_1^2\, e^{Q^*}}{2\, d}-\frac{d}{2\, (1-A^*)}\, \left(\frac{b_1}{d}-\frac{b_2}{\tan \theta_t^d}\right)\, \left[2\, A^*+\frac{B^*\, \cos 2\theta_t^d}{\sin^2 \theta_t^d}\right]~,\label{b1-eqn-1}\\
        b_1'-a_1 &= \frac{(m_1+m_1')^2\, e^{Q^*}}{2\, d}-\frac{d}{2\, (1-A^*)}\, \left(\frac{b_1'-a_1}{d}-\frac{b_2'-a_2}{\tan \theta_t^d}\right)\, \left[2\, A^*+\frac{B^*\, \cos 2\theta_t^d}{\sin^2 \theta_t^d}\right]~,\label{b1p-a1-eqn-1}
    \end{align}
\end{subequations}
where the superscript `*' indicates that $A(r)$, $B(r)$, and $Q(r)$ are evaluated at $r = d_t$. Likewise, applying the ansatz to the left-set of equations in \eqref{d-square}, \eqref{c-square}, and \eqref{c-prime-square} also yields
 \begin{subequations}
     \begin{align}
              b_1 &= \frac{2\, b_2\, d}{\sin 2\theta_t^d}-d\, \left(\frac{A^*-B^*}{1-A^*}\right)\, \left(\frac{b_1}{d}-\frac{b_2}{\tan \theta_t^d}\right)~, \label{b1-eqn-2}\\
              b_1'-a_1 &= \frac{2\, (b_2'-a_2)\, d}{\sin 2\theta_t^d}-d\, \left(\frac{A^*-B^*}{1-A^*}\right)\, \left(\frac{b_1'-a_1}{d}-\frac{b_2'-a_2}{\tan \theta_t^d}\right)~. \label{b1p-a1-eqn-2}
     \end{align}
 \end{subequations}
Additionally, applying the ansatz $\overline{x}_3^{-\infty'} = \overline{x}_3^{-\infty} + \varepsilon \, \left(\upDelta \hat{x}_3\right)$ to the left-set of equations in \eqref{d-prime-square} and \eqref{d-square} gives
\begin{equation}
    \upDelta\hat{x}_3 = \overline{x}_3^{-\infty}\, \left[\left(\frac{1-B^*}{1-A^*}\right)\, \left(\frac{a_1}{d}-\frac{a_2}{\tan \theta_t^d}\right)-a_2\, \tan \theta_t^d\right]~.
\end{equation}
By solving \eqref{a1-eqn-1} together with the first equation in \eqref{c-square}, we obtain expressions for $a_1$ and $a_2$ as 
\begin{subequations}
    \begin{align}
        a_1 &= -(\upDelta\hat{x}_3)\, \frac{\left[A^*-(A^*-B^*)\, \cos 2\theta_t^d\right]\, e^{Q^*/2}}{(2-B^*)\, \tan \theta_t^d}~,\\
        a_2 &= -(\upDelta\hat{x}_3)\, \frac{\left[1-(1-B^*)\, \cos 2\theta_t^d\right]\, e^{Q^*/2}}{(2-B^*)\, d}~. 
        \label{solutions-a2}
    \end{align}
    \label{solutions-a1-a2}
\end{subequations}
Similarly, by equating \eqref{b1-eqn-1} with \eqref{b1-eqn-2}, and \eqref{b1p-a1-eqn-1} with \eqref{b1p-a1-eqn-2}, other unknown parameters can be obtained as
\begin{subequations}
    \begin{align}
        b_1 &= \frac{m_1^2\, e^{Q^*}}{2\, d}\, \frac{\left[2-(A^*+B^*)+(A^*-B^*)\, \cos 2\theta_t^d\right]}{(2-B^*)}~,\\
        b_2 &= \frac{m_1^2\, e^{Q^*}}{2\, d^2}\, \left(\frac{1-B^*}{2-B^*}\right)\,\sin 2\theta_t^d~,\label{b2-equation-solution}\\
        b_1' &= a_1+\frac{(m_1+m_1')^2\, e^{Q^*}}{2\, d}\, \frac{\left[2-(A^*+B^*)+(A^*-B^*)\, \cos 2\theta_t^d\right]}{(2-B^*)}~, \\
        b_2' &= a_2+\frac{(m_1+m_1')^2\, e^{Q^*}}{2\, d^2}\, \left(\frac{1-B^*}{2-B^*}\right)\,\sin 2\theta_t^d~. 
        \label{b2p-equation-solution}
    \end{align}
\end{subequations}
Now, we match the negative and positive branches of the $\varepsilon > 0$ trajectories at $\phi = \pi/2$, i.e., imposing $r_{\pi/2}^{-} = r_{\pi/2}^{+}$, where the LHS corresponds to the perturbed $(c, \theta_t^c)$ branch and the RHS is perturbed about $(\tilde{c}, \theta_t^{\tilde{c}})$. Accounting for the respective expressions, this condition leads to the following matching equation:
\begin{equation}
c_t+\varepsilon\, k(c,\theta_t^c) = c_t'-\varepsilon\, k(\tilde{c},\theta_t^{\tilde{c}}) \Rightarrow c_t' = c_t + 2\, \varepsilon\, k(d,\theta_t^d) + \textit{o}(\varepsilon)~.
\end{equation}
Using the left-set of equations from \eqref{c-square} and \eqref{c-prime-square}, the matching condition can be rewritten as
\begin{equation}
    \frac{\overline{x}_3^{-\infty'}\, e^{Q(c_t')/2}}{\cos \theta_t^{\tilde{c}}} = \frac{\overline{x}_3^{-\infty}\, e^{Q(c_t)/2}}{\cos \theta_t^{c}}+2\, \varepsilon\, k(d,\theta_t^d)~,
\end{equation}
which is correct up to $\textit{o}(\varepsilon)$. Using again the ansatz $\overline{x}_3^{-\infty'} = \overline{x}_3^{-\infty} + \varepsilon\, \left( \upDelta \hat{x}_3\right)$, and the expansions for $\theta_t^{\tilde{c}}$, $\theta_t^c$, $c_t$ and $c_t'$ about $d_t$, we can solve for the $\mathcal{O}(\varepsilon)$ constant as
 \begin{eqnarray}
          d_t\, \left[\frac{\upDelta \hat{x}_3\, e^{Q^*/2}}{d_t\, \cos \theta_t^d}+(b_2'-b_2)\, \tan \theta_t^d+\left(\frac{A^*-B^*}{1-A^*}\right)\, \left(\frac{b_1-b_1'}{d}-\frac{b_2-b_2'}{\tan \theta_t^d}\right)\right] &=& 2\, k^* ~,\nonumber \\
   \Rightarrow \frac{\upDelta \hat{x}_3\, e^{Q^*/2}}{\cos \theta_t^d}  +\frac{d_t\, \tan \theta_t^d}{d\, (1-B^*)}\, \left[(1-A^*)\, (b_2'-b_2)\, d-(A^*-B^*)\, (\upDelta \hat{x}_3)\, e^{Q^*/2}\right] &=& 2\, k^*~,
 \end{eqnarray}
where $k^* = k(d,\theta_t^d)$), and this can be simplified to the same form as previously obtained in \eqref{expression-for-k-1} as
 \begin{equation}
     k^* := k(d,\theta_t^d) =  \frac{2\, (1-A^*)}{d_t\, (2-B^*)}\, M_I(d_t)\, e^{Q(d_t)}~,
 \end{equation}
From \eqref{b1-eqn-2} and \eqref{b1p-a1-eqn-2}, by substituting $b_1$ and $b_1'$ in terms of $a_1$, $a_2$,$b_2$, and $b_2'$, and then using the expressions for $a_1$ and $a_2$ from \eqref{solutions-a1-a2}, we obtain
 \begin{equation}
     (b_2'-b_2)\, d+(\upDelta \hat{x}_3)\, e^{Q^*/2} = 2\, k^*\, \left(\frac{1-B^*}{1-A^*} \right)\, \cos \theta_t^d~.
     \label{b2p-b2-eqn-1}
 \end{equation}
Similarly, by matching the polar angle coordinates of the +ve and -ve branches, we get another relation as
\begin{eqnarray}
    \theta_{\pi/2}^{-} = \theta_{\pi/2}^{+} & \Rightarrow & \theta_t^c +\varepsilon\, \theta_f(c,\theta_t^c) = \theta_t^{\tilde{c}} -\varepsilon\, \theta_f(\tilde{c},\theta_t^{\tilde{c}}) \nonumber\\
    & \Rightarrow  &\theta_t^{\tilde{c}} = \theta_t^c +2\, \varepsilon\, \theta_f(d,\theta_t^d)+\textit{o}(\varepsilon)~.
\end{eqnarray}
Using the expansion ansatz in \eqref{expansion-ansatz}, this simplifies to $b_2'-b_2 = 2\, \theta_f(d,\theta_t^d)$, correct up to $\textit{o}(\varepsilon)$. Combining this result with \eqref{b2p-b2-eqn-1} and \eqref{theta_f_star}, we can further simplify to obtain $\upDelta \hat{x}_3$ as
\begin{equation}
    \upDelta \hat{x}_3 = 2\,  N_I(d_t)~, 
\end{equation}
This can then be transformed back to the original coordinates to obtain $\upDelta \overline{x}_3=\varepsilon\, \upDelta \hat{x}_3$, as in \eqref{Deltax3-off-plane-2}. Now, to determine $\upDelta \overline{x}_2$, we rewrite \eqref{b2p-b2-eqn-1} to express $b_2'$, which contains the unknown $m_1'$, in terms of the other known quantities as
\begin{equation}
       b_2' = b_2 -\frac{\upDelta \hat{x}_3}{d}\, e^{Q^*/2}+\frac{2\, k^*}{d}\, \left(\frac{1-B^*}{1-A^*} \right)\, \cos \theta_t^d~.
\end{equation}
By substituting \eqref{b2-equation-solution} and \eqref{b2p-equation-solution} into this expression, and then using the left equation in \eqref{d-square}, along with \eqref{solutions-a2}, we obtain a quadratic equation for the unknown $m_1'$ as
\begin{eqnarray}
    m_1'^2+2\, m_1\, m_1' &=& 2\, d_t\, k^*\, \left(\frac{2-B^*}{1-A^*}\right)\, e^{-Q^*}-2\, (\upDelta \hat{x}_3)\,\overline{x}_3^{-\infty}  \nonumber \\
    &=& 4\, M_I(d_t)-2\,(\upDelta \hat{x}_3)\,\overline{x}_3^{-\infty}~. 
\end{eqnarray}
Solving for $m_1'$ gives $m_1' = -m_1+\sqrt{m_1^2+4\, M_I(d_t)-2\, (\upDelta \hat{x}_3)\,\overline{x}_3^{-\infty}}$. Using $\upDelta \overline{x}_2 = m_1'\, \varepsilon^{1/2}$, and $\upDelta \overline{x}_3 = \varepsilon\, \upDelta \hat{x}_3$ we get the final expression for $\upDelta \overline{x}_2$ as in \eqref{Deltax2-off-plane-2}. For the limiting trajectory, $m_1$ should be zero, so that $\upDelta \overline{x}_2 = m_1^c\, \varepsilon^{1/2}$, where $m_1^c = m_1' = \sqrt{4\, M_I(d_t)-2\, (\upDelta \overline{x}_3)\, \overline{x}_3^{-\infty}}$. This expression also reduces to the in-plane result we got when $\overline{x}_3^{-\infty} = 0$. 

\bibliographystyle{jfm}
\bibliography{jfm}






\end{document}